%% file: continuity_constraints.tex
\documentclass[11pt]{article}

\pdfoutput=1

\usepackage{verbatim,amsmath,amssymb}
\usepackage{epsfig,float,color}
\usepackage{geometry}
\usepackage{setspace}
\usepackage[export]{adjustbox}
\usepackage{subfig}
\usepackage{natbib}
\usepackage{placeins}
\usepackage{amsthm,mathrsfs,amsfonts,dsfont} 
\usepackage{tikz}
\usetikzlibrary{arrows,matrix,positioning,angles,quotes}
\usetikzlibrary{calc,math}
\usepackage{nicefrac}
\usepackage{pgfplots}
\pgfplotsset{compat=newest}
\usepackage{framed}
\usepackage[abs]{overpic}
\usepackage{tabularx}
\usepackage{scrextend}

\usepackage{enumitem}
\usepackage{longtable}
\usepackage{boldline}
\newlist{introlist}{itemize}{1}
\setlist[introlist]{label=\textbullet,
                 nosep, wide, leftmargin=*,
                 before=\vspace*{-\dimexpr\baselineskip},
                 after =\vspace*{-\dimexpr\baselineskip}
                 }
\newcommand\pro{\item[$+$]}
\newcommand\con{\item[$-$]}

\newcolumntype{L}[1]{>{\raggedright\arraybackslash}p{#1}}
\newcolumntype{C}[1]{>{\centering\arraybackslash}p{#1}}
\newcolumntype{R}[1]{>{\raggedleft\arraybackslash}p{#1}}
\newcolumntype{P}[1]{>{\raggedright\let\newline\\\arraybackslash\hspace{0pt}}p{#1}}

\usepackage{mathtools}
\DeclarePairedDelimiterX{\normadjust}[1]{\lVert}{\rVert}{#1}

\newcommand\tightdots{\hbox to 1em{.\hss.\hss.}}

\definecolor{dblue2}{rgb}{0,0.75,0.9}
\usepackage{tcolorbox}
\usepackage{adjustbox}

\definecolor{darkblue}{rgb}{0,0,1}
\definecolor{dgreen}{rgb}{0,0.5,0}

\usepackage{hyperref}
\hypersetup{colorlinks=true, breaklinks=true, linkcolor=darkblue, menucolor=darkblue,  citecolor=darkblue, urlcolor=darkblue}

\input{neco.tex}

\begin{document}

\begin{center}
\Large{\bf{Isogeometric continuity constraints for multi-patch shells governed by fourth-order deformation and phase field models}}\\
\end{center}

\begin{center}
\large{Karsten Paul$^\ast$, Christopher Zimmermann$^\ast$, Thang X. Duong$^\ast$, Roger A. Sauer$^{\ast,\dagger}$\footnote{corresponding author, email: sauer@aices.rwth-aachen.de}}
\vspace{4mm}

\small{\textit{$^\ast$Aachen Institute for Advanced Study in Computational Engineering Science (AICES), \\ RWTH Aachen University, Templergraben 55, 52062 Aachen, Germany}}

\small{\textit{$^\dagger$Department of Mechanical Engineering, Indian Institute of Technology Kanpur, UP 208016, India}}
\vspace{4mm}

Published\footnote{This pdf is the personal version of an article whose final publication is available at \href{https://www.sciencedirect.com/science/article/pii/S0045782520304047?dgcid=rss_sd_all}{www.sciencedirect.com}.} 
in \textit{Comput. Methods Appl. Mech. Eng.}, 
\href{https://www.sciencedirect.com/science/article/pii/S0045782520304047?dgcid=rss_sd_all}{DOI: 10.1016/j.cma.2020.113219} \\
Submitted on 5.~March 2020, Revised on 10.~June 2020, Accepted on 14.~June 2020 

\end{center}

\vspace{3mm}

\rule{\linewidth}{.15mm}
{\bf Abstract}\\
This work presents numerical techniques to enforce continuity constraints on multi-patch surfaces for three distinct problem classes. The first involves structural analysis of thin shells that are described by general Kirchhoff-Love kinematics. Their governing equation is a vector-valued, fourth-order, nonlinear, partial differential equation (PDE) that requires at least $C^1$-continuity within a displacement-based finite element formulation. The second class are surface phase separations modeled by a phase field. Their governing equation is the Cahn-Hilliard equation – a scalar, fourth-order, nonlinear PDE – that can be coupled to the thin shell PDE. The third class are brittle fracture processes modeled by a phase field approach. In this work, these are described by a scalar, fourth-order, nonlinear PDE that is similar to the Cahn-Hilliard equation and is also coupled to the thin shell PDE. Using a direct finite element discretization, the two phase field equations also require at least a $C^1$-continuous formulation. Isogeometric surface discretizations – often composed of multiple patches – thus require constraints that enforce the $C^1$-continuity of displacement and phase field. For this, two numerical strategies are presented: A Lagrange multiplier formulation and a penalty method. The curvilinear shell model including the geometrical constraints is taken from \cite{duong2017} and it is extended to model the coupled phase field problems on thin shells of \cite{zimmermann2019} and \cite{paul2020} on multi-patches. Their accuracy and convergence are illustrated by several numerical examples considering deforming shells, phase separations on evolving surfaces, and dynamic brittle fracture of thin shells.

{\bf Keywords:} Isogeometric analysis, multi-patch discretization, Kirchhoff-Love shells, phase field methods, Cahn-Hilliard equation, brittle fracture

\vspace{-4mm}
\rule{\linewidth}{.15mm}

\section{Introduction} \label{s:intro}
Thin-walled structures commonly appear in engineering design since they combine the advantages of low weight and high strength. Kirchhoff-Love (KL) theory is a suitable choice to model these, especially if the slenderness ratio is high. This theory contains higher order derivatives, so that the geometric discretization requires higher continuity than standard finite elements. Likewise, diffusion problems are also often described with higher order operators, which necessitates a higher continuous approximation space for the test and trial functions. The same is true for higher-order fracture models. Isogeometric analysis (IGA) offers the possibility of high smoothness in the geometry and solution. IGA works simplest on a single patch of elements that is discretized with Non-Uniform Rational B-Splines (NURBS). But in real-world problems, single patches are often not sufficient to represent arbitrarily complex shapes or topologies. Instead, multi-patch descriptions are used to discretize these geometries, but the higher continuity is not automatically preserved across the patch interfaces. This work presents a general framework to model coupled problems, in which diffusion and fracture processes take place on deforming thin shells, using multi-patch NURBS discretizations and corresponding patch constraints. The latter enforce the continuity of the surface normal and phase field gradient in order to accurately transfer stresses, moments, mass fluxes and damage gradients at patch interfaces. 

In IGA, introduced by \cite{hughes2005}, splines are used to describe the geometry and the solution. This offers the possibility of discretizations with high continuity. Several models have been proposed to describe isogeometric KL shells, starting with \cite{kiendl2009}. KL shell theory only uses displacement degrees-of-freedom (dofs), which distinguishes it from thick shell theories, such as the Reissner-Mindlin shell theory. The pure displacement formulation results in a partial differential equation (PDE) that contains fourth-order derivatives. The required $C^1$-continuity in a Galerkin-type finite element formulation is then obtained by means of isogeometric shape functions.

The current work studies two coupled models on deforming thin shells. These models represent two different physical processes but their resulting finite element formulation exhibits many similarities.\\
The first model is a phase field formulation for phase separations based on the Cahn-Hilliard theory \citep{cahn1958,cahn1961}. The theory of coupling in-plane phase transitions and surface deformations is taken from \cite{sahu2017}. Phase separation in shells occurs in a wide range of applications, including chemical, biological, thermo- and electro-mechanical problems. For instance, lipid membranes may separate into two immiscible phases, often linked to the formation of rafts. The latter are assumed to play a crucial role in the regulation of protein activity, which might affect biological processes such as signaling and trafficking, see \cite{elson2010} for a review on theoretical and experimental models. Phase transformations in electro-mechanical devices and the resulting change of kinematics and mechanical behavior are for instance investigated by \cite{tang2010} and \cite{ebner2013}.\\
The second model investigates brittle fracture based on an adaptive phase field framework. Here, the stored elastic energy can be seen as a driving force for crack evolution\footnote{Note that this viewpoint is questionable, see \cite{gerasimov2019}.} and material parameters are degraded in regions of fracture. As for the Cahn-Hilliard theory, this model resembles a diffuse interface model in which the transition zone between different phases is smeared out. The foundation for brittle fracture has been established by \cite{griffith1921}, reformulated by \cite{francfort1998} and first implemented within a finite element method by \cite{bourdin2000}. The structural shell formulation for both models is taken from the work of \cite{duong2017} and the individual coupling is described in detail in \cite{zimmermann2019} and \cite{paul2020}. Both models lead to a coupled formulation of two nonlinear fourth-order PDEs that are defined on an evolving two-dimensional manifold. Those generally require continuity constraints for multi-patch discretizations. In \tabref{t:intro}, existing techniques for patch coupling in isogeometric analysis are listed together with their main properties and references.\footnote{It is noted that the choice of categories reflects the author's interpretation, which is not unique. Categories can in principle be further subdivided, added or merged.}
\pagebreak
\begin{center}
	\begin{longtable}{ L{0.17\textwidth} L{0.37\textwidth} L{0.38\textwidth} }
	\caption{Existing techniques for patch coupling in isogeometric analysis.}\label{t:intro}\\
		\textbf{Technique} & \textbf{Properties} & \textbf{References}\tabularnewline*\hlineB{3} 
			Penalty method & \begin{introlist}
				\pro No additional dofs
				\pro Easy implementation
				\pro Non-conforming meshes
				\con Possible ill-conditioning
				\con Constraint violated
				\con Needs user-specified parameter
			\end{introlist} \vspace*{-\baselineskip}
				& \cite{apostolatos2014,apostolatos2019}, \cite{lei2015}, \cite{duong2017}, \cite{horger2019}, \cite{herrema2019}, \cite{leidinger2019}, \cite{bauer2020} \tabularnewline \hlineB{1}
		Bending strip method & \begin{introlist}
					\pro Easy implementation
					\con Possible ill-conditioning
					\con Only conforming meshes
			\end{introlist}
				& \cite{kiendl2010}, \cite{goyal2017} \tabularnewline\hlineB{1}
		Lagrange multiplier method & \begin{introlist}
					\pro Non-conforming meshes
					\pro No user-specified parameter
					\con Additional dofs
					\con Need to ensure LBB-stability
			\end{introlist}
				& \cite{dornisch2011}, \cite{apostolatos2014,apostolatos2019}, \cite{bouclier2016}, \cite{duong2017}, \cite{sommerwerk2017}, \cite{mi2018} \tabularnewline\hlineB{1}
		Primal mortar method & \begin{introlist}
					\pro Non-conforming meshes
					\pro Increased robustness due to averaging
					\con Additional dofs
					\con Need to ensure LBB-stability
			\end{introlist}
				&  \cite{hesch2012}, \cite{brivadis2015}, \cite{bouclier2017}, \cite{dittmann2019,dittmann2020}, \cite{hirschler2019a,hirschler2019b}, \cite{horger2019}, \cite{schuss2019} \tabularnewline\hlineB{1}
		Dual mortar method & \begin{introlist}
					\pro Non-conforming meshes
					\pro Increased robustness due to averaging
					\pro No additional dofs
					\con Need to ensure LBB-stability
			\end{introlist}
				&  \cite{dornisch2015,dornisch2017}, \cite{zou2018}, \cite{wunderlich2019} \tabularnewline\hlineB{1}
		Nitsche's method & \begin{introlist}
					\pro Non-conforming meshes
					\pro Variational consistency
					\pro Stiffness matrix well conditioned and semi-positive-definite
					\pro No additional dofs			
					\con Formulation for general problems difficult
			\end{introlist}
				& \cite{nguyen2014}, \cite{apostolatos2014,apostolatos2019}, \cite{ruess2014}, \cite{du2015}, \cite{guo2015a}, \cite{bouclier2017}, \cite{nguyenthanh2017}, \cite{zhao2017}, \cite{bouclier2018}, \cite{gu2018}, \cite{hu2018}, \cite{du2019}, \cite{liu2019}, \cite{yin2019} \tabularnewline \hlineB{1}
		Direct elimination & \begin{introlist}
					\pro Fewer dofs
					\con Large implementation effort
			\end{introlist}
				& \cite{lei2015}, \cite{coox2017a,coox2017b}, \cite{duong2017} \tabularnewline \hlineB{1}
		Continuous multi-patch discretization & \begin{introlist}
					\pro Optimal approximation properties
					\con Construction difficult
			\end{introlist}
				& \cite{kapl2015,kapl2017a}, \cite{collin2016}, \cite{bracco2019}, \cite{kapl2020}
	\end{longtable}
\end{center} 
Penalty and Lagrange multiplier methods can be derived as special cases of mortar methods. But since they can also be (and have been in the past) derived without introducing the averaging notion of mortar methods, they are listed separately. The dual mortar method constitutes a particularly attractive version that allows to eliminate the unknown Lagrange multipliers through the use of so-called dual mortar shape functions, see also \cite{wohlmuth2000} and \cite{seitz2016}.

There are also references that lie outside the categories of \tabref{t:intro}. \cite{kiendl2009} construct $G^1$-continuous KL shell parametrizations by coupling the first and second rows of control points at patch interfaces with each other. This method is restricted to conforming meshes and smooth interfaces. \cite{beiraodaveiga2011} study the approximation properties of T-splines on non-conforming two-patch geometries. The patches are coupled by inserting knots at the interface and overlapping the underlying T-meshes. 

Cahn-Hilliard phase transitions on fixed multi-patch domains have recently been investigated by \cite{chan2018} and \cite{dittmann2020}. In the present work, this is exceeded by considering deforming multi-patch domains. \cite{chan2018} also propose a method to overcome $C^1$-locking by local degree elevation along patch interfaces. $C^1$-locking refers to restricted convergence rates due to overconstraining of the solution space \citep{collin2016}. Their work is further advanced to a more efficient two- and three-dimensional coupling of multi-patch domains for general geometries by \cite{chan2019}.

The present work extends the general geometrical constraint formulation of \cite{duong2017} to coupled problems describing Cahn-Hilliard-type phase transitions on deforming thin shells \citep{zimmermann2019}, and phase field based fracture of brittle shells \citep{paul2020}. Patch constraints that enforce the required $C^1$-continuity of the phase field across patch interfaces are proposed, and their accuracy and convergence behavior are studied. Summarizing, the proposed formulation includes following highlights:
\begin{itemize}
	\item It proposes a new constraint for $C^1$-continuous phase fields on multi-patch NURBS.	
	\item It is fully formulated in a curvilinear coordinate system.
	\item It uses these constraints to model coupled problems on multi-patch NURBS.
	\item It exhibits excellent accuracy compared to single patch discretizations.
	\item It is capable to describe fracture branching across kinked surfaces within a higher-order phase field model.
	\item The proposed constraints are validated by several nonlinear benchmark problems.
\end{itemize}
The remainder of this paper is structured as follows: \secref{s:s_tot} summarizes the employed thin shell framework. The computational formulation for the Cahn-Hilliard phase separation model is summarized in \secref{s:c_tot}. In \secref{s:f_tot}, the computational phase field model of brittle shells is outlined. The proposed formulation for the required patch constraints is discussed in \secref{s:p_tot}. Numerical examples that highlight the effect of these constraints on the numerical solution are shown in \secref{s:n_tot}. This paper concludes with a summary and an outlook in \secref{s:cncl}.

\section{Thin shell theory} \label{s:s_tot}
\begin{figure}[!ht]
	\centering
		\subfloat[Phase separation of a binary mixture\label{f:s_mtvtn1}]{\includegraphics[scale=1]{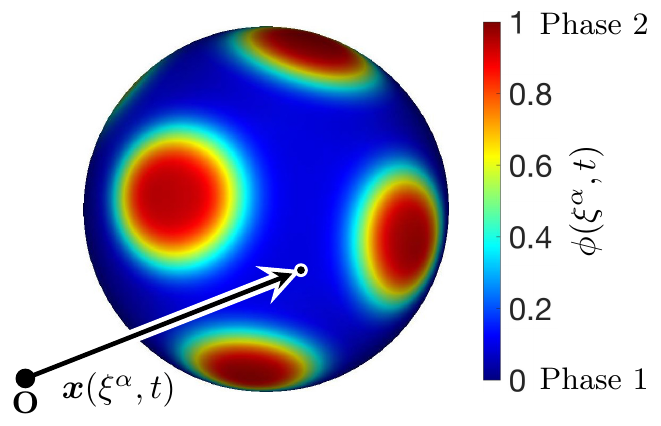}}
	\qquad
		\subfloat[Brittle fracture\label{f:s_mtvtn2}]{\includegraphics[scale=1]{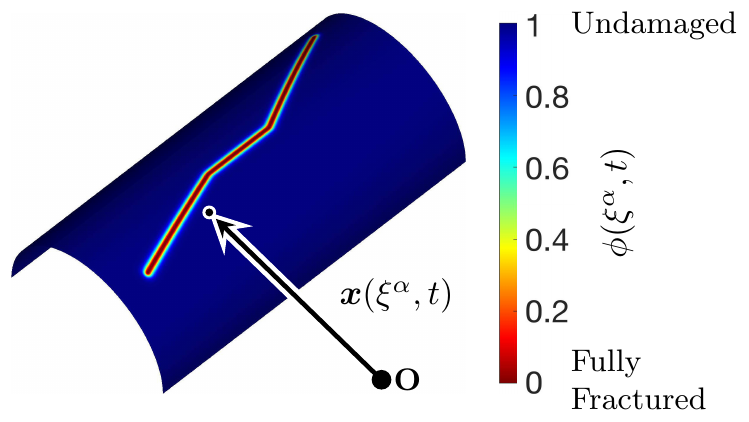}}
	\caption{Two cases of phase fields on curved shells. (a) Phase separation of a binary mixture, where $\phi$ describes the mixture of two phases, and (b) fracture, where $\phi$ distinguishes between undamaged and fully fractured material.} \label{f:s_mtvtn}
\end{figure}

This paper is concerned with thin shell theory and its coupling to two phase field models, see \figref{f:s_mtvtn}. This section summarizes the computational description of thin shells, while binary mixtures and fracture follow in Secs.~\ref{s:c_tot}--\ref{s:f_tot}.
At first, a short introduction to the surface description is given, which is then followed by the discretized weak form and constitutive relations. Further mathematical background and details of thin shell theory can be found for example in \cite{sauer2018}.

\subsection{Surface description} \label{s:s_descr}
Following the framework of differential geometry, a curved surface $\sS$ is described by a convected coordinate system and the mapping 
\eqb{l}
\bx = \bx(\xi^\alpha,t)\,, \label{e:map}
\eqe
between the parameter domain and the physical surface. Here, $\xi^\alpha$, $\alpha=1,2$, denote the convective coordinates and $t$ denotes time.\footnote{All Greek indices run from $1$ to $2$ and obey Einstein's summation convention.} The phase field is denoted $\phi(\xi^\alpha,t)$ and ranges from $0$ to $1$, see \figref{f:s_mtvtn}. The mapping in \eqsref{e:map} then defines a set of surface points $\bx\in\sS$ to represent the surface. A co-variant basis $\{\ba_\alpha,\bn\}$ is associated with each of these surface points with co-variant tangent vectors $\ba_\alpha$ and surface normal $\bn$ defined by
\eqb{l}
\ba_\alpha:=\ds\pa{\bx}{\xi^\alpha}\,,\quad\mathrm{and}\quad\bn := \ds\frac{\ba_1\times\ba_2}{\norm{\ba_1\times\ba_2}}\,. \label{e:s_cobas}
\eqe
The co-variant surface metric with components
\eqb{l}
a_{\alpha\beta} := \ba_\alpha\cdot\ba_\beta \label{e:s_aab}
\eqe
then follows. Orthonormality in the basis is achieved by the introduction of a contra-variant basis $\{\ba^\alpha,\bn\}$, such that $\ba_\alpha\cdot\ba^\beta=\delta_\alpha^\beta$, with Kronecker delta $\delta_\alpha^\beta$. The contra-variant tangent vectors are given by
\eqb{l}
\ba^\alpha = a^{\alpha\beta}\,\ba_\beta\,, \label{e:s_abcontra}
\eqe
where the contra-variant surface metric $a^\ab$ follows from the inverse of the co-variant surface metric, i.e. $[a^{\alpha\beta}] = [a_{\alpha\beta}]^{-1}$. The second parametric derivative $\ba_{\alpha,\beta} := \partial\ba_\alpha/\partial\xi^\beta$ is introduced in order to determine the curvature components and the mean curvature, i.e.
\eqb{l}
b_{\alpha\beta} := \ba_{\alpha,\beta}\cdot\bn\,,\quad\mathrm{and}\quad H := a^{\alpha\beta}\,b_{\alpha\beta}/2\,. \label{e:s_bab}
\eqe
Based on this geometric representation of surface $\sS$, surface operators can be defined, namely the surface gradient and surface Laplacian of a general scalar $\phi$, and the surface divergence of a general vector $\bv$,
\eqb{lllll}
\mathrm{grad}_\mrs\phi \dis \nablas\phi \dis \phi_{;\alpha}\,\ba^\alpha\,, \\[2mm]
\mathrm{div}_\mrs\bv \dis \nablas\cdot\bv \dis \bv_{;\alpha}\cdot\ba^\alpha\,, \\[2mm]
\Delta_\mrs\phi \dis \nablas\cdot\nablas\phi \is \phi_{;\alpha\beta}\,a^{\alpha\beta}\,. \label{e:s_diff}
\eqe
The subscript `;' is used to denote the co-variant derivative defined by
\eqb{lll}
\phi_{;\alpha} \is \phi_\ca\,, \\[2mm]
\bv_{;\alpha} \is \bv_\ca\,, \\[2mm]
\phi_{;\alpha\beta} \is \phi_{,\alpha\beta} - \Gamma^\gamma_{\alpha\beta}\,\phi_{,\gamma}\,, \label{e:s_phiab}
\eqe
with $\tightdots_{,\alpha}:=\partial\tightdots/\partial\xi^\alpha$ and the Christoffel symbols of the second kind $\Gamma^\gamma_{\alpha\beta} := \ba_{\alpha,\beta}\cdot\ba^\gamma$ on $\sS$.

\subsection{Discretization of primary fields} \label{s:s_prmflds}
Subsequently, the finite element (FE) approximations of the primary fields, the deformation and phase field, are outlined. The $n_e$ spline basis functions on element $\Omega^e$ are numbered with global indices $i_1,\dots,i_{n_e}$. The FE approximations of $\bx$ and its variation $\delta\bx$ on element $\Omega^e$ can then be written as
\eqb{l}
	\bx\approx\mN\,\mx_e\,,\quad\mathrm{and}\quad\delta\bx\approx\mN\,\delta\mx_e\,, \label{e:s_x}
\eqe
with $\mx_e$ and $\delta\mx_e$ describing the element-level vectors of nodal values and their variation, respectively. Further, the shape function array in \eqsref{e:s_x} is given by
\eqb{l}
	\mN = [N_{i_1}\bone,\,N_{i_2}\bone,\,...,\,N_{i_{n_e}}\bone]\,, \label{e:s_shpN}
\eqe
with $\bone$ denoting the $(3\times3)$-identity matrix. The FE approximations of the phase field $\phi$ and its variation $\delta\phi$ on element $\Omega^e$ follow in analogy, i.e.
\eqb{l}
	\phi\approx\bar\mN\,\bphi_e\,,\quad\mathrm{and}\quad\delta\phi\approx\bar\mN\,\delta\bphi_e\,, \label{e:s_phi}
\eqe
with shape function array
\eqb{l}
	\bar\mN := [N_{i_1},\,N_{i_2},\,...,\,N_{i_{n_e}}]\,, \label{e:c_shpN}
\eqe
and with the element-level vector of the nodal phase values and its variation $\bphi_e$ and $\delta\bphi_e$.

\subsection{Discretized mechanical weak form} \label{s:s_dscr}
The discretized mechanical weak form can be written as \citep{duong2017}
\eqb{l}
\delta\mx^\mrT\,\big[\mf_\mathrm{in} + \mf_\mathrm{int} - \mf_\mathrm{ext}\big] = 0\,, \quad \forall\,\delta\mx\in\sU^h\,, \label{e:s_dscrw}
\eqe
where the global force vectors $\mf_\mathrm{in}$, $\mf_\mathrm{int}$ and $\mf_\mathrm{ext}$ are assembled from their respective element-level contributions
\eqb{lll}
\mf^e_\mathrm{in} \dis \mm_e\,\ddot\mx_e~,\quad\mm_e := \ds\int_{\Omega^e}\rho(\phi)\,\mN^\mrT\mN\,\dif a\,,  \\[4mm]
\mf^e_\mathrm{int} \dis \ds\int_{\Omega^e}\sig^{\alpha\beta}\!(\phi)\,\mN_{\!,\alpha}^\mrT\,\ba_\beta\,\dif a
+ \int_{\Omega^e}M^{\alpha\beta}\!(\phi)\,\mN^\mrT_{\!;\alpha\beta}\,\bn\,\dif a\,, \\[4mm] 
\mf^e_\mathrm{ext} \dis \ds\int_{\Omega^e}\mN^\mrT\,p(\phi)\,\bn\,\dif a + \ds\int_{\Omega^e}\mN^\mrT\,f^\alpha\!(\phi)\,\ba_\alpha\,\dif a\,, \label{e:s_f}
\eqe
with shape function array $\mN$ from \eqsref{e:s_shpN}. Here, $\mathcal{U}^h$ denotes the corresponding discrete space, and $\mN^\mrT_{\!;\alpha\beta}$ follows in analogy to Eq.~(\ref{e:s_phiab}.3). The density is $\rho(\phi)$, the prescribed body forces are $\bff(\phi) = f^\alpha\!(\phi)\,\ba_\alpha + p(\phi)\,\bn$, and the stress and moment components are $\sig^\ab\!(\phi)$ and $M^\ab\!(\phi)$, respectively.\footnote{$\sigma^\ab$ is a 2D stress (measured as force per length) that is also referred to the stress resultant, i.e. the thickness integral of the 3D stress, e.g. see \cite{simo1989} and \cite{simo1990}.\label{footnote_s_stress}} Note that boundary loads acting on $\partial\sS$ are assumed to be zero in the above expression of $\mf^e_\mathrm{ext}$. The corresponding extension to boundary loads is described in \cite{duong2017}. From \eqsref{e:s_dscrw} follows the equation of motion at the free nodes (where no Dirichlet boundary conditions are prescribed)
\eqb{l}
\mf(\mx,\bphi) = \mM\,\ddot\mx + \mf_\mathrm{int}(\mx,\bphi) - \mf_\mathrm{ext}(\mx,\bphi) = \mathbf{0}\,, \label{e:s_ode}
\eqe
where $\mx$ and $\bphi$ are the global unknowns, similar to the element-level unknowns $\mx_e$ and $\bphi_e$. The global mass matrix $\mM$ follows from the assembly of the matrices $\mm_e$. \eqsref{e:s_ode} can be solved by itself in case $\bphi$ is not an unknown, as in the examples of \secref{s:n_s_tot}.

\subsection{Constitutive relations} \label{s:s_cnst}
The constitutive behavior follows from the Helmholtz free energy function 
\eqb{l}
	\Psi = \Psi_\mathrm{el} + \Psi_\mathrm{phase}\,. \label{e:s_hlmhltz}
\eqe
The second term is discussed in Secs.~\ref{s:c_cnst} and \ref{s:f_cnst}. For the first term, a hyperelastic material behavior with an elastic energy density $\Psi_\mathrm{el}=\Psi_\mathrm{el}(a_\ab,b_\ab,\phi)$ is assumed. It is composed of membrane and bending contributions, i.e. $\Psi_\mathrm{el}=\Psi_\mathrm{dil}+\Psi_\mathrm{dev}+\Psi_\mathrm{bend}$. The first is given by a Neo-Hookean membrane model \citep{sauer2017a}
\eqb{l}
\Psi_\mathrm{dil} = \ds\frac{K}{4}\big(J^2-1-2\,\ln J\big)\,,\quad\mathrm{and}\quad\Psi_\mathrm{dev} = \ds\frac{G}{2}\big(I_1/J-2\big)\,, \label{e:s_psimem}
\eqe
and the second by the Koiter bending model \citep{ciarlet1993} 
\eqb{l}
\Psi_\mathrm{bend} = \ds\frac{c}{2}\big(b_{\alpha\beta}-B_{\alpha\beta}\big)\big(b^{\alpha\beta}_0-B^{\alpha\beta}\big)\,,\quad\mathrm{with}\quad b^{\alpha\beta}_0:=A^{\alpha\gamma}b_{\gamma\delta}A^{\beta\delta}\,. \label{e:s_psibnd}
\eqe
Note that quantities defined on the reference surface are either indicated by a capital symbol, or the subscript `$0$', whereas quantities on the current surface are indicated by lowercase symbols. The two invariants in \eqsref{e:s_psimem} are
\eqb{l}
I_1 := A^{\alpha\beta}\,a_{\alpha\beta}\,,\quad\text{and}\quad J := \sqrt{\det[A^{\alpha\beta}]\det[a_{\alpha\beta}]}\,, \label{e:s_invrnts}
\eqe
with contra-variant metric $A^\ab$ in the reference configuration. The 2D bulk, 2D shear, and bending moduli are denoted $K$, $G$, and $c$, respectively, and they are generally a function of the phase variable $\phi$. The stress and moment components follow from\footref{footnote_s_stress}
\eqb{lll}
\sigma^{\alpha\beta} \is \ds\df{2}{J}\ds\pa{\Psi}{a_{\alpha\beta}} - \eta\,\dot a^{\alpha\beta}\,, \\[4mm]
M^{\alpha\beta} \is \ds\df{1}{J}\ds\pa{\Psi}{b_{\alpha\beta}}\,. \label{e:s_sigM}
\eqe
Eq.~(\ref{e:s_sigM}.1) represents a Kelvin model to account for viscous in-plane stresses. In this model, a spring and dashpot act in parallel and thus, their stresses are added.\footnote{In \eqsref{e:s_sigM}, $\dot a^{\alpha\beta} = -a^{\alpha\gamma}\,\dot a_{\gamma\delta}\,a^{\delta\beta}$ are the components of the symmetric surface velocity gradient, e.g. see \cite{sauer2018}.} The dynamic surface viscosity is denoted $\eta$. More complex constitutive models require a general strain decomposition, which has been recently formulated for Kirchhoff-Love shells in \cite{sauer2019}. For the above material model, the individual contributions to the stress components are given by
\eqb{lll}
\sigma_\mathrm{dil}^\ab \is \dfrac{K}{2J}\bigl(J^2-1\bigr)\,a^\ab\,,\\[4mm]
\sigma_\mathrm{dev}^\ab \is \dfrac{G}{2J^2}\bigl(2A^\ab-I_1\,a^\ab\bigr)\,,\\[4mm]
\sigma_\mathrm{visc}^\ab \is -\eta\,\dot{a}^\ab\,. \label{e:s_sigcontr}
\eqe
Likewise, the moment components stemming from \eqsref{e:s_psibnd} are
\eqb{l}
M^\ab = \dfrac{c}{J}\bigl(b_0^\ab-B^\ab\bigr)\,. \label{e:s_Mab}
\eqe
Note that Eqs.~\eqref{e:s_sigcontr}--\eqref{e:s_Mab} do not contain the stresses and moments coming from the phase field model. These are discussed in Secs.~\ref{s:c_cnst} and \ref{s:f_cnst}.

\section{Phase transitions on deforming surfaces} \label{s:c_tot}
This section briefly summarizes the discretized weak form and constitutive relations describing phase transitions on deforming surfaces according to the computational model of \cite{zimmermann2019}. The Cahn-Hilliard model is essentially a temperature dependent diffusion model with an additional term that penalizes phase interfaces. For low temperatures, the phases are driven towards separation (considered here), while for high temperatures, they are driven towards mixture. In the following, the dimensionless concentration field $\phi(\xi^\alpha,t)$ is used to describe the local density fractions of binary mixtures, see \figref{f:s_mtvtn1}.\footnote{The concentration field $\phi$ is also referred to as the \textit{order parameter field} or \textit{phase field} in the literature.}

\subsection{Discretized weak form} \label{s:c_dscr}
The discretized weak form of the surface Cahn-Hilliard equation can be written as
\eqb{l}
\delta\bphi^\mrT\,\big[\bar\mf_\mathrm{in} + \bar\mf_\mathrm{int} - \bar\mf_\mathrm{ext}\big] = 0\,, \quad\forall\,\delta\bphi\in\sV^h\,. \label{e:c_dscrw}
\eqe 
The global vectors $\bar\mf_\mathrm{in}$, $\bar\mf_\mathrm{int}$ and $\bar\mf_\mathrm{ext}$ are assembled from their respective element-level contributions
\eqb{lll}
\bar\mf^e_\mathrm{in} \dis \bar\mm_e\,\dot\bphi_e\,,\quad\bar\mm_e := \ds\int_{\Omega^e}\rho\,\bar\mN^\mrT\bar\mN\,\dif a\,, \\[4mm]
\bar\mf^e_\mathrm{int} \dis \bar\mk_e\,\bphi_e-\bar\mf^e_\mathrm{el}\,,\quad\!
\bar\mk_e :=\! \ds\int_{\Omega^e_0}\!\Big[\bar\mN_{\!,\alpha}^\mrT\,a^{\alpha\beta}\Big(\!M \mu'_\phi - M'\big(\mu_\mri+\mu_\mathrm{el}\big)\!\Big)\,\bar\mN_{\!,\beta}+\Delta_\mrs\bar\mN^\mrT J\ell^2 M\Delta_\mrs\bar\mN\Big]\,\dif A\,, \\[3mm]
\bar\mf^e_\mathrm{el} \dis \ds\int_{\Omega^e_0}\Delta_\mrs\bar\mN^\mrT M\mu_\mathrm{el}\,\dif A\,, \\[4mm]
\bar\mf^e_\mathrm{ext} \dis \mathbf{0}\,. \label{e:c_f}
\eqe
Here, $(\tightdots)':=\partial\tightdots/\partial\phi$, while the corresponding discrete space is denoted $\mathcal{V}^h$, and $\Delta_\mrs\bar\mN$ follows in analogy to Eq.~(\ref{e:s_diff}.3). Further, $M=D\,\phi\,(1-\phi)$, $D=\mathrm{const.}$, denotes the degenerate mobility, $\ell$ represents the length scale of the phase interface and $J$ is the surface stretch, see Eq.~(\ref{e:s_invrnts}.2). The remaining terms in \eqsref{e:c_f} are described in \secref{s:c_cnst}. Note that $\bar\mf^e_\mathrm{ext}$ is assumed to be zero in the above expressions. The extension to non-zero $\bar\mf^e_\mathrm{ext}$ is described in \cite{zimmermann2019}. From \eqsref{e:c_dscrw} follows the evolution equation for $\bphi$ at the free nodes (after application of Dirichlet boundary conditions)
\eqb{l}
\bar\mf(\mx,\bphi) = \bar\mM\,\dot\bphi + \bar\mf_\mathrm{int}(\mx,\bphi) - \bar\mf_\mathrm{ext}(\mx) = \mathbf{0}\,, \label{e:c_ode}
\eqe
with global mass matrix $\bar\mM$ being assembled from $\bar\mm_e$. Together, Eqs.~\eqref{e:s_ode} and \eqref{e:c_ode} describe a coupled chemo-mechanical problem. It is discretized and integrated in time with an implicit, monolithic and adaptive method \citep{zimmermann2019} based on the generalized-$\alpha$ scheme by \cite{chung93}.

\subsection{Constitutive relations} \label{s:c_cnst}
The material parameters are assumed to depend on the concentration field $\phi$ via the following mixture rules
\eqb{lllllll}
K(\phi) \is K_1\,f(\phi) + K_0\,(1-f(\phi))\,,&G(\phi) \is G_1\,f(\phi) + G_0\,(1-f(\phi))\,, \\[4mm]
c(\phi) \is c_1\,f(\phi) + c_0\,(1-f(\phi))\,, &\eta(\phi) \is \eta_1\,f(\phi) + \eta_0\,(1-f(\phi))\,, \label{e:c_mix}
\eqe
with the interpolation function
\eqb{lll}
f(\phi) = \ds \frac{1}{2}\Big(1+\text{tanh}\left(-\rho_\mathrm{sh}\,\pi +4\,\pi\,\phi \right) \Big)\,. \label{e:c_intf}
\eqe
Here, $K_i$, $G_i$, $c_i$ and $\eta_i$ ($i=0,1$) are the material parameters of the two components. The parameter $\rho_\mathrm{sh}\in\mathbb{R}$ specifies whether a small or a large portion of the phase interface is characterized by material properties at $\phi=1$. The force vectors in \eqsref{e:c_f} include the chemical potential $\mu$ that has the contributions $\mu_\mrb$, $\mu_\mri$, $\mu_\mathrm{el}$, $\mu_\phi$ associated with the bulk, interface, elastic, and mixing energy, given by
\eqb{lllllll}
\mu_\mrb \is \mu_\phi + \mu_\mathrm{el}\,, 	&\mu_{\phi} \is N\,k_\mathrm{B}\,T\,\ln{\ds\frac{\phi}{1-\phi}} + N\omega\,(1-2\phi)\,, \\[4mm]
\mu_\mathrm{el}\is \Psi'_\mathrm{el}\,, 			&\mu_\mri \is -J\,N\,\omega\,\ell^2\,\Delta_\mrs\phi\,. \label{e:c_pot}
\eqe
Here, $N$ denotes the number of molecules per reference area, $k_\mathrm{B}$ is Boltzmann's constant and $\omega=2\,k_\mathrm{B}\,T_\mathrm{c}$ is a bulk energy that is related to the critical temperature $T_\mathrm{c}$ at which phase separation occurs. 

The Cahn-Hilliard energy in \eqsref{e:s_hlmhltz} is given by
\eqb{l}
	\Psi_\mathrm{phase}=N\,\omega\,\phi\,(1-\phi)+T\,N k_\mathrm{B}\bigl(\phi\,\ln\phi+(1-\phi)\,\ln(1-\phi)\bigr)+J\,N\,\omega\,\dfrac{\ell^2}{2}\,\nablas\phi\cdot\nablas\phi\,. \label{e:c_psich}
\eqe
The stresses stemming from $\Psi_\mathrm{phase}$ are known as Korteweg stresses and they are given by
\eqb{l}
	\sigma_\mathrm{CH}^\ab=N\omega\dfrac{\ell^2}{2}\Bigl( a^\ab a^\gd-2\,a^\ag a^\btad \Bigr)\,\phi_{;\gamma}\,\phi_{;\delta}\,. \label{e:c_sigch}
\eqe
Note that there are no bending moments induced by the Cahn-Hilliard energy.

\section{Brittle fracture of deforming thin shells} \label{s:f_tot}
This section summarizes the phase field model for brittle fracture of thin shells by \cite{paul2020}. In phase field methods, cracks are smeared out and described by a field $\phi(\xi^\alpha,t)\in[0,1]$, see \figref{f:s_mtvtn2}.\footnote{The phase field $\phi$ is also referred to as the \textit{fracture field} in the literature.} It indicates fully fractured ($\phi=0$) and undamaged ($\phi=1$) material. A length scale parameter $\ell_0\,[\mathrm{m}]$ gives control over the support width of the phase field profile, i.e. $\mathrm{supp}(\phi)\sim\ell_0$.

\subsection{Discretized weak form} \label{s:f_dscr}
The phase field evolution equation is determined from the minimization of the Helmholtz free energy that consists of elastic and fracture contributions, see \secref{s:f_cnst}. Based on this, the discretized weak form of the phase field evolution equation becomes
\eqb{l}
\delta\bphi^\mrT\,\big[\bar\mf_\mathrm{el} + \bar\mf_\mathrm{frac}\big] = 0\,, \quad\forall\,\delta\bphi\in\sV^h\,, \label{e:f_dscrw}
\eqe 
with the corresponding finite-dimensional space $\sV^h$ and the element-level contributions
\eqb{lll}
\bar\mf^e_\mathrm{el} \dis \ds\int_{\Omega_0^e}\bar\mN^\mrT\ds\frac{2\ell_0}{\sG_\mrc}g'(\phi)\,\sH\,\dif A\,, \\[4mm]
\bar\mf^e_\mathrm{frac} \dis \ds\int_{\Omega_0^e}\bigg[\bar\mN^\mrT\bar\mN + \bar\mN_{,\alpha}^\mrT\,2\,\ell_0^2\,A^{\alpha\beta}\,\bar\mN_{,\beta}+\Delta_\mrS\bar\mN^\mrT\,\ell_0^4\,\Delta_\mrS\bar\mN\bigg]\bphi_e\,\dif A-\ds\int_{\Omega_0^e}\bar\mN^\mrT\,\dif A\,. \label{e:f_f}
\eqe
The shape function array for the phase field is denoted $\bar\mN$, see \eqsref{e:c_shpN}, and $\Delta_\mrS\bar\mN$ follows in analogy to Eq.~(\ref{e:s_diff}.3). Note that here, the surface Laplacian is defined on the reference configuration. $\sG_\mrc$ denotes the fracture toughness. The degradation function \citep{borden2016}
\eqb{l}
	g(\phi) = (3-s)\,\phi^2 - (2-s)\,\phi^3\,,\quad s=10^{-4}\,, \label{e:f_degrfct}
\eqe
controls the loss of material stiffness in regions of fracture and $\sH$ is the history field that is motivated and discussed in the next section. The degradation function in \eqsref{e:f_degrfct} couples the elastic energy density in \eqsref{e:s_hlmhltz} with the phase field, see \eqsref{e:f_hlmhltz1}. The phase field force vector is coupled to the deformation by the history field $\sH$, see Eq.~(\ref{e:f_f}.1). The resulting equations at the free nodes (where no Dirichlet boundary conditions are imposed) simplify to
\eqb{l}
\bar\mf(\mx,\bphi) =  \bar\mf_\mathrm{el}(\mx,\bphi)+\bar\mf_\mathrm{frac}(\bphi) = \mathbf{0}\,. \label{e:f_ode}
\eqe
To obtain a highly resolved mesh in regions of damage, a local refinement strategy based on LR NURBS is employed \citep{paul2020}. As for the phase separation process, the temporal discretization is based on the generalized-$\alpha$ scheme by \cite{chung93} using adaptive time-stepping.

\subsection{Constitutive relations} \label{s:f_cnst}
Crack evolution exhibits anisotropic behavior since it does not occur in compression. Thus, an additive split of the elastic energy density is required,
\eqb{l}
\Psi_\mathrm{el}=\Psi_\mathrm{el}^++\Psi_\mathrm{el}^-\,, \label{e:f_splt}
\eqe
that distinguishes between a positive part that contributes to crack evolution, and a negative part, which does not. The elastic part of the Helmholtz free energy in \eqsref{e:s_hlmhltz} is thus, changed to
\eqb{l}
	\Psi_\mathrm{el} = g(\phi)\,\sH+\Psi_\mathrm{el}^-\,, \label{e:f_hlmhltz1}
\eqe
with the history field $\sH(\bx,t)=\max_{\tau\in[0,t]}\Psi_\mathrm{el}^+(\bx,\tau)$ that enforces an irreversible fracture process and the degradation function $g(\phi)$ from \eqsref{e:f_degrfct}. The fracture energy is given by \citep{borden2014,paul2020}
\eqb{l}
	\Psi_\mathrm{phase} = \ds\frac{\sG_\mrc}{4\ell_0}\Big[(\phi-1)^2+2\ell_0^2\,\nablao\phi\cdot\nablao\phi + \ell_0^4\,(\Delta_\mrS\phi)^2\Big]\,. \label{e:f_hlmhltz2}
\eqe

The membrane and bending contributions of $\Psi_\mathrm{el}$ are split separately and the split of \cite{amor2009} is adopted for the membrane part, i.e.
\eqb{l}
	\Psi_\mathrm{mem}^{+}=\begin{cases}\Psi_\mathrm{dev}+\Psi_\mathrm{dil}\,, &J\geq1 \\ \Psi_\mathrm{dev}\,, &J<1 \end{cases}\,, \quad\mathrm{and}\quad
	\Psi_\mathrm{mem}^{-}=\begin{cases}0\,, &J\geq1 \\ \Psi_\mathrm{dil}\,, & J<1 \end{cases}\,, \label{e:f_memsplt}
\eqe
with dilatational and deviatoric energies taken from \eqsref{e:s_psimem} and surface stretch $J$ from Eq.~(\ref{e:s_invrnts}.2). Now, the total stresses in the system are given by
\eqb{l}
	\sigma^\ab=g(\phi)\,\sigma_+^\ab+\sigma_-^\ab\,, \label{e:f_sig}
\eqe
with
\eqb{l}
	\sigma^\ab_+=\begin{cases}\sigma_\mathrm{dev}^\ab+\sigma_\mathrm{dil}^\ab\,,&J\geq1\\\sigma_\mathrm{dev}^\ab\,,&J<1\end{cases}\,,\quad\mathrm{and}\quad\sigma_-^\ab=\begin{cases}0\,,&J\geq1\\\sigma_\mathrm{dil}^\ab\,,&J<1\end{cases}\,, \label{e:f_sig2}
\eqe
and contributions from Eqs.~(\ref{e:s_sigcontr}.1)--(\ref{e:s_sigcontr}.2). The degradation function $g(\phi)$ is given in \eqsref{e:f_degrfct}. The brittle fracture model does not presently include viscosity, such that $\eta=0$, see Eq.~(\ref{e:s_sigM}.1).

The bending part of the elastic energy density $\Psi_\mathrm{bend}$ is split based on thickness integration \citep{paul2020}. It is constructed from the three-dimensional Saint Venant-Kirchhoff model \citep{duong2017}
\eqb{l}
\tilde\Psi_\mathrm{bend}(\bK,\xi,T)=\xi^2\ds\df{12}{T^3}\df{c}{2}\,\mathrm{tr}\!\lr{\bK^2}\,, \label{e:f_psibnd3}
\eqe
with the relative curvature tensor $\bK=(b_\ab-B_\ab)\,\bA^\alpha\otimes\bA^\beta$. The energy split is then performed on this three-dimensional constitutive law. Afterwards, it is integrated over the thickness to obtain the energy split in surface energy form, i.e.
\eqb{l}
\Psi_\mathrm{bend}^\pm=\ds\int_{-\frac{T}{2}}^{\frac{T}{2}}\tilde\Psi_\mathrm{bend}^\pm(\xi)\,\dif\xi\,. \label{e:f_bndsplt}
\eqe
In analogy to \eqsref{e:f_memsplt}, the split of \eqsref{e:f_psibnd3} is based on the surface stretch $\tilde{J}$ of the shell layer at position $\xi\in[-T/2,T/2]$, i.e.
\eqb{l}
	\tilde\Psi_\mathrm{bend}^{+}(\xi)=\begin{cases}\xi^2\dfrac{12}{T^3}\dfrac{c}{2}\,\mathrm{tr}\!\lr{\bK^2}\,, &\tilde{J}(\xi)\geq1 \\ 0\,, & \tilde{J}(\xi)<1 \end{cases}\,, \quad\!\mathrm{and}\quad
	\tilde\Psi_\mathrm{bend}^{-}(\xi)=\begin{cases}0\,, &\tilde{J}(\xi)\geq1 \\ \xi^2\dfrac{12}{T^3}\dfrac{c}{2}\,\mathrm{tr}\!\lr{\bK^2}\,, & \tilde{J}(\xi)<1 \end{cases}\,. \label{e:f_bnd3splt}
\eqe
In analogy to Eq.~(\ref{e:s_invrnts}.2), the surface stretch $\tilde{J}$ is computed from
\eqb{l}
	\tilde{J}=\sqrt{\det[G^{\alpha\beta}]\det[g_{\alpha\beta}]}\,, \label{e:f_jtild}
\eqe 
where the metrics $G^{\alpha\beta}$ and $g_{\alpha\beta}$ follow from the tangent vectors $\bG_{\alpha}$ and $\bg_{\alpha}$ of the shell layer at points $\bX+\xi\,\bN$ and $\bx+\xi\,\bn$, respectively \citep{duong2017}. The resulting moment is given by
\eqb{l}
	M^\ab=g(\phi)\,M^\ab_++M^\ab_-\,, \label{e:f_m}
\eqe
with degradation function $g(\phi)$ from \eqsref{e:f_degrfct} and the contributions
\eqb{l}
	M_\pm^\ab=\ds\int_{-\frac{T}{2}}^{\frac{T}{2}}\tilde{M}_\pm^\ab(\xi)\,\dif\xi\,.
\eqe
Here, $\tilde{M}^\ab_\pm$ refers to the split of the moment components stemming from the three-dimensional constitutive law. The split is given by
\eqb{l}
	\tilde{M}^\ab_+=\begin{cases}\tilde{M}^\ab\,,&\tilde{J}(\xi)\geq1\\0\,,&\tilde{J}(\xi)<1\end{cases}\,,\quad\mathrm{and}\quad \tilde{M}^\ab_-=\begin{cases}0\,,&\tilde{J}(\xi)\geq1\\\tilde{M}^\ab\,,&\tilde{J}(\xi)<1\end{cases}\,, \label{e:f_m2}
\eqe
with
\eqb{l}
	\tilde{M}^\ab=\xi^2\dfrac{12}{T^3}\,\dfrac{c}{J}\,\Bigl(b_0^\ab-B^\ab\Bigr)\,,\quad\mathrm{and}\quad b_0^\ab=A^\ag\,b_\gd\,A^\btad\,. \label{e:f_m3}
\eqe

\section{Continuity constraints for patch interfaces} \label{s:p_tot}
The force vectors of the discretized weak forms in Eqs.~\eqref{e:s_f}, ~\eqref{e:c_f} and ~\eqref{e:f_f} include second-order operators. This necessitates an at least $C^1$-continuous discretization\footnote{Unless other methods, like mixed (e.g. displacement and rotation) methods, are used.}. Here, this is obtained by using an isogeometric surface discretization. In general, especially for complex engineering structures, it is not possible to represent the geometry with a single NURBS patch. Instead, multiple patches are required. At the interfaces between these patches, the $C^1$-continuity is lost and needs to be recovered. For the mechanical shell equations in the weak form from Eqs.~\eqref{e:s_dscrw}--\eqref{e:s_f}, it is sufficient to impose a geometric $G^1$-continuity constraint to transfer bending moments across the patch interfaces. \cite{duong2017} introduce two methods to enforce this constraint for five different edge rotation conditions: enforcing $G^1$-continuity, maintaining surface folds, enforcing symmetry conditions on flat sheets, enforcing symmetry conditions on folded sheets and prescribing boundary rotations. Their formulation is summarized and simplified in \secref{s:p_g_tot}. For the surface concentration and the fracture field, a new $C^1$-continuity constraint is proposed in \secref{s:p_c_tot}. It is used to obtain continuous surface gradients $\nablas\phi$ and $\nablao\phi$  across patch interfaces. In the present work, conforming meshes are considered and thus, the $G^0$- and $C^0$-continuity of $\bx$ and $\phi$ are automatically satisfied at the patch interfaces.

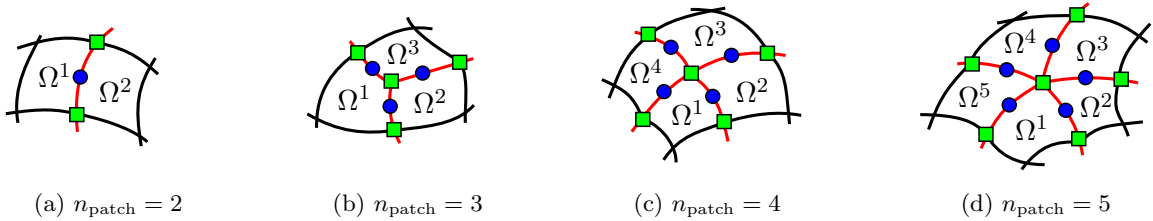
\begin{figure}[!ht]
	\newcommand{\cursize}{0.14}
	\centering
		\subfloat[$n_\mathrm{patch}=2$]{
			\begin{tikzpicture}
				\node at (0,0) {\includegraphics[height=\cursize\textwidth]{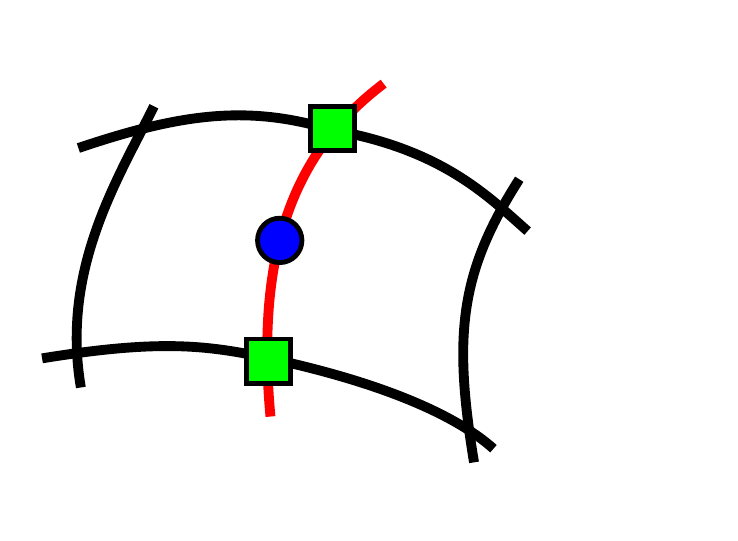}};
				\node at (-.7,.15) {$\Omega^1$};
				\node at (.1,-.05) {$\Omega^2$};
			\end{tikzpicture}
		}
		\quad
		\subfloat[$n_\mathrm{patch}=3$]{
			\begin{tikzpicture}
				\node at (0,0) {\includegraphics[height=\cursize\textwidth]{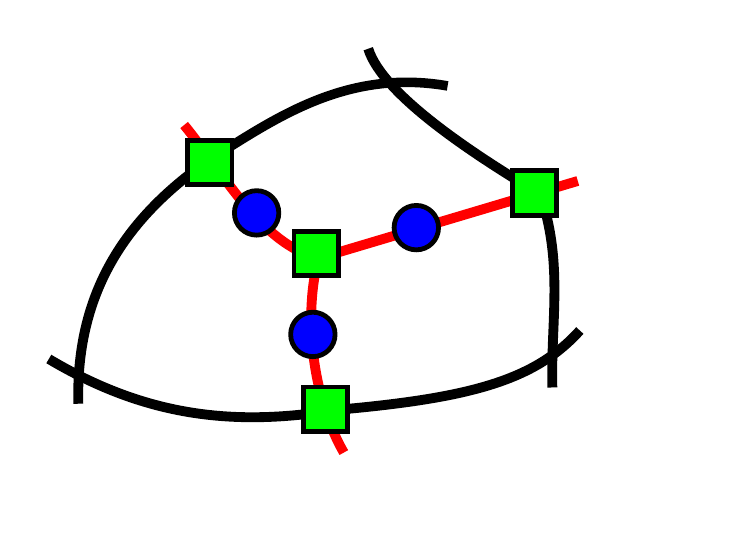}};
				\node at (-.7,-.15) {$\Omega^1$};
				\node at (.3,-.2) {$\Omega^2$};
				\node at (-.05,.45) {$\Omega^3$};
			\end{tikzpicture}
		}
		\quad
		\subfloat[$n_\mathrm{patch}=4$]{
			\begin{tikzpicture}
				\node at (0,0) {\includegraphics[height=\cursize\textwidth]{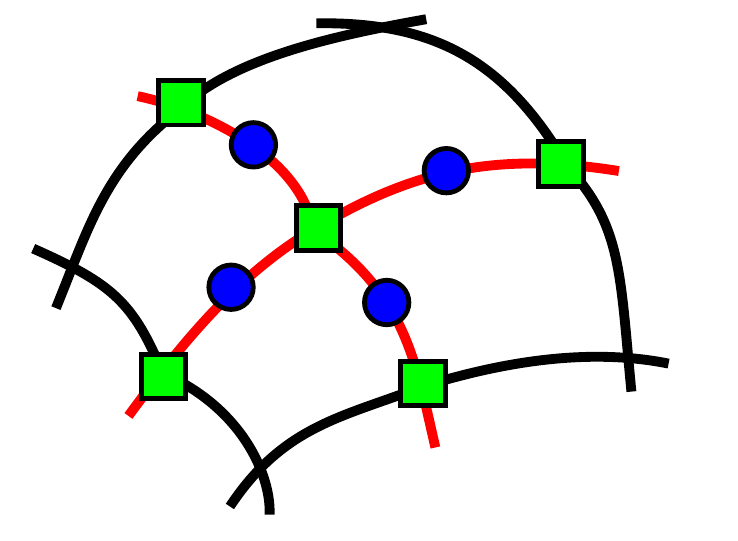}};
				\node at (-.25,-.35) {$\Omega^1$};
				\node at (.6,-.05) {$\Omega^2$};
				\node at (.05,.7) {$\Omega^3$};
				\node at (-.8,.2) {$\Omega^4$};
			\end{tikzpicture}
		}
		\qquad
		\subfloat[$n_\mathrm{patch}=5$]{
			\begin{tikzpicture}
				\node at (0,0) {\includegraphics[height=\cursize\textwidth]{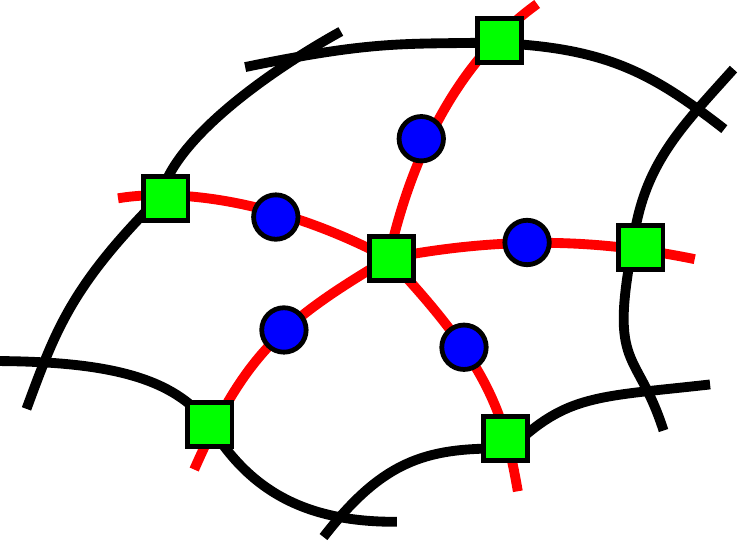}};
				\node at (-.05,-.55) {$\Omega^1$};
				\node at (.8,-.25) {$\Omega^2$};
				\node at (.75,.5) {$\Omega^3$};
				\node at (-.2,.6) {$\Omega^4$};
				\node at (-.8,-.12) {$\Omega^5$};
			\end{tikzpicture}
		}
	\caption{Illustration of different element-level patch constellations. From left to right, the number of patch elements that share at least one common control point increases from two to five. The red lines mark the patch interfaces and the elements are denoted $\Omega^e$, $e=1,\dots,n_\mathrm{patch}$. The circular blue and square green symbols mark the position of constant and linear Lagrange multipliers, respectively, along the patch interfaces.} \label{f:p_css}
\end{figure}

Here, the four different types of element-level patch interfaces depicted in \figref{f:p_css} are considered. From left to right, the number of elements from different patches that meet at a point on the surface increases from two to five. The following theory is also valid for other cases, but the examples in \secref{s:n_tot} are restricted to the cases depicted in \figref{f:p_css}. In the present formulation, a potential that enforces the constraints is added to the total potential of the system, see the subsequent sections. Thus, numerical integration along the patch interfaces is required. In this section, these interfaces are shown by red lines in the figures. The elements meeting at these interfaces are denoted $\Omega^e$, $e=1,\dots,n_\mathrm{patch}$, where $n_\mathrm{patch}$ denotes the number of patches.

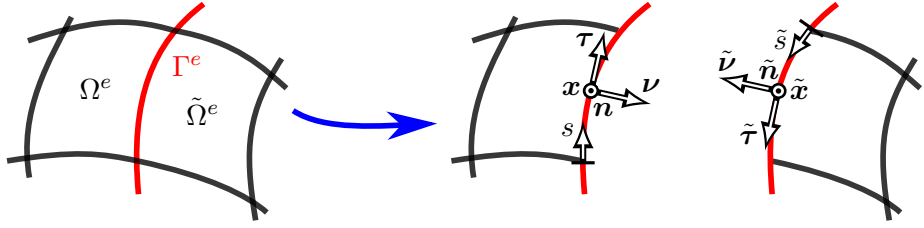
\begin{figure}[!ht]
	\centering
		\begin{tikzpicture}
			\node at (0,0) {\includegraphics[height=0.2\textwidth]{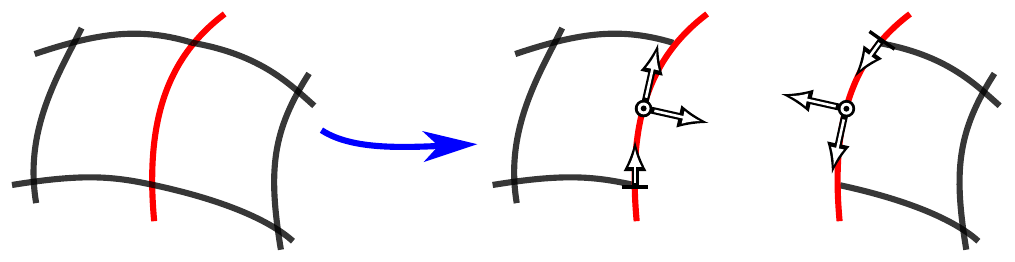}};
				\node at (-3.655,.6) {\color{red}$\Gamma^e$};
				\node at (-4.855,.3) {$\Omega^e$};
				\node at (-3.45,0) {$\tilde\Omega^e$};
				\node at (1.35,-.25) {$s$};
				\node at (1.4,.3) {$\bx$};
				\node at (1.82,0.02) {$\bn$};
				\node at (1.58,0.95) {$\btau$};;
				\node at (2.45,.34) {$\bnu$};
				\node at (4.15,0.95) {$\tilde{s}$};
				\node at (4.4,.3) {$\tilde\bx$};
				\node at (4.0,.56) {$\tilde\bn$};
				\node at (3.74,-.3) {$\tilde{\btau}$};
				\node at (3.45,.67) {$\tilde\bnu$};
		\end{tikzpicture}
	\caption{Local coordinate systems $\{\bnu,\btau,\bn\}$ and $\{\tilde\bnu,\tilde\btau,\tilde\bn\}$ at the patch interface $\Gamma^e$. Note that the surface normals $\bn$ and $\tilde\bn$ point out of the paper plane.} \label{f:p_vcs}
\end{figure}

The patch interfaces $\Gamma$ are discretized into line elements denoted $\Gamma^e$ that conform to surface elements. \figref{f:p_vcs} illustrates the local coordinate systems for a given element-level patch interface $\Gamma^e$ that is parameterized by the coordinate $s$. Note that the boundary quantities on two adjacent patches are distinguished by a tilde.
Both the interface tangent and normal vector are orthogonal to the surface normal. They follow from
\eqb{l}
	\btau=\ds\frac{\partial\bx}{\partial s}\,,\qquad \bnu=\btau\times\bn\,,\qquad \tilde\btau=\ds\frac{\partial\tilde\bx}{\partial \tilde{s}}\,,\qquad \tilde\bnu=\tilde\btau\times\tilde\bn\,, \label{e:p_crd}
\eqe
with the relations $\tilde{\bx}=\bx$ and $\tilde{s}=-s$ for the special case of conforming meshes that is studied in the present work. Hence, $\tilde\btau=-\btau$. For the reference configuration, these vectors are defined in analogy to \eqsref{e:p_crd}.

The elements $\Omega^e$ and $\tilde\Omega^e$ share $\hat{n}_\mathrm{CP}$ control points along the interface $\Gamma^e$. These control points are denoted $\hat\mx_e$ and are used to compute the discretized tangent vector along $\Gamma^e$, 
\eqb{l}
	\hat{\ba}_\xi\approx\hat\mN_{,\xi}\,\hat\mx_e\,, \label{e:p_axi}
\eqe
with the shape function array
\eqb{l}
	\hat\mN =[\hat{N}_{i_1}\bone,\,\hat{N}_{i_2}\bone,\,...,\,\hat{N}_{i_{\hat{n}_\mathrm{CP}}}\bone]\,. \label{e:s_shpNxi}
\eqe
Given $\hat{\ba}_\xi$, the tangent vector $\btau$ is obtained from $\btau=\hat{\ba}_\xi/||\hat{\ba}_\xi||$. Its variation is given in Eq.~(\ref{e:v_varntau}.3). Note that other surface quantities on element $\Omega^e$, like the surface normal $\bn$, are obtained from the shape function array $\mN$ and the nodal values $\mx_e$, see \secref{s:s_prmflds}. Likewise, these quantities on $\tilde\Omega^e$ are obtained from $\tilde\mN$ and $\tilde\mx_e$.

\subsection{\texorpdfstring{$G^1$}{G1}-continuity constraint for the surface deformation} \label{s:p_g_tot}
The geometric continuity constraint of \cite{duong2017} is summarized and simplified in this section. It can be used to enforce the four\footnote{A fifth possibility is to prescribe boundary rotations, see \cite{duong2017}.} different inter-patch conditions shown in \tabref{t:p_g_css}. The table distinguishes between the general case of a kink ($\theta\neq\pi$) and the special case of a planar interface ($\theta=\pi$). In both cases either the continuity at patch interfaces or the symmetry at patch boundaries can be enforced. The vectors $\tilde\bN$ and $\tilde\bn$ denote the normal of either the neighboring patch or the symmetry plane.
\begin{table}[!ht]
	\centering
	\caption{Overview of different inter-patch conditions.} \label{t:p_g_css}
	\begin{tabular}{ >{\centering\arraybackslash} m{2.1cm} || C{5cm} | C{5cm} }
		 	& \textbf{Kink} & \textbf{Planar} \\ \hline\hline
			 \textbf{Patch interface} & 
			\adjustbox{valign=c}{
			 	\begin{tikzpicture}
			 		\draw[line width=2,black] (0,0) coordinate (orig) -- (-2,1.5) coordinate (topleft);
			 		\draw[line width=2,densely dotted,blue] (0,0) -- (2,1.5) coordinate(topright);
			 		\fill[red] (0,0) circle (0.1);
			 		\path (topright) -- (orig) -- (topleft) pic["$\theta$",draw=black,>=latex,,angle eccentricity=1.3,angle radius=20,->] {angle=topright--orig--topleft};
			 		\draw[>=latex,->,line width=1] (-1,.75) -- ({-1+1.5/3},{.75+2/3});
			 		\draw[>=latex,->,line width=1] (1,.75) -- ({1-1.5/3},{.75+2/3});
			 		\node[anchor=north] at (0,-.05) {\color{red}$\Gamma^e$};
			 		\node[anchor=south east] at (-1,.2) {\color{black}$\Omega^e$};
			 		\node[anchor=south west] at (1,.2) {\color{blue}$\tilde\Omega^e$};
			 		\node[anchor=north] at (-.9,1.7) {\color{black}$\bn$};
			 		\node[anchor=north] at (.9,1.7) {\color{black}$\tilde\bn$};
			 		\node at (0,1.9) {};
			 	\end{tikzpicture}} & 		 	
			\adjustbox{valign=c}{
			 	\begin{tikzpicture}
			 		\draw[line width=2,black] (0,0) coordinate (orig) -- (-2,0) coordinate (topleft);
			 		\draw[line width=2,densely dotted,blue] (0,0) -- (2,0) coordinate(topright);
			 		\fill[red] (0,0) circle (0.1);
			 		\draw[>=latex,->,line width=1] (-1,0) -- (-1,0.833);
			 		\draw[>=latex,->,line width=1] (1,0) -- (1,0.833);
			 		\node[anchor=north] at (0,-.05) {\color{red}$\Gamma^e$};
			 		\node[anchor=south] at (-1,-.6) {\color{black}$\Omega^e$};
			 		\node[anchor=south] at (1,-.6) {\color{blue}$\tilde\Omega^e$};
			 		\node[anchor=east] at (-1,.416) {\color{black}$\bn$};
			 		\node[anchor=west] at (1,.416) {\color{black}$\tilde\bn$};
			 		\node at (0,1.9) {};
			 	\end{tikzpicture}} \\ \hline
			 \textbf{Symmetry condition at a patch boundary} & 
			\adjustbox{valign=c}{
			 	\begin{tikzpicture}
			 		\draw[line width=2,black] (0,0) coordinate (orig) -- (-2,1.5) coordinate (topleft);
			 		\draw[line width=2,densely dotted,blue] (0,0) -- (2,1.5) coordinate(topright);
			 		\draw[line width=1,densely dotted,black!70] (0,-.5) -- (0,1.8) coordinate(topmid);
			 		\fill[red] (0,0) circle (0.1);
			 		\path (topmid) -- (orig) -- (topleft) pic["$\theta/2$",draw=black,>=latex,,angle eccentricity=1.3,angle radius=20,->] {angle=topmid--orig--topleft};
			 		\draw[>=latex,->,line width=1] (-1,.75) -- ({-1+1.5/3},{.75+2/3});
			 		\draw[>=latex,->,line width=1] (0,1.6) -- (-.833,1.6);
			 		\node[anchor=north west] at (0,0) {\color{red}$\Gamma^e$};
			 		\node[anchor=south east] at (-1,.2) {\color{black}$\Omega^e$};
			 		\node[anchor=south west] at (1,.2) {\color{blue}$\tilde\Omega^e$};
			 		\node[anchor=north] at (-.9,1.5) {\color{black}$\bn$};
			 		\node[anchor=south] at (-.416,1.6) {\color{black}$\tilde\bn$};
			 		\node[anchor=west,black!70] at (0,1.9) {\footnotesize symmetry plane};
			 		\node at (0,1.9) {};
			 	\end{tikzpicture}} &
			\adjustbox{valign=c}{
			 	\begin{tikzpicture}
			 		\draw[line width=2,black] (0,0) coordinate (orig) -- (-2,0) coordinate (topleft);
			 		\draw[line width=2,densely dotted,blue] (0,0) -- (2,0) coordinate(topright);
			 		\draw[line width=1,densely dotted,black!70] (0,-.5) -- (0,1.8) coordinate(topmid);
			 		\fill[red] (0,0) circle (0.1);
			 		\draw[>=latex,->,line width=1] (-1,0) -- (-1,0.833);
			 		\draw[>=latex,->,line width=1] (0,1) -- (-0.833,1);
			 		\node[anchor=north west] at (0,0) {\color{red}$\Gamma^e$};
			 		\node[anchor=south] at (-1,-.6) {\color{black}$\Omega^e$};
			 		\node[anchor=south] at (1,-.6) {\color{blue}$\tilde\Omega^e$};
			 		\node[anchor=east] at (-1,.416) {\color{black}$\bn$};
			 		\node[anchor=south] at (-.416,1) {\color{black}$\tilde\bn$};
			 		\node[anchor=west,black!70] at (0,1.9) {\footnotesize symmetry plane};
			 		\node at (0,1.9) {};
			 	\end{tikzpicture}}
	\end{tabular}
\end{table}

In the following three subsections, the constraint enforcement for the various cases using the penalty and Lagrange multiplier method are summarized. Their linearizations are reported in \appref{s:l_g_tot}.

\subsubsection{Constraint formulation} \label{s:p_g_cnstr}

\begin{figure}[!ht]
	\centering
	 	\begin{tikzpicture}
	 		\draw[line width=2,black] (0,0) coordinate (orig) -- (-2,1.5) coordinate (topleft);
	 		\draw[line width=2,densely dotted,blue] (0,0) -- (2,1.5) coordinate(topright);
	 		\fill[red] (0,0) circle (0.1);
	 		\path (topright) -- (orig) -- (topleft) pic["$\theta$",draw=black,->,>=latex,,angle eccentricity=1.3,angle radius=20] {angle=topright--orig--topleft};
	 		\draw[>=latex,->,line width=1] (-1,.75) -- ({-1+1.5/3},{.75+2/3});
	 		\draw[>=latex,->,line width=1] (1,.75) -- ({1-1.5/3},{.75+2/3});
	 		\node[anchor=north] at (0,-.05) {\color{red}$\Gamma^e$};
	 		\node[anchor=south east] at (-1,.2) {\color{black}$\Omega^e$};
	 		\node[anchor=south west] at (1,.2) {\color{blue}$\tilde\Omega^e$};
	 		\node[anchor=north] at (-.9,1.7) {\color{black}$\bn$};
	 		\node[anchor=north] at (.9,1.7) {\color{black}$\tilde\bn$};
	 		\node at (0,1.9) {};
	 		\node at (0,-0.9) {$\bn\cdot\tilde\bn=\cos\alpha$};
	 	\end{tikzpicture}
 	\qquad\qquad
	 	\begin{tikzpicture}
	 		\draw[line width=2,black] (0,0) coordinate (orig) -- (-2,-1.5) coordinate (topleft);
	 		\draw[line width=2,densely dotted,blue] (0,0) -- (2,-1.5) coordinate(topright);
	 		\fill[red] (0,0) circle (0.1);
	 		\path (topleft) -- (orig) -- (topright) pic["$-\theta$",<-,draw=black,>=latex,,angle eccentricity=1.3,angle radius=20] {angle=topleft--orig--topright};
	 		\draw[>=latex,->,line width=1] (-1,-.75) -- ({-1-1.5/3},{-.75+2/3});
	 		\draw[>=latex,->,line width=1] (1,-.75) -- ({1+1.5/3},{-.75+2/3});
	 		\node[anchor=south] at (0,.05) {\color{red}$\Gamma^e$};
	 		\node[anchor=south west] at (-1,-1.25) {\color{black}$\Omega^e$};
	 		\node[anchor=south east] at (1,-1.25) {\color{blue}$\tilde\Omega^e$};
	 		\node[anchor=south] at (-1,-.3) {\color{black}$\bn$};
	 		\node[anchor=south] at (1,-.3) {\color{black}$\tilde\bn$};
	 		\node at (0,-1.9) {};
	 		\node at (0,-2.35) {$\bn\cdot\tilde\bn=\cos\alpha$};
	 	\end{tikzpicture}
	\caption{Ambiguity of constraint \eqref{e:p_g_coscnstr}. Both configurations equally fulfill the constraint, which leads to an ambiguity in the numerical solution. If the constraint in \eqsref{e:p_g_sincnstr} is considered in addition, the desired configuration will be uniquely defined.} \label{f:p_g_amb}
\end{figure}
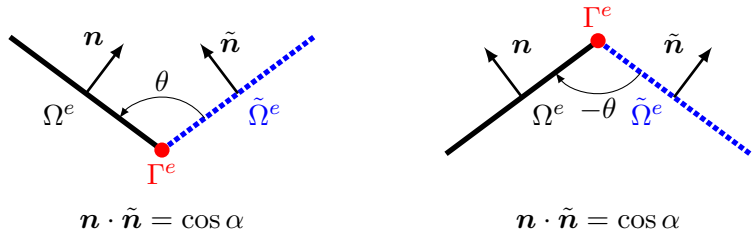

As shown in \tabref{t:p_g_css}, the general case requires to maintain a fixed angle $\theta$ between two adjacent patches. This implies that the angle between the surface normals, which is denoted $\alpha$, has to be constant for all load or time steps, i.e.
\eqb{l}
	\alpha-\alpha_0=0\,,\quad\forall\,\bx\in\Gamma\,.
\eqe
To avoid the computation and differentiation of the $\arccos$-function, the constraint is reformulated as
\eqb{l}
	\cos\alpha-\cos\alpha_0=0\,,\quad\forall\,\bx\in\Gamma\,, \label{e:p_g_coscnstr}
\eqe
with
\eqb{lll}
	\cos\alpha:=\bn\cdot\tilde\bn\,,\qquad\cos\alpha_0:=\bN\cdot\tilde\bN\,. \label{e:p_g_cos}
\eqe
Note that \eqsref{e:p_g_cos} makes use of the fact that the surface normals have unit length. The constraint in \eqsref{e:p_g_coscnstr} can uniquely handle angles within the range $[0,\pi]$. For the range $[0,2\pi)$, however, there is an ambiguity in the solution as illustrated in \figref{f:p_g_amb}.
To avoid this ambiguity, a second constraint is incorporated, i.e.
\eqb{l}
	\sin\alpha-\sin\alpha_0=0\,,\quad\forall\,\bx\in\Gamma\,, \label{e:p_g_sincnstr}
\eqe
where
\eqb{l}
	\sin\alpha:=\bigl(\bn\times\tilde\bn\bigr)\cdot\btau\,,\qquad\sin\alpha_0:=\bigl(\bN\times\tilde\bN\bigr)\cdot\btau_0\,. \label{e:p_g_sin}
\eqe
Thus, the following set of constraints needs to be enforced
\eqb{lllllll}
	g_\mathrm{c}\dis\cos\alpha_0-\cos\alpha \is 0\,,\quad&\forall\,\bx\in\Gamma\,,\\[2mm]
	g_\mathrm{s}\dis\sin\alpha_0-\sin\alpha \is 0\,,\quad&\forall\,\bx\in\Gamma\,. \label{e:p_g_cnstr}
\eqe
The constraints in Eq.~\eqref{e:p_g_cnstr} can then uniquely enforce any angle $\alpha_0,\alpha\in[0,2\pi]$.

\textbf{Remark:} For the planar setting that is illustrated in the third column of \tabref{t:p_g_css}, the simplification $\alpha_0=\alpha=0$ holds true. Thus, $\cos\alpha_0=1$ and $\sin\alpha_0=0$, and \eqsref{e:p_g_cnstr} turns into
\eqb{lllllll}
	g_\mathrm{c}^\mathrm{planar}\dis1-\cos\alpha \is 0\,,\quad&\forall\,\bx\in\Gamma\,,\\
	g_\mathrm{s}^\mathrm{planar}\dis-\sin\alpha \is 0\,,\quad&\forall\,\bx\in\Gamma\,. \label{e:p_g_cnstrsmpl_tmp}
\eqe
Note that the constraint $g_\mrs^\mathrm{planar}\!=0$ in \eqsref{e:p_g_cnstrsmpl_tmp} is automatically fulfilled if $g_\mrc^\mathrm{planar}\!=0$ holds. The two constraints thus simplify to
\eqb{l}
	1-\cos\alpha=0\,,\quad\forall\,\bx\in\Gamma\,,
\eqe
which can also be expressed as
\eqb{lll}
	\bg_\mrn^\mathrm{planar}\dis\bn-\tilde{\bn}=\boldsymbol{0}\,,\quad\forall\,\bx\in\Gamma\,. \label{e:p_g_cnstrsmpl}
\eqe

\subsubsection{Penalty method} \label{s:p_g_pen}

For the penalty method, the constraints in \eqsref{e:p_g_cnstr} are enforced by adding the potential
\eqb{l}
	\Pi_\mrn = \ds\int_{\Gamma_0}\frac{\eps_\mathrm{n}}{2}\,g_\mrn\,\dif S = \ds\int_{\Gamma_0}\eps_n\bigl(1-c_0\cos\alpha-s_0\sin\alpha\bigr)\,\dif S\,, \label{e:p_g_pipen}
\eqe
to the total potential of the system. Here, $g_\mrn:=g_\mrc^2+g_\mrs^2$, $c_0:=\cos\alpha_0$ and $s_0:=\sin\alpha_0$.  The penalty parameter $\eps_n\in(0,\infty)$ controls how well the constraint will be fulfilled. It has units of moment per length $[\mathrm{N}\mathrm{m}/\mathrm{m}=\mathrm{N}]$. The expression $\eps_\mrn\,g_\mrn$ remains finite as $\eps_\mrn\rightarrow\infty$ and $g_\mrc,g_\mrs\rightarrow0$. The derivative of $g_\mrn$ with respect to $\alpha$ is $g_\mrn'=2\,\sin(\alpha-\alpha_0)$. The positions of the maxima of $g_\mrn$, where $g_\mrn'=0$, fulfill $\alpha-\alpha_0=\pm\pi$. The minimum is found at $\alpha-\alpha_0=0$. A unique solution with the Newton-Raphson method can thus be obtained provided that the initial guess $\alpha_\mri$ fulfills $|\alpha_\mri-\alpha_0|<\pi$.

Now, the variation of the penalty potential from \eqsref{e:p_g_pipen} is given by (see \appref{s:l_hlp})
\eqb{l}
	\delta\Pi_\mrn=-\ds\int_{\Gamma_0}\eps_\mrn\bigl(\delta\btau\cdot\btheta+\delta\bn\cdot\tilde\bd+\delta\tilde\bn\cdot\bd\bigr)\,\dif S\,, \label{e:p_g_dpipen}
\eqe
with
\eqb{l}
	\btheta:=s_0\,\bn\times\tilde\bn=s_0\,\sin(\alpha)\,\btau\,,\quad\tilde\bd:=c_0\,\tilde\bn+s_0\,\tilde\bnu\,,\quad\bd:=c_0\,\bn+s_0\,\bnu\,. \label{e:p_g_hlp}
\eqe
Using the variation of $\btau$ given in Eq.~(\ref{e:v_varntau}.3), the first term in the expression for $\delta\Pi_\mrn$ vanishes as $\btau\cdot\delta\btau=0$. \eqsref{e:p_g_dpipen} then simplifies to\footnote{This is a simplification compared to the formulation of \cite{duong2017}.\label{footnote_p_g_smpl}}
\eqb{l}
	\delta\Pi_\mrn=-\ds\int_{\Gamma_0}\eps_\mrn\bigl(\delta\bn\cdot\tilde\bd+\delta\tilde\bn\cdot\bd\bigr)\,\dif S\,. \label{e:p_g_dpipensmpl}
\eqe
\eqsref{e:p_g_dpipensmpl} implies that the bending moments are transmitted exactly across the interface. They are given by $m_\tau=m_{\tilde\tau}=\eps_\mrn\sin(\alpha-\alpha_0)$, see \appref{s:b_pen}.\footnote{Note that the bending moments have been changed compared to the journal version.} The missing variations $\delta\bn$ and $\delta\tilde\bn$ in \eqsref{e:p_g_dpipensmpl} are given in \appref{s:l_hlp}. Inserting the FE approximations yields the element-level approximation of \eqsref{e:p_g_dpipensmpl}
\eqb{l}
	\delta\Pi^e_\mrn:=\delta\mx_e^\mrT\,\mf_\mrn^e+\delta\tilde{\mx}_e^\mrT\,\mf_{\tilde{n}}^e\,. \label{e:p_g_dpipene}
\eqe
The element-level force vectors in \eqsref{e:p_g_dpipene} are given by
\eqb{lll}
	\mf_\mrn^e \dis \ds\int_{\Gamma_0^e}\eps_\mrn\,\mN_{,\alpha}^\mrT\bigl(\tilde\bd\cdot\ba^\alpha\bigr)\,\bn\,\dif S\,,\\[4mm]
	\mf_{\tilde\mrn}^e \dis \ds\int_{\Gamma_0^e}\eps_\mrn\,\tilde\mN_\ca^\mrT\bigl(\bd\cdot\tilde\ba^\alpha\bigr)\,\tilde\bn\,\dif S\,, \label{e:p_g_fpen}
\eqe
with shape function array $\mN$ from \eqsref{e:s_shpN} and $\Gamma_0^e$ denoting a finite line element along the patch interface. The force vectors $\mf_\mrn^e$ and $\mf_{\tilde\mrn}^e$ have dimension $3\,n_e\times1$ and $3\,n_{\tilde{e}}\times1$, respectively. Here, $n_e$ and $n_{\tilde{e}}$ refer to the number of control points associated with the elements $\Omega^e$ and $\tilde\Omega^e$, respectively.

\textbf{Remark:} Plugging $\alpha_0=0$ into the force vectors in \eqsref{e:p_g_fpen}, leads to the simplified force vectors
\eqb{lllll}
	\mf_\mrn^{e,\mathrm{planar}} \dis \ds\int_{\Gamma_0^e}\eps_\mrn\,\mN_{,\alpha}^\mrT\bigl(\tilde\bn\cdot\ba^\alpha\bigr)\,\bn\,\dif S \is \ds\int_{\Gamma_0^e}\eps_\mrn\,\mN_{,\alpha}^\mrT\bigl(\bn\otimes\tilde\bn\bigr)\,\ba^\alpha\,\dif S\,,\\[4mm]
	\mf_{\tilde\mrn}^{e,\mathrm{planar}} \dis \ds\int_{\Gamma_0^e}\eps_\mrn\,\tilde\mN_\ca^\mrT\bigl(\bn\cdot\tilde\ba^\alpha\bigr)\,\tilde\bn\,\dif S \is \ds\int_{\Gamma_0^e}\eps_\mrn\,\tilde\mN_\ca^\mrT\bigl(\tilde\bn\otimes\bn\bigr)\,\tilde\ba^\alpha\,\dif S\,, \label{e:p_g_fpensmpl}
\eqe
for the penalty method.

\subsubsection{Lagrange multiplier method} \label{s:p_g_lag}

For the Lagrange multiplier approach, the constraints in \eqsref{e:p_g_cnstr} are reformulated as follows
\eqb{lll}
	\bar{g}_\mrc\dis1-\cos(\alpha-\alpha_0)=0\,,\\[2mm]
	\bar{g}_\mrs\dis\sin(\alpha-\alpha_0)=0\,,\label{e:p_g_cnstr2}
\eqe
to guarantee unique solutions with the Newton-Raphson method as long as $|\alpha_\mri-\alpha_0|<\pi/4$, with initial guess $\alpha_\mri$ \citep{duong2017}. The new potential then is
\eqb{l}
	\Pi_\mrn=\ds\int_{\Gamma_0} q\,\bigl(\bar{g}_\mrc+\bar{g}_\mrs\bigr)\,\dif S=\ds\int_{\Gamma_0} q\,\bigr(1-\cos(\alpha-\alpha_0)+\sin(\alpha-\alpha_0)\bigr)\,\dif S\,, \label{e:p_g_pilag}
\eqe
with Lagrange multiplier $q\in L^2(\Gamma_0)$. Using Eqs.~\eqref{e:p_g_dc2}--\eqref{e:p_g_ds3}, the variation of the Lagrange multiplier potential in \eqsref{e:p_g_pilag} is given by
\eqb{l}
	\delta\Pi_\mrn = \ds\int_{\Gamma_0}\delta q\,\big(\bar{g}_\mrc+\bar{g}_\mrs\big)\,\dif S - \ds\int_{\Gamma_0} q\,\bigl(\delta\btau\cdot\btheta+\delta\bn\cdot\tilde\bd+\delta\tilde\bn\cdot\bd\bigr)\,\dif S\,, \label{e:p_g_dpilag}
\eqe
with the redefinitions
\eqb{l}
	\btheta:=(s_0-c_0)\,\bn\times\tilde\bn\,,\quad\bd:=(s_0+c_0)\,\bn+(s_0-c_0)\,\bnu\,,\quad\tilde\bd:=(s_0+c_0)\,\tilde\bn+(s_0-c_0)\,\tilde\bnu\,, \label{e:p_g_hlp2}
\eqe
and the variation of the Lagrange multiplier $\delta q\in L^2(\Gamma_0)$. \eqsref{e:p_g_dpilag} simplifies to\textsuperscript{\ref{footnote_p_g_smpl}}
\eqb{l}
	\delta\Pi_\mrn = \ds\int_{\Gamma_0}\delta q\,\big(\bar{g}_\mrc+\bar{g}_\mrs\big)\,\dif S - \ds\int_{\Gamma_0} q\,\bigl(\delta\bn\cdot\tilde\bd+\delta\tilde\bn\cdot\bd\bigr)\,\dif S\,, \label{e:p_g_dpilagsmpl}
\eqe
since $\btau\cdot\delta\btau=0$, see previous section. \eqsref{e:p_g_dpilagsmpl} implies that the bending moment $m_\tau=m_{\tilde\tau}=q$ is transmitted exactly across the patch interface, see \appref{s:b_lag}.\footnote{Note that the bending moments have been changed compared to the journal version.} The FE approximation of the Lagrange multiplier along $\Gamma^e$ can be written as
\eqb{l}
	q\approx\hat\mN_\mrq\,\hat\mq_e\,, \label{e:p_g_qfe}
\eqe
where $\hat\mN_\mrq$ and $\hat\mq_e$ are the corresponding shape function array and nodal values of line element $\Gamma^e$, respectively, similar to Eqs.~\eqref{e:p_axi}--\eqref{e:s_shpNxi}. The discretized form of the system with Lagrange multipliers is a saddle-point problem that has a zero-tangent block in the stiffness matrix, see Eq.~(\ref{e:l_g_lag_f3}.3). The chosen finite-dimensional subspace for the Lagrange multiplier thus must satisfy the LBB-condition in order to obtain a stable solution. In the present work, the displacement field $\bx$ and its variation $\delta\bx$ are interpolated by $C^1$-continuous, bi-quadratic NURBS and the Lagrange multiplier $q$ with either $C^{-1}$- or $C^0$-continuous functions. The latter are obtained by either a constant interpolation on each line element along the patch interfaces, or linear interpolation, see \figref{f:p_css}. \cite{brivadis2015} investigate suitable Lagrange multiplier spaces depending on the approximation order of the primal variable for a displacement constraint used for non-conforming meshes. For this, the element-wise constant interpolation of $q$ is shown to be LBB-stable for the used bi-quadratic approximation of $\bx$. The authors note that there is no proof on LBB-stability for the normal constraints used for conforming meshes, but the results in \cite{duong2017} and \secref{s:n_s_tot} indicate a stable method for the considered examples. Note that for the element-wise linear interpolation, the Lagrange multipliers need to be repeated at patch junctions with valence greater than or equal to three in order to avoid over-constraining. A more detailed mathematical investigation on this is left open for future work, also see \secref{s:n_evl}. The discretized element-level variation of the Lagrange multiplier potential is given by
\eqb{l}
	\delta\Pi_\mrn^e=\delta\mx_e^\mrT\,\bar\mf_\mrn^e+\delta\tilde\mx_e^\mrT\,\bar\mf_{\tilde\mrn}^e+\delta\mq_e^\mrT\,\bar\mf_\mrq^e\,, \label{e:p_g_dpilage}
\eqe
with the force vectors
\eqb{lll}
	\bar\mf_\mrn^e\dis \ds\int_{\Gamma_0^e}q\,\mN_\ca^\mrT\,\bigl(\tilde\bd\cdot\ba^\alpha\bigr)\,\bn\,\dif S\,,\\[4mm]
	\bar\mf_{\tilde\mrn}^e \dis \ds\int_{\Gamma_0^e}q\,\tilde\mN_\ca^\mrT\,\bigl(\bd\cdot\tilde\ba^\alpha\bigr)\,\tilde\bn\,\dif S\,,\\[4mm]
	\bar\mf_\mrq^e \dis \ds\int_{\Gamma_0^e}\hat\mN_\mrq^\mrT\,\bigl(\bar{g}_\mrc+\bar{g}_\mrs\bigr)\,\dif S\,. \label{e:p_g_flag}
\eqe
Here, the variables $\bd$ and $\tilde\bd$ are defined in \eqsref{e:p_g_hlp2}. The force vectors $\bar\mf_\mrn^e$ and $\bar\mf_{\tilde\mrn}^e$ have dimension $3\,n_e\times1$ and $3\,n_{\tilde{e}}\times1$, respectively. Here, $n_e$ and $n_{\tilde{e}}$ refer to the number of control points associated with the elements $\Omega^e$ and $\tilde\Omega^e$, respectively. Further, $\bar\mf_\mrq^e$ has dimension $n_\mrq\times1$ with $n_\mrq$ denoting the number of Lagrange multipliers on the line element $\Gamma^e$, see \figref{f:p_css}.

\textbf{Remark:} Note that the Lagrange multiplier method only converges correctly if it is integrated consistently. So for constant interpolation of the Lagrange multiplier, one should not use the trapezoidal rule, while for linear interpolation, the midpoint rule should not be used.

\textbf{Remark:} For the simplified constraint in \eqsref{e:p_g_cnstrsmpl}, the first two force vectors in \eqsref{e:p_g_flag} are simplified by substituting $\bd\leftarrow\bd^\mathrm{\,planar}$ and $\tilde\bd\leftarrow\tilde\bd^\mathrm{\,planar}$, where
\eqb{l}
	\bd^\mathrm{\,planar} := \bn-\bnu\,,\qquad\tilde\bd^\mathrm{\,planar}:=\tilde\bn-\tilde\bnu\,.
\eqe
The force vector $\bar\mf_\mrq^e$ in \eqsref{e:p_g_flag} stays the same.

\subsection{\texorpdfstring{$C^1$}{C1}-continuity constraint for the phase field}\label{s:p_c_tot}
This section discusses formulations to enforce $C^1$-continuity for the phase field $\phi$ within the weak forms \eqref{e:c_f} and \eqref{e:f_f}. Their linearization is outlined in \appref{s:l_c_tot}.

\subsubsection{Constraint formulation} \label{s:p_c_cnstr}

\begin{figure}[!ht]
	\centering
		\begin{tikzpicture}
			\node at (0,0) {\includegraphics[height=0.2\textwidth]{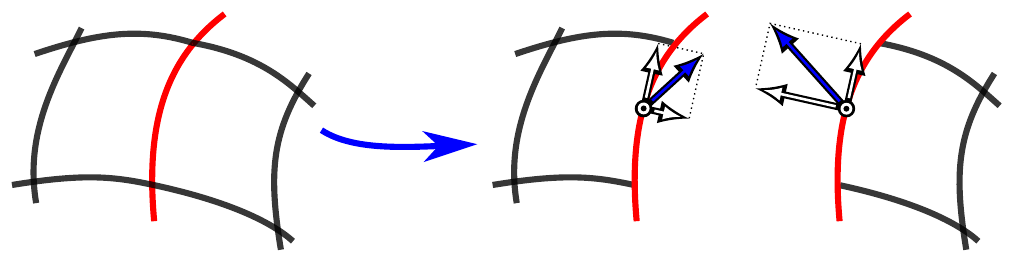}};
				\node at (-3.655,.6) {\color{red}$\Gamma^e$};
				\node at (-4.855,.3) {$\Omega^e$};
				\node at (-3.45,0) {$\tilde\Omega^e$};
				\node at (1.4,.3) {$\bx$};
				\node at (2.6,1.075) {\color{blue}$\nablas\phi$};
				\node[anchor=east] at (1.75,.65) {\scriptsize $\nablas\phi\cdot\btau$};
				\node[anchor=north] at (2.1,.12) {\scriptsize $\nablas\phi\cdot\bnu$};
				\node at (4.4,.3) {$\tilde\bx$};
				\node at (3.15,1.6) {\color{blue}$\nablas\phitilde$};
				\node[anchor=west] at (4.28,.7) {\scriptsize $\nablas\phitilde\cdot\btau$};
				\node[anchor=north] at (3.55,.4) {\scriptsize $\nablas\phitilde\cdot\tilde\bnu$};
		\end{tikzpicture}
	\caption[]{Surface gradients $\nablas\phi$ and $\nablas\phitilde$ at points $\bx$ and $\tilde\bx$, respectively. Due to conforming meshes at the interface $\Gamma^e$, the parametrization along the interface is the same for both patches.{\footnotemark} Thus, the tangential component of the surface gradients is equal by construction, i.e. $\nablas\phi\cdot\btau=\nablas\phitilde\cdot\btau$. Only their normal components need to be enforced to be equal, see \eqsref{e:p_c_cnstr3}.} \label{f:p_c_vcs}
\end{figure}

As \figref{f:p_c_vcs} illustrates, the surface gradients of the phase field, $\nablas\phi$ and $\nablas\phitilde$, need to be equal to ensure $C^1$-continuity of $\phi$. The constraint can thus be written as
\eqb{l}
	\nablas\phi-\nablas\phitilde=\mathbf{0}\,,\quad\forall\,\bx\in\Gamma\,. \label{e:p_c_cnstr1}
\eqe
The parametrization of the elements $\tilde\Omega^e$ and $\Omega^e$ is the same along the interface because here, conforming meshes are considered, see \figref{f:p_vcs}.
\footnotetext{Note that the parametrizations of elements $\Omega^e$ and $\tilde\Omega^e$ might be different. But for the line integrals, only the parametrization of the patch interface $\Gamma^e$ is of importance.}
Thus, by construction, the tangential component of $\nablas\phi$ is already $C^1$-continuous across the patch interface. Hence, only its normal component needs to be enforced to be equal. The constraint in \eqsref{e:p_c_cnstr1} then simplifies to
\eqb{lll}
	g_\nabla\dis\nablas\phi\cdot\bnu+\nablas\phitilde\cdot\tilde\bnu=0\,,\quad\forall\,\bx\in\Gamma\,. \label{e:p_c_cnstr3}
\eqe
It has the unit $[\mathrm{m}^{-1}]$. In the case of fracture, the gradients are evaluated in the reference configuration and hence, constraint \eqref{e:p_c_cnstr3} is rewritten as
\eqb{lll}
	g_\nabla^\mathrm{frac}\dis\nablao\phi\cdot\bnu_0+\nablao\phitilde\cdot\tilde\bnu_0=0\,,\quad\forall\,\bx\in\Gamma\,. \label{e:p_c_cnstr2}
\eqe
Based on Eq.~(\ref{e:s_diff}.1), \eqsref{e:p_c_cnstr3} can be rewritten as
\eqb{l}
	g_\nabla=\phi_\ca\,\ba^\alpha\cdot\bnu+\phitilde_\ca\,\tilde\ba^\alpha\cdot\tilde\bnu=0\,,\quad\forall\,\bx\in\Gamma\,. \label{e:p_c_cnstr4}
\eqe
Keeping $\bx$ fixed, the variation of \eqsref{e:p_c_cnstr4} is given by
\eqb{l}
	\delta g_\nabla = \delta\phi_{,\alpha}\,\ba^\alpha\cdot\bnu+\delta\tilde\phi_\ca\,\tilde\ba^\alpha\cdot\tilde\bnu\,. \label{e:p_c_dgnbl}
\eqe
Keeping $\bx$ fixed is sufficient for building a numerical method, even though it introduces a variational inconsistency as is noted below. The reason is that constraint \eqref{e:p_c_cnstr4} is used for the phase field, and not for the displacement field. Thus, the latter can be fixed in the variation and hence, \eqsref{e:p_c_dgnbl} does not include the variations of the tangent vectors $\ba^\alpha$, $\tilde\ba^\alpha$ or the interface normals $\bnu$, $\tilde\bnu$, which simplifies the numerical formulations greatly. The linearization of $g_\nabla$ and $\delta g_\nabla$, however, needs to include the linearizations with respect to $\bx$, see \appref{s:l_hlp}, in order to ensure quadratic convergence of the Newton-Raphson method. Based on the FE approximations in \eqsref{e:s_phi} and the analogous quantities for $\phitilde$, the variation of the constraint in \eqsref{e:p_c_dgnbl} is discretized as
\eqb{l}
	\delta g_\nabla \approx \delta\bphi_e^\mrT\,\bar\mN_\ca^\mrT\,\bigl(\ba^\alpha\cdot\bnu\bigr) + \delta\tilde\bphi_e^\mrT\,\tilde{\bar\mN}_\ca^\mrT\,\bigl(\tilde\ba^\alpha\cdot\tilde\bnu\bigr)\,, \label{e:p_c_dgnblfe}
\eqe
where the tangent vectors and interface normals are also understood to be discrete.

\textbf{Remark:} In the case of planar connections of patches, the relation $\bn=\tilde\bn$ can be used, see \eqsref{e:p_g_cnstrsmpl}. From \eqsref{e:p_crd} then follows that $\bnu=-\tilde\bnu$.
The general constraint in \eqsref{e:p_c_cnstr4} now simplifies to
\eqb{l}
	g_\nabla^\mathrm{planar}=\Bigl(\phi_\ca\,\ba^\alpha-\phitilde_\ca\,\tilde\ba^\alpha\Bigr)\cdot\bnu=0\,,\quad\forall\,\bx\in\Gamma\,. \label{e:p_c_cnstr4smpl}
\eqe
Its variation is then given by
\eqb{l}
	\delta g_\nabla^\mathrm{planar} =\Bigl( \delta\phi_{,\alpha}\,\ba^\alpha-\delta\tilde\phi_\ca\,\tilde\ba^\alpha\Bigr)\cdot\bnu\,. \label{e:p_c_dgnblsmpl}
\eqe

The following sections discuss a penalty and a Lagrange multiplier approach to enforce these constraints. Both methods are based on a potential $\Pi_\nabla$ that is added to the system's total potential. Its variation and discretization will be outlined. Their linearization is derived in \appref{s:l_c_tot}.

\textbf{Remark:} In the present work, only planar connections are considered for the phase separation examples in \secref{s:n_c_tot}. In contrast, the fracture examples in \secref{s:n_f_tot} include the more general case of non-smooth connections between patches, but the phase field gradient is based on the reference configuration. To cover all cases, the following derivations only contain the general case.

\subsubsection{Penalty method} \label{s:p_c_pen}
In analogy to the penalty potential for the $G^1$-constraint, see \eqsref{e:p_g_pipen}, the penalty potential for the phase field constraint in \eqsref{e:p_c_cnstr4} is given by
\eqb{l}
	\Pi_\nabla = \ds\int_{\Gamma_0}\frac{\eps_\phi}{2}g_\nabla^2\,\dif S\,, \label{e:p_c_pipen}
\eqe
where the penalty parameter $\eps_\phi\in(0,\infty)$ with units $[\mathrm{kg}\,\mathrm{m}^3/\mathrm{s}^2]$ controls how well the constraint will be fulfilled. The variation of $\Pi_\nabla$ is
\eqb{l}
	\delta\Pi_\nabla = \ds\int_{\Gamma_0} \eps_\phi\,g_\nabla\,\delta g_\nabla\,\dif S\,, \label{e:p_c_dpipen}
\eqe
where $\delta g_\nabla$ is given in \eqsref{e:p_c_dgnbl}.\footnote{Since $\bx$ is kept fixed in $\delta g_\nabla$,  \eqsref{e:p_c_dpipen} is not the full variation of $\Pi_\nabla$.} Based on \eqsref{e:p_c_dgnblfe}, the discretized variation of the penalty potential becomes
\eqb{l}
	\delta\Pi_\nabla^e = \delta\bphi_e^\mrT\,\mf_\phi^e+\delta\tilde\bphi_e^\mrT\,\mf_\phitilde^e\,. \label{e:p_c_dpipene}
\eqe
The general force vectors in \eqsref{e:p_c_dpipene} are given by\footnote{Note that the strong form is multiplied with the factor $2\,\ell_0/\sG_\mrc$ in the phase field model for fracture \citep{paul2020}. The force vectors are thus, implicitly scaled with this factor.}
\eqb{lll}
	\mf_\phi^e \dis \ds\int_{\Gamma_0^e}\eps_\phi\,\bar\mN_\ca^\mrT\,g_\nabla\,\bigl(\ba^\alpha\cdot\bnu\bigr)\,\dif S\,,\\[4mm]
	\mf_\phitilde^e \dis \ds\int_{\Gamma_0^e}\eps_\phi\,\tilde{\bar\mN}_\ca^\mrT\,g_\nabla\,\bigl(\tilde\ba^\alpha\cdot\tilde\bnu\bigr)\,\dif S\,, \label{e:p_c_fpen}
\eqe
with $g_\nabla$ from \eqsref{e:p_c_cnstr4}.

\paragraph{Choice of the penalty parameter} ~\\[2mm]
Based on numerical investigations, the following penalty parameter for phase separations is proposed
\eqb{l}
	\eps_\phi=\eps_\phi^0\,2^{d(p-1)}\,, \label{e:p_c_epsch}
\eqe
with $\eps_\phi^0=1000\,N\omega L_0^3$. Here, $L_0$ refers to a reference length stemming from the non-dimen\-siona\-lization of \cite{zimmermann2019}, see \secref{s:n_c_tot}. The polynomial order of the discretization is $p$ and the refinement depth is denoted $d$. According to \eqsref{e:p_c_epsch}, finer meshes will lead to a larger penalty parameter.

For brittle fracture, the following problem-independent penalty parameter is proposed
\eqb{l}
	\eps_\phi=\eps_\phi^0\,\dfrac{\Delta t_{\max}}{\Delta t}\,\dfrac{1}{\Delta x_{\min}^\Gamma}\,\dfrac{\Delta x_{\max}^{\Gamma}}{\Delta x^{\Gamma}}\,, \label{e:p_c_epsf}
\eqe
with $\eps_\phi^0=0.1\,E_0\,L_0^4$.\footnote{Note that for fracture, $\eps_\phi$ and $\eps_\phi^0$ have different units since the first one is scaled by $1/\Delta x^\Gamma_{\min}$, which introduces the units of $[\mathrm{m}^{-1}]$.} The reference stiffness $E_0$ ($[\mathrm{N}/\mathrm{m}]=[\mathrm{kg}/\mathrm{s}^2]$) and length $L_0$ ($[\mathrm{m}]$) stem from the non-dimensionization described in \cite{paul2020}. Further, $\Delta x_\mathrm{min}^\Gamma$ and $\Delta x_\mathrm{max}^\Gamma$ refer to the minimum and maximum length of the line elements along the patch interfaces, respectively, whereas $\Delta x^\Gamma$ refers to the one from the currently considered line element. The current and maximum time steps are $\Delta t$ and $\Delta t_{\max}$, respectively. The proposed penalty parameter in \eqsref{e:p_c_epsf} leads to the following behavior:
\begin{itemize}
	\item In case of crack evolution, the adaptive time stepping scheme from \cite{paul2020} can lead to small time steps. The scaling with $1/\Delta t$ then ensures a high penalty parameter in case of crack evolution, whereas the penalty parameter is kept smaller in case of no crack propagation (when the phase field does not change significantly).
	\item In cases of uniform meshes, the scaling factor $1/\Delta x_{\min}^{\Gamma}$ will ensure that the penalty parameter increases with mesh refinement. A similar behavior is obtained with the penalty parameter for the phase transition model, see \eqsref{e:p_c_epsch}.
	\item In cases of locally refined meshes, which are often used in phase field fracture models, the scaling factor $\Delta x_{\max}^{\Gamma}/\Delta x^{\Gamma}$ increases the penalty parameter in regions of highly resolved meshes. Thus, the constraint will be fulfilled more accurately in regions of fracture since these are the areas of primary interest, also see \secref{s:n_f_vld}.
\end{itemize}
The value $\eps_\phi^0=0.1\,E_0\,L_0^4$ seems small, but it yields sufficiently accurate results while preventing ill-conditioning, see \secref{s:n_f_vld}.

\subsubsection{Lagrange multiplier method} \label{s:p_c_lag}
In the following, a Lagrange multiplier approach to enforce constraint \eqref{e:p_c_cnstr4} is described. Since the constraint is scalar, the additional potential reads
\eqb{l}
	\Pi_\nabla = \ds\int_{\Gamma_0}\lambda\,g_\nabla\,\dif S\,, \label{e:p_c_pilag}
\eqe
with Lagrange multiplier $\lambda\in L^2(\Gamma_0)$. The variation of potential \eqref{e:p_c_pilag} is
\eqb{l}
	\delta\Pi_\nabla = \ds\int_{\Gamma_0}\delta\lambda\,g_\nabla\,\dif S + \ds\int_{\Gamma_0}\lambda\,\delta g_\nabla\,\dif S\,, \label{e:p_c_dpilag}
\eqe
where $\delta g_\nabla$ is given in \eqsref{e:p_c_dgnbl} and $\delta\lambda\in L^2(\Gamma_0)$ is the variation of the Lagrange multiplier. LBB-stability applies analogously to \secref{s:p_g_lag}. The FE approximations of the Lagrange multiplier and its variation are
\eqb{l}
	\lambda\approx\hat\mN_\lambda\,\hat\blam_e\,,\quad\mathrm{and}\quad\delta\lambda\approx\hat\mN_\lambda\,\delta\hat\blam_e\,, \label{e:p_c_lfe}
\eqe
with shape function array
\eqb{lllll}
	\hat\mN_\lambda:= [N_{i_1}]\,,\quad\mathrm{or}\quad\hat\mN_\lambda:= [N_{i_1},\,N_{i_2}]\,, \label{e:p_shpNlag}
\eqe
depending on the order of interpolation (constant or linear, respectively). The nodal values of the Lagrange multipliers and their variation are denoted $\hat\blam_e$ and $\delta\hat\blam_e$, respectively. These and the shape function array in \eqsref{e:p_shpNlag} are defined on the patch interfaces, in analogy to Eqs.~\eqref{e:p_axi}--\eqref{e:s_shpNxi}. Using the FE approximations of $\delta\lambda$ from \eqsref{e:p_c_lfe} and of $\delta g_\nabla$ from \eqsref{e:p_c_dgnblfe}, the discretized variation of the Lagrange multiplier potential becomes
\eqb{l}
	\delta\Pi_\nabla^e=\delta\bphi_e^\mrT\,\bar\mf_\phi^e+\delta\tilde\bphi_e^\mrT\,\bar\mf_\phitilde^e+\delta\hat\blam_e^\mrT\,\bar\mf_\lambda^e\,. \label{e:p_c_dpilage}
\eqe
The general force vectors in \eqsref{e:p_c_dpilage} are given by
\eqb{lll}	
	\bar\mf_\phi^e \dis \ds\int_{\Gamma^e_0} \bar\mN_\ca^\mrT\,\lambda\,\bigl(\ba^\alpha\cdot\bnu\bigr)\,\dif S\,,\\[4mm]
	\bar\mf_\phitilde^e \dis \ds\int_{\Gamma^e_0} \tilde{\bar\mN}_\ca^\mrT\,\lambda\,\bigl(\tilde\ba^\alpha\cdot\tilde\bnu\bigr)\,\dif S\,,\\[4mm]
	\bar\mf_\lambda^e \dis \ds\int_{\Gamma^e_0} \hat\mN_\lambda^\mrT\,g_\nabla\,\dif S\,, \label{e:p_c_flag}
\eqe
with $g_\nabla$ from \eqsref{e:p_c_cnstr4}.

\section{Numerical examples} \label{s:n_tot}
This section shows several numerical examples that illustrate the influence of the patch constraints from \secref{s:p_tot} on the numerical solution. At first, \secref{s:n_s_tot} focuses on the pure shell framework and the influence of the $G^1$-continuity constraint from \secref{s:p_g_tot}. Secondly, phase transitions on thin shells, which are based on a multi-patch description, are investigated in \secref{s:n_c_tot}. Similar investigations are made for brittle fracture in \secref{s:n_f_tot}.

In the subsequent sections, the following notation is used: The surface consists of $n_\mathrm{patch}$ patches, which are discretized by $n_\mathrm{sel}$ surface elements and $n_\mathrm{CP}$ control points. The penalty method is abbreviated by $\mathrm{PM}$ and the Lagrange multiplier method with constant/linear interpolation by constant/linear $\mathrm{LMM}$. For all examples, bi-quadratic NURBS are used for the interpolation of $\bx$, $\delta\bx$, $\phi$ and $\delta\phi$.

\textbf{Remark:} Numerical integration of surface integrals $\int_{\Omega_0^e}\tightdots\,\dif S$ is performed using $(p+1)\times(q+1)$ Gaussian quadrature points. Here, $p$ and $q$ refer to the polynomial orders in the two parametric directions $\xi^\alpha$, $\alpha=1,2$. In contrast, the integration of $n_\mathrm{lel}$ line integrals $\int_{\Gamma_0^e}\tightdots\,\dif S$ is performed using a different number of quadrature points $n_\mathrm{qp}$. If not stated otherwise, Gaussian quadrature with $n_\mathrm{qp}=3$ is used. For $n_\mathrm{qp}=1$ this corresponds to the midpoint rule, whereas for $n_\mathrm{qp}=2$,  the trapezoidal rule (`$n_\mathrm{qp}=2$, Trapezoidal') is used instead of Gaussian quadrature.

\subsection{Deforming shells} \label{s:n_s_tot}
This section investigates the $G^1$-continuity constraint from \secref{s:p_g_tot} within the pure mechanical shell framework (where $\phi$ is not unknown). The discussion includes comparisons to given solutions and convergence rates. The examples in this section are based on quasi-static conditions. In contrast to the work of \cite{duong2017}, here, a simpler formulation is derived and used, see Secs.~\ref{s:p_g_pen}--\ref{s:p_g_lag}. Further, patch junctions with valence three and five are investigated  since these kinds of junctions are of special interest for spline parametrizations.

\subsubsection{Pure bending} \label{s:n_s_bnd}
\begin{figure}[!ht]
	\centering
		\subfloat[Setup for $n_\mathrm{patch}=5$\label{f:n_s_bnd_stp1left}]{
			\begin{tikzpicture}
				\node at (0,0) {\includegraphics[height=.18\textwidth]{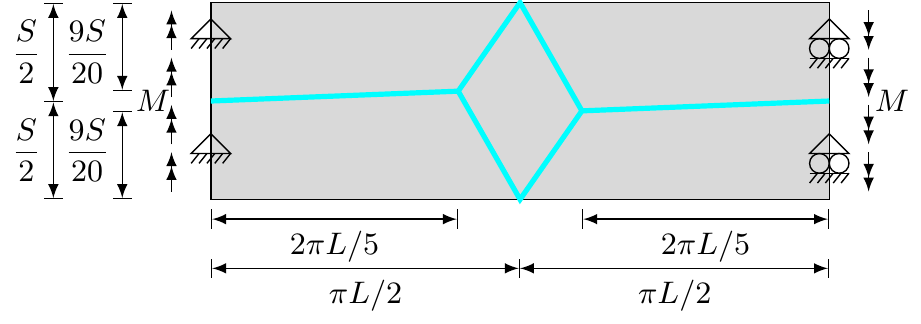}};
				\draw[>=latex,->,draw=black] (-0.5,2) -- (-0.5,0.65);
				\node[align=center,fill=white,draw=blue!0.5!cyan] at (-0.5,1.8) {$C^0$-lines};
			\end{tikzpicture}
		}
	\qquad
		\subfloat[Relative error ${[-]}$ in $H$ for linear $\mathrm{LMM}$, $n_\mathrm{sel}=52$\label{f:n_s_bnd_stp1right}]{\includegraphics[trim=630 135 105 15,clip,height=0.25\textwidth]{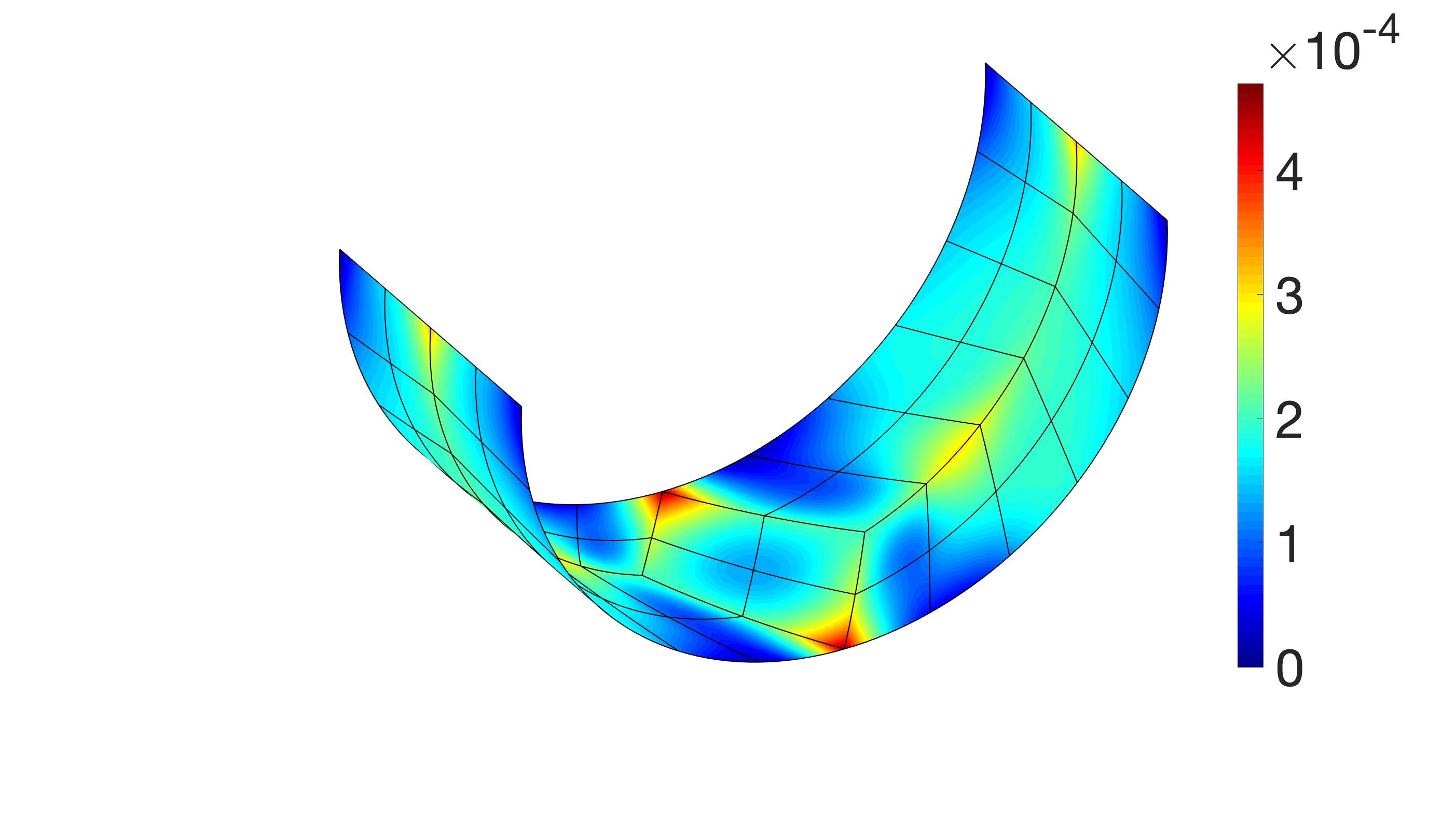}}
\caption{Pure bending: (a) Five-patch geometry (the out-of-plane movement is constrained at the left and right edge) and (b) relative error in the mean curvature $H$ plotted over the deformed configuration.} \label{f:n_s_bnd_stp1}
\end{figure}

\begin{figure}[!ht]
	\centering
		\subfloat[Setup for $n_\mathrm{patch}=7$\label{f:n_s_bnd_stp2left}]{
			\begin{tikzpicture}
				\node at (0,0) {\includegraphics[height=.18\textwidth]{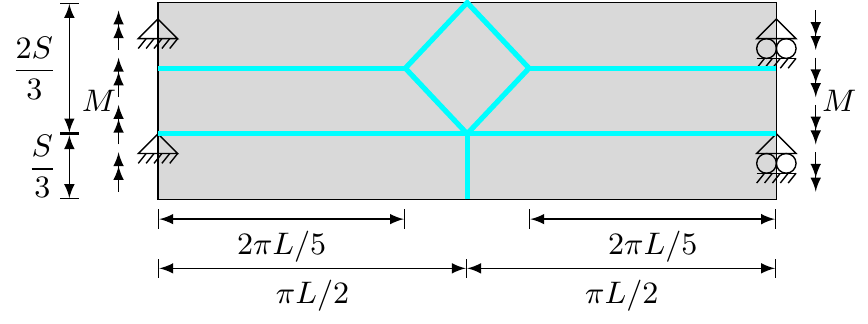}};
				\draw[>=latex,->,draw=black] (-0.5,2) -- (-0.5,0.875);
				\node[align=center,fill=white,draw=blue!0.5!cyan] at (-0.5,1.8) {$C^0$-lines};
			\end{tikzpicture}
		}
	\qquad
		\subfloat[Relative error ${[-]}$ in $H$ for linear $\mathrm{LMM}$, $n_\mathrm{sel}=76$\label{f:n_s_bnd_stp2right}]{\includegraphics[trim=630 135 105 15,clip,height=0.25\textwidth]{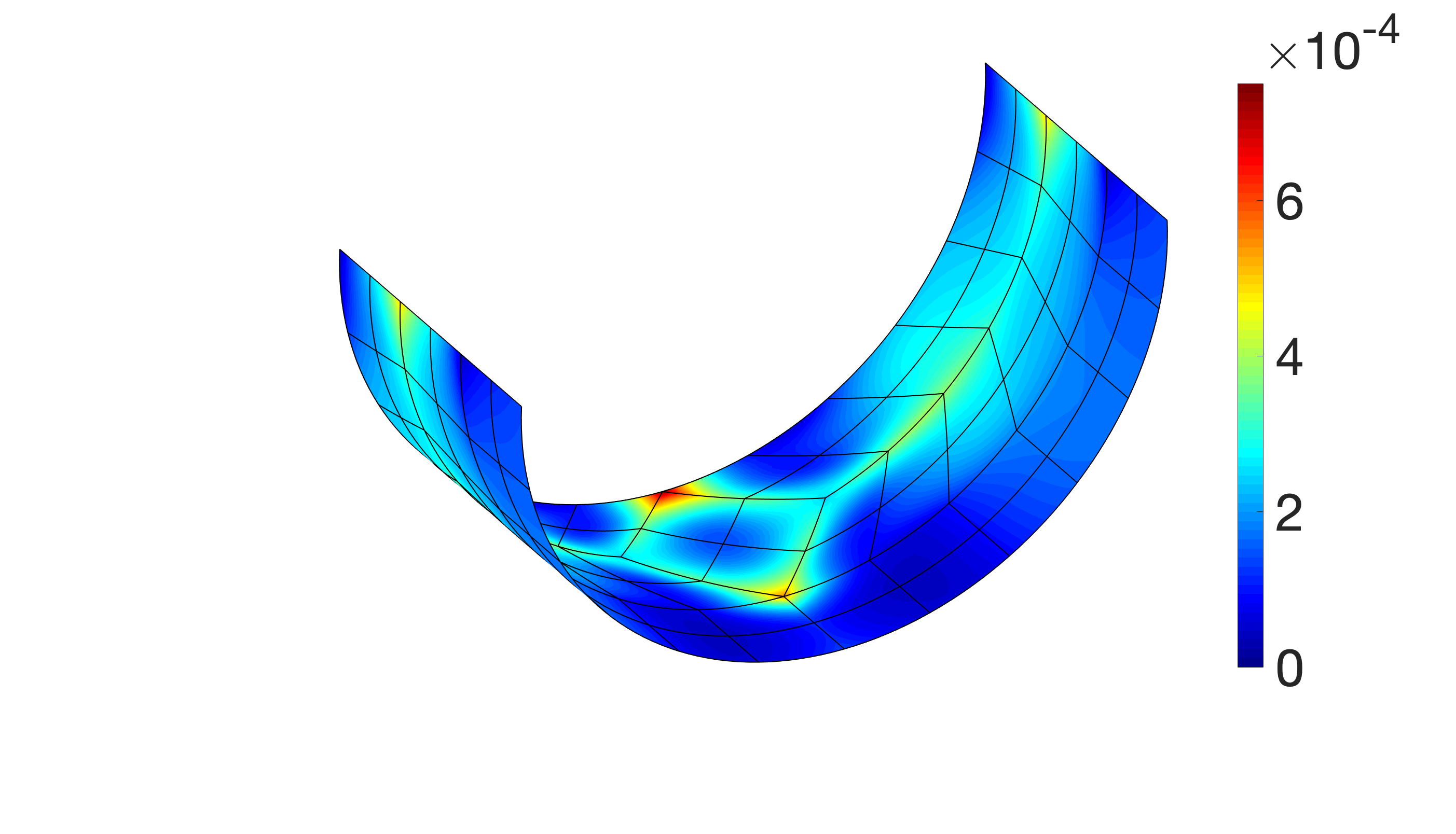}}
\caption{Pure bending: (a) Seven-patch geometry (the out-of-plane movement is constrained at the left and right edge) and (b) relative error in the mean curvature $H$ plotted over the deformed configuration.} \label{f:n_s_bnd_stp2}
\end{figure}

First, the pure bending of an initially flat shell is considered. Here, the Canham material model is used, see \cite{duong2017}. The same material parameters as in their work are used. The problem is normalized by the unit length $L_0$ and the unit stiffness $c$. The setup is depicted in Figs.~\ref{f:n_s_bnd_stp1left} and \ref{f:n_s_bnd_stp2left} for different number of patches.
As shown, a distributed bending moment $M=1\,c_0\,L_0^{-1}$ is applied at the two opposite edges to obtain pure bending. Figs.~\ref{f:n_s_bnd_stp1right} and \ref{f:n_s_bnd_stp2right} show the relative error of the mean curvature $H$ in comparison to the analytical solution given in \cite{sauer2017a}.
\begin{figure}[!ht]
	\centering
	\begin{tikzpicture}
		\pgfplotsset{width=1\textwidth,height=0.2\textwidth,compat=newest,}
		\begin{axis}[hide axis,xmin=0,xmax=0.00001,ymin=0,ymax=0.00001,legend cell align={left},
					 legend columns=2,legend style={/tikz/every even column/.append style={column sep=2ex}}]
    			\addlegendimage{black,line width=1,mark=square,mark size=2,mark options=solid}
   			 	\addlegendentry{$n_\mathrm{patch}=1$};
  			  	\addlegendimage{blue,densely dotted,line width=1,mark=square,mark size=3,mark options=solid}
 			   	\addlegendentry{$n_\mathrm{patch}=5$, constant $\mathrm{LMM}$};
  			  	\addlegendimage{blue,densely dotted,line width=1,mark=o,mark size=2,mark options=solid}
  			  	\addlegendentry{$n_\mathrm{patch}=5$, linear $\mathrm{LMM}$};
  			  	\addlegendimage{red,densely dotted,line width=1,mark=square,mark size=2,mark options=solid}
  			  	\addlegendentry{$n_\mathrm{patch}=5$, $\mathrm{PM}$: $\eps_\mrn=1/3200\,n_\mathrm{sel}^2\,c\,L_0^{-1}$};
  			  	\addlegendimage{red,densely dotted,line width=1,mark=x,mark size=4,mark options=solid}
  			  	\addlegendentry{$n_\mathrm{patch}=5$, $\mathrm{PM}$: $\eps_\mrn=1/320\,n_\mathrm{sel}^2\,c\,L_0^{-1}$};
   			 	\addlegendimage{red,solid,line width=1,mark=triangle,mark size=5,mark options=solid}
   			 	\addlegendentry{$n_\mathrm{patch}=7$, $\mathrm{PM}$: $\eps_\mrn=1/3200\,n_\mathrm{sel}^2\,c\,L_0^{-1}$};
   			 	\addlegendimage{blue,solid,line width=1,mark=triangle,mark size=4,mark options=solid}
   			 	\addlegendentry{$n_\mathrm{patch}=7$, linear $\mathrm{LMM}$};
  			 \end{axis}
	\end{tikzpicture}
	\\ \vspace{-3mm}
	\subfloat[]{%
		\begin{tikzpicture}
			\def\cdot{\times}
			\begin{axis}[xmode=log,ymode=log,grid=both,xlabel={Number of elements $n_\mathrm{sel}\:[-]$},ylabel={$L_2$-error of the displacement $[-]$},width=0.45\textwidth,xmin=10,xmax=1e5,ymin=1e-7,ymax=1e-1,legend cell align={left},legend pos=north east,legend style={nodes={scale=0.75, transform shape}},tick label style={font=\footnotesize},ytick={1e-1,1e-2,1e-3,1e-4,1e-5,1e-6,1e-7}]
				\addplot[black,line width=1,mark=square,mark size=2,mark options=solid] table [x index = {0}, y index = {1},col sep=comma,]{fig/s/bnd/u_l2.csv};
				\addplot[blue,densely dotted,line width=1,mark=square,mark size=3,mark options=solid] table [x index = {2}, y index = {3},col sep=comma,]{fig/s/bnd/u_l2.csv};
				\addplot[blue,densely dotted,line width=1,mark=o,mark size=2,mark options=solid] table [x index = {4}, y index = {5},col sep=comma,]{fig/s/bnd/u_l2.csv};
				\addplot[red,densely dotted,line width=1,mark=square,mark size=2,mark options=solid] table [x index = {6}, y index = {7},col sep=comma,]{fig/s/bnd/u_l2.csv};
				\addplot[red,densely dotted,line width=1,mark=x,mark size=4,mark options=solid] table [x index = {8}, y index = {9},col sep=comma,]{fig/s/bnd/u_l2.csv};
				\addplot[red,solid,line width=1,mark=triangle,mark size=5,mark options=solid] table [x index = {10}, y index = {11},col sep=comma,]{fig/s/bnd/u_l2.csv};
				\addplot[blue,solid,line width=1,mark=triangle,mark size=4,mark options=solid] table [x index = {12}, y index = {13},col sep=comma,]{fig/s/bnd/u_l2.csv};
				\addplot[black,solid,line width=.5,mark=none,domain=100: 3.499236449108733e+02,samples=2]{ 0.047495352225498*x^-1.838325556374762};
				\draw[black,solid,line width=.5] (100,1e-5) -- (100, 1e-6); \node[anchor=east] at (110,3e-6) {\scriptsize $\boldsymbol{1.84}$};
				\draw[black,solid,line width=.5] (100, 1e-6) -- ( 3.499236449108733e+02, 1e-6); \node[anchor=north] at (160,1.2e-6) {\scriptsize $\boldsymbol{1}$};
				\addplot[black,solid,line width=.5,mark=none,domain=2.904359501167351e+03:1e4,samples=2]{2.815064543204646e+03*x^-1.862372089178393};
				\draw[black,solid,line width=.5] (1e4,1e-3) -- (1e4, 1e-4); \node[anchor=west] at (9000,4.5e-4) {\scriptsize $\boldsymbol{1.86}$};
				\draw[black,solid,line width=.5] (2.904359501167351e+03,1e-3) -- (1e4, 1e-3); \node[anchor=south] at (7000,8.8e-4) {\scriptsize $\boldsymbol{1}$};
			\end{axis}
		\end{tikzpicture}
	}
	\quad
	\subfloat[]{%
		\begin{tikzpicture}
			\def\cdot{\times}
			\begin{axis}[xmode=log,ymode=log,grid=both,xlabel={Number of elements $n_\mathrm{sel}\:[-]$},ylabel={Maximum displacement error $[-]$},width=0.45\textwidth,xmin=10,xmax=1e5,ymin=1e-5,ymax=1e-1,legend cell align={left},legend pos=south west,legend style={nodes={scale=0.75, transform shape}},tick label style={font=\footnotesize},ytick={1e-1,1e-2,1e-3,1e-4,1e-5}]
				\addplot[black,line width=1,mark=square,mark size=2,mark options=solid]table [x index = {0}, y index = {1},col sep=comma,]{fig/s/bnd/u_max.csv};
				 \addplot[blue,densely dotted,line width=1,mark=square,mark size=3,mark options=solid] table [x index = {2}, y index = {3},col sep=comma,]{fig/s/bnd/u_max.csv};
			 	\addplot[blue,densely dotted,line width=1,mark=o,mark size=2,mark options=solid] table [x index = {4}, y index = {5},col sep=comma,]{fig/s/bnd/u_max.csv};
				 \addplot[red,densely dotted,line width=1,mark=square,mark size=2,mark options=solid] table [x index = {6}, y index = {7},col sep=comma,]{fig/s/bnd/u_max.csv};
				\addplot[red,densely dotted,line width=1,mark=x,mark size=4,mark options=solid] table [x index = {8}, y index = {9},col sep=comma,]{fig/s/bnd/u_max.csv};
				\addplot[red,solid,line width=1,mark=triangle,mark size=5,mark options=solid] table [x index = {10}, y index = {11},col sep=comma,]{fig/s/bnd/u_max.csv};
				\addplot[blue,solid,line width=1,mark=triangle,mark size=4,mark options=solid] table [x index = {12}, y index = {13},col sep=comma,]{fig/s/bnd/u_max.csv};
				\addplot[black,solid,line width=.5,mark=none,domain=40:150,samples=2]{300e-4*x^-1};
				\draw[black,solid,line width=.5] (40,7.5e-4) -- (40,2e-4); \node[anchor=east] at (45,4e-4) {\scriptsize $\boldsymbol{1}$};
				\draw[black,solid,line width=.5] (40,2e-4) -- (150,2e-4); \node[anchor=north] at (70,2.2e-4) {\scriptsize $\boldsymbol{1}$};
			\end{axis}
		\end{tikzpicture}
	}
	\caption{Pure bending: Convergence of the proposed method with respect to (a) the $L_2$-error of the displacement and (b) the maximum displacement error.} \label{f:n_s_bnd_cnv}
\end{figure}
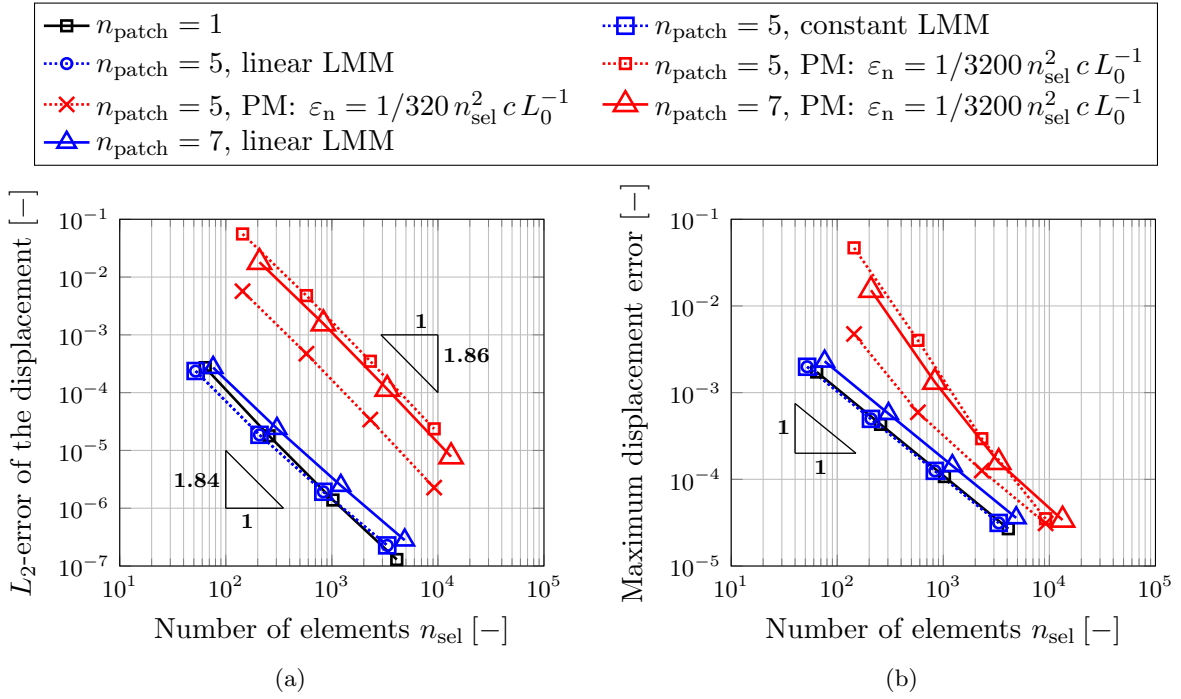
\figref{f:n_s_bnd_cnv} shows the $L_2$- and maximum displacement error for an increasing number of elements. For reference, the error decay for an increasing number of elements using a single patch is shown. The convergence rates for the penalty and Lagrange multiplier method are nearly the same.\footnote{The convergence behavior of the penalty method strongly depends on the penalty parameter. Here, uniform refinement is considered such that $n_\mathrm{sel}$ is inversely proportional to the area of the surface elements adjacent to the patch interfaces. This area is the relevant quantity affecting the constraint behavior and convergence rate.} No difference between constant and linear interpolation of the Lagrange multipliers can be observed. This is due to the fact that the bending moment is constant everywhere. Hence, a constant Lagrange multiplier interpolation is sufficient.

\subsubsection{Pinching of a hemisphere} \label{s:n_s_hmsphr}
\begin{figure}[!ht]
	\centering
		\subfloat[$n_\mathrm{patch}=1$\label{f:n_s_hmsphr_stpleft}]{\includegraphics[scale=0.45]{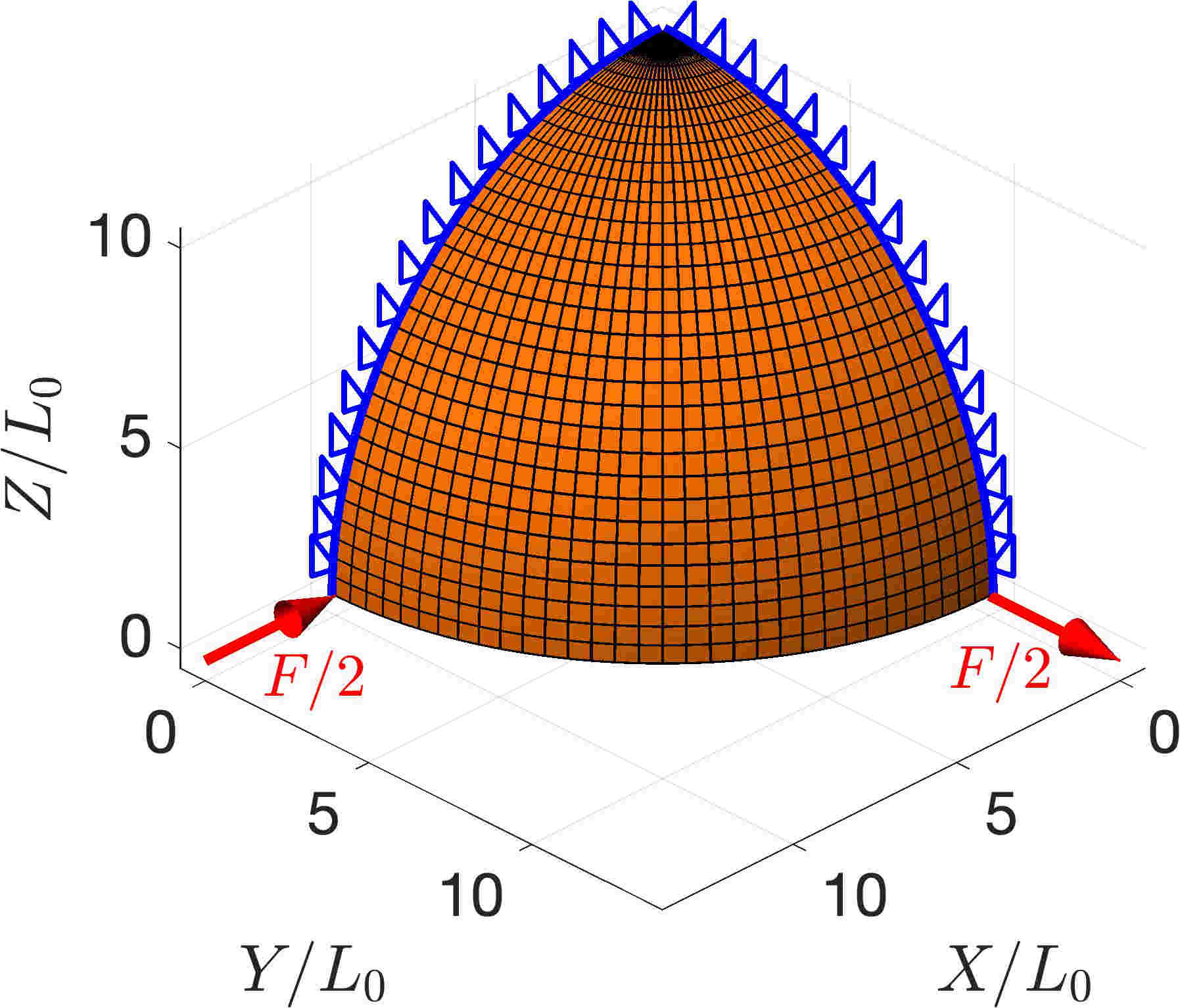}}
	\qquad
		\subfloat[$n_\mathrm{patch}=3$\label{f:n_s_hmsphr_stpright}]{\includegraphics[scale=0.45]{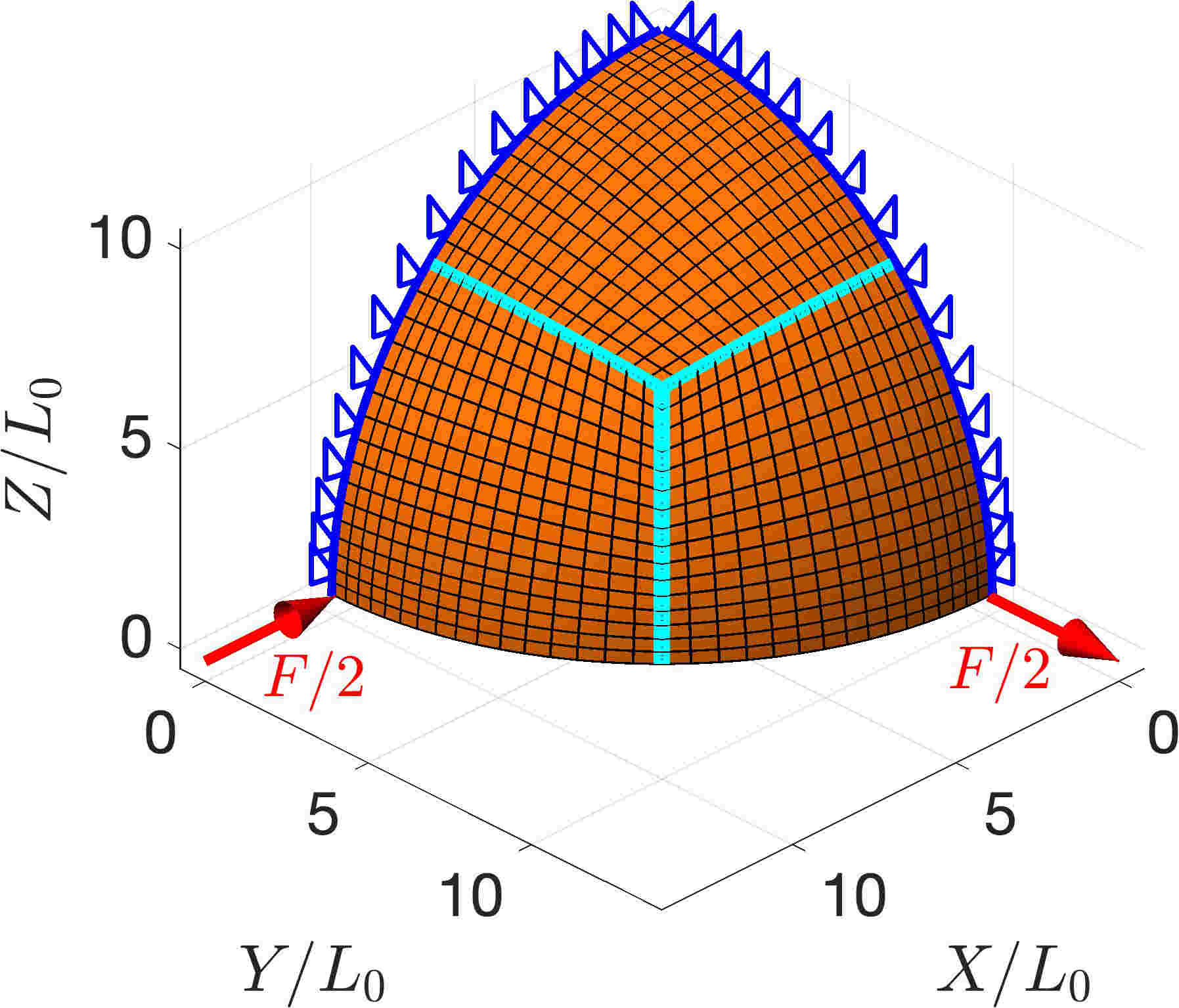}}
	\caption{Pinching of a hemisphere: Discretizations of a quarter model using (a) a single-patch or (b) three patches. The bold cyan-colored lines mark the patch interfaces. The dark blue lines mark the patch boundaries where symmetry is enforced.} \label{f:n_s_hmsphr_stp}
\end{figure}

This section investigates the pinching of a hemisphere. Here, a three-dimensional Saint Venant-Kirchhoff material model is used, see \cite{duong2017}, using the same material parameters as \cite{belytschko1985}. The problem is normalized by the unit length $L_0$ and the 3D unit modulus $\tilde{E}_0$ (measured as force/area). The setup consists either of a single degenerated patch or three patches, see Fig.~\ref{f:n_s_hmsphr_stp}. Along the vertical edges, the symmetry constraint for smooth patch connections (see \tabref{t:p_g_css}) is enforced by the Lagrange multiplier method using constant interpolation. External forces of magnitude $F=2\,\tilde{E}_0\,L_0^2$ act on the lower corners, see \figref{f:n_s_hmsphr_stp}.
\begin{figure}[!ht]
	\centering
	\begin{tikzpicture}
		\def\cdot{\times}
		\begin{axis}[grid=both,xlabel={Number of elements $n_\mathrm{sel}\:[-]$},ylabel={Normalized displacement $[-]$},width=0.6\textwidth,height=0.4\textwidth,xmin=0,xmax=2e4,ymin=0,ymax=1.02,legend cell align={left},legend pos= outer north east,legend style={nodes={scale=1, transform shape}},minor xtick={2000,6000,10000,14000,18000}, xtick={0,4000,8000,12000,16000,20000}]
			\addplot[black,line width=1,mark=square,mark size=2,mark options=solid]table [x index = {0}, y index = {1},col sep=comma,]{fig/s/hmsphr/displ1.csv};
			\addlegendentry{$n_\mathrm{patch}=1$};
			\addplot[blue,densely dotted,line width=1,mark=o,mark size=2,mark options=solid] table [x index = {2}, y index = {3},col sep=comma,]{fig/s/hmsphr/displ1.csv};
			\addlegendentry{$n_\mathrm{patch}=3$, linear $\mathrm{LMM}$};
			\addplot[red,densely dashed,line width=1,mark=x,mark size=4,mark options=solid] table [x index = {0}, y index = {1},col sep=comma,]{fig/s/hmsphr/displ2.csv};
			\addlegendentry{$n_\mathrm{patch}=3$, $g_\mrn$ not enforced};
			\addplot[red,solid,line width=1,mark=triangle,mark size=3,mark options=solid] table [x index = {2}, y index = {3},col sep=comma,]{fig/s/hmsphr/displ2.csv};
			\addlegendentry{$n_\mathrm{patch}=3$, $\mathrm{PM}$: $\eps_\mrn=100\,\tilde{E}_0\,L_0^2$};
		\coordinate (insertPosition) at (rel axis cs:0.6,0.75);
		\end{axis}
			\begin{axis}[at={(insertPosition)},anchor={north},axis background/.style={fill=white},xmin=1e4,xmax=2e4,ymin=0.99,ymax=1.015,width=0.4\textwidth,height=0.25\textwidth,tick label style={font=\footnotesize},grid=both]
			\addplot[black,line width=1,mark=square,mark size=2,mark options=solid]table [x index = {0}, y index = {1},col sep=comma,]{fig/s/hmsphr/displ1.csv};
			\addplot[blue,densely dotted,line width=1,mark=o,mark size=2,mark options=solid] table [x index = {2}, y index = {3},col sep=comma,]{fig/s/hmsphr/displ1.csv};
			\addplot[red,densely dashed,line width=1,mark=x,mark size=4,mark options=solid] table [x index = {0}, y index = {1},col sep=comma,]{fig/s/hmsphr/displ2.csv};
			\addplot[red,solid,line width=1,mark=triangle,mark size=3,mark options=solid] table [x index = {2}, y index = {3},col sep=comma,]{fig/s/hmsphr/displ2.csv};
			\end{axis}
	\end{tikzpicture}
	\caption{Pinching of a hemisphere: Normalized displacement over the mesh refinement.} \label{f:n_s_hmsphr_cmprsn}
\end{figure}
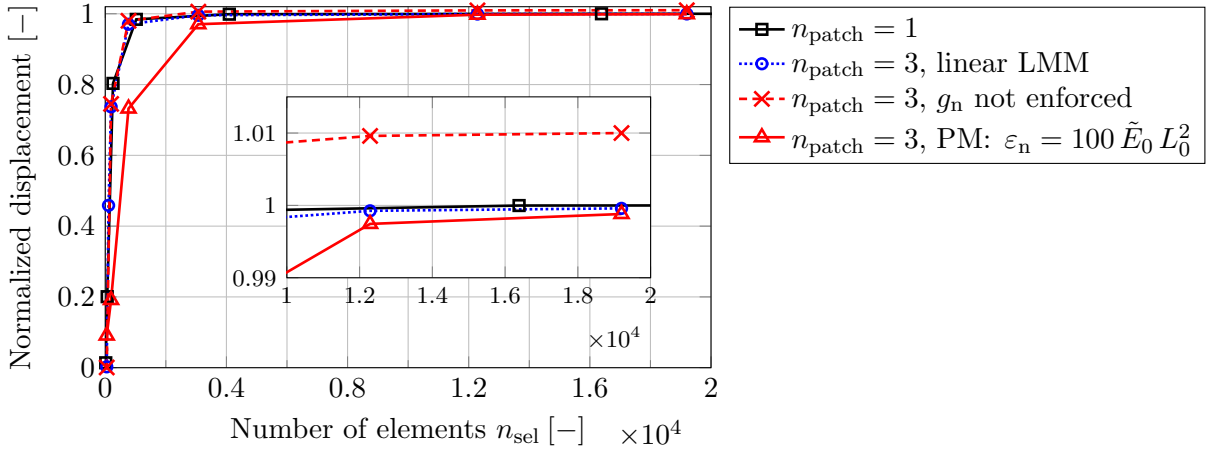

\figref{f:n_s_hmsphr_cmprsn} shows the normalized displacement over the number of elements. It is measured under the points loads and the reference displacement is $0.0924129\,L_0$ \citep{dedoncker2018}. The black line marks the reference solution using a single-patch discretization. Good convergence is obtained, except for the case, when the constraint is not enforced. In that case, the solution converges to a slightly larger displacement.

\subsection{Phase separation on deforming shells} \label{s:n_c_tot}
This section investigates phase separation according to the theory from \secref{s:c_tot}. The color coding of the phase field follows a blue ($\phi=0$) to red ($\phi=1$) transition, see \figref{f:s_mtvtn1}. For the examples in this section, the simplified phase field constraint for the special case of planar patch connections is used, see \eqsref{e:p_c_cnstr4smpl}. The lines of reduced continuity are referred to as $C^0$-lines.
In all examples, the initial concentration field is set to
\eqb{l}
\phi(\xi^\alpha,0)=\bar{\phi}+\phi_\mrr(\xi^\alpha)\,, \label{e:n_c_initphi}
\eqe
with $\bar{\phi}=1/3$ and random distribution $\phi_\mrr \in [-0.05,0.05]$, if not stated otherwise. The mechanical material model of Eqs.~\eqref{e:s_psimem}--\eqref{e:s_psibnd} is used together with the parameters given in \tabref{t:n_c_mat} with  2D Young's modulus $E=N\omega$ and Poisson's ratio $\nu=0.3$.
\begin{table}[ht]
	\caption{Material parameters for the phase separation examples.} \label{t:n_c_mat}
	\centering
	\begin{tabular}{L{1cm} | L{6cm} | L{6cm} }
	    $\,$ & Pure phase state $\phi=0$ (blue color)  & Pure phase state $\phi= 1$ (red color)\\ \hline
		\rule{0pt}{4ex}$\ds K_i$ & $K_0 = 1.25\,\dfrac{E_0\,\nu}{(1+\nu)(1-2\,\nu)}$ & $K_1 = 0.0375\,\ds \frac{E_0\,\nu}{(1+\nu)(1-2\,\nu)}$\vspace{1mm}  \\ \hline
	    \rule{0pt}{4ex}$G_i$ & $G_0 = 6.25\,\ds \frac{E_0}{2\,(1+\nu)}$ & $G_1 = 0.375\,\ds \frac{E_0}{2\,(1+\nu)}$\vspace{1mm}  \\ \hline
	    \rule{0pt}{3ex}$c_i$ & $c_0=0.01\,E_0\,L_0^2$ & $c_1 = 0.0001875\,E_0\,L_0^2$\vspace{1mm}  \\ \hline
	    \rule{0pt}{3ex}$\eta_i$ & $\eta_0 = 1.5\,K_0\,T_0$ & $\eta_1 = 1.5\,K_0\,T_0$
	\end{tabular}
\end{table}
All quantities are dimensionless by the introduction of a reference length $L_0$, time $T_0$ and energy density $\Psi_0$. From this, a reference modulus $E_0=E$ and density $\rho_0=T_0 ^2\,\Psi_0/L_0^2$ follow \citep{zimmermann2019}. Further, $N\omega=\Psi_0$ and $N k_\mrB\,T=\Psi_0/3$ are chosen.

\subsubsection{Pressurized torus}
\begin{figure}[!ht]
		\subfloat[$t\approx0\,T_0$\label{f:n_c_trs_evlm0_1}]{\includegraphics[scale=1]{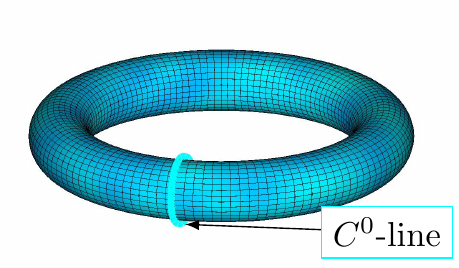}}
	\quad	
		\subfloat[$t\approx112.7344\,T_0$\label{f:n_c_trs_evlm0_2}]{\includegraphics[scale=1]{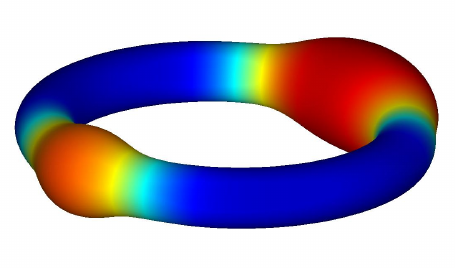}}
	\quad
		\subfloat[$t\approx356.5112\,T_0$\label{f:n_c_trs_evlm0_3}]{\includegraphics[scale=1]{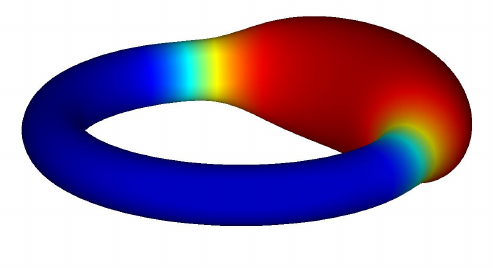}}
	\caption{Pressurized torus: (a) Geometry with a $C^0$-line (marked by the bold cyan-colored line), and (b)--(c) evolution of the phase field and mechanical deformation for the case that the constraints are fulfilled with the Lagrange multiplier method using constant interpolation.} \label{f:n_c_trs_evlm0}
\end{figure}

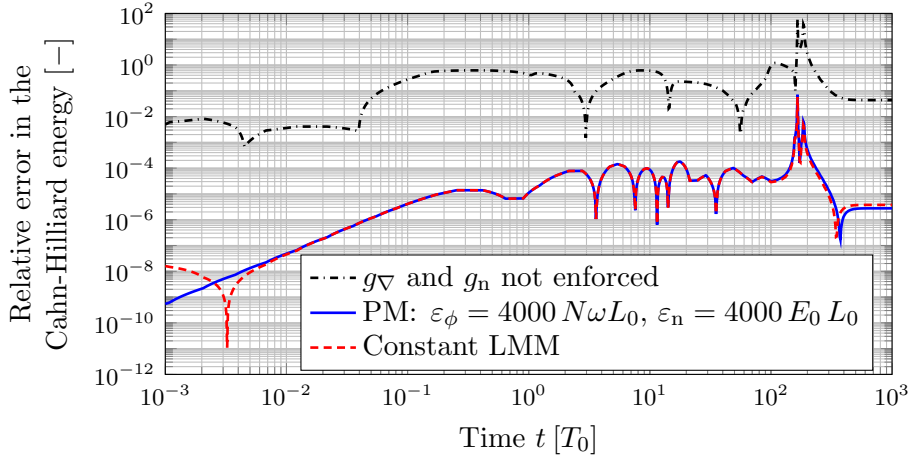
\begin{figure}[!ht]
	\centering
	\begin{tikzpicture}
		\def\cdot{\times}
		\begin{axis}[xmode=log,ymode=log,grid=both,xlabel={Time $t\:[T_0]$},ylabel style={align=center},ylabel={Relative error in the\\Cahn-Hilliard energy $[-]$},width=0.7\textwidth,height=0.4\textwidth,xmin=1e-3,xmax=1e3,ymin=1e-12,ymax=1e2,legend cell align={left},legend style={nodes={scale=1, transform shape},at={(0.58,0.01)},anchor=south},tick label style={font=\footnotesize},
					ytick={1e-12,1e-11,1e-10,1e-9,1e-8,1e-7,1e-6,1e-5,1e-4,1e-3,1e-2,1e-1,1e0,1e1,1e2},
					yticklabels={$10^{-12}$,,$10^{-10}$,,$10^{-8}$,,$10^{-6}$,,$10^{-4}$,,$10^{-2}$,,$10^{0}$,,$10^{2}$},]
			\addplot[black,dash dot,line width=1]table [x index = {0}, y index = {1},col sep=comma,]{fig/ch/trs/timeError.csv};
			\addlegendentry{$g_\nabla$ and $g_\mrn$ not enforced};
			\addplot[blue,solid,line width=1] table [x index = {0}, y index = {2},col sep=comma,]{fig/ch/trs/timeError.csv};
			\addlegendentry{$\mathrm{PM}$: $\eps_\phi=4000\,N\omega L_0$, $\eps_\mrn=4000\,E_0\,L_0$};
			\addplot[red,densely dashed,line width=1] table [x index = {0}, y index = {3},col sep=comma,]{fig/ch/trs/timeError.csv};
			\addlegendentry{Constant $\mathrm{LMM}$};
		\end{axis}
	\end{tikzpicture}
	\caption[]{Pressurized torus: Relative error of the Cahn-Hilliard energy over time. The results from an overall $C^1$-continuous discretization with $n_\mathrm{CP}=4096$ control points are used as a reference solution.\textsuperscript{\ref{footnote_c_trs_err}} The used penalty parameter $\eps_\phi$ is equal to the one proposed in \eqsref{e:p_c_epsch}.} \label{f:n_c_trs_err}
\end{figure}

This section investigates phase separation on a torus, which is subjected to the constant internal pressure $p=0.09\,E_0\,L_0^{-1}$. The geometry is shown in \figref{f:n_c_trs_evlm0_1}.  The major and minor radii are $1.3\,L_0$ and $0.2\,L_0$, respectively. The length scale parameter is $\ell=\sqrt{0.075}\,L_0$. The cyan-colored line shows a patch interface, where the surface discretization is only $C^0$-continuous. Given a continuous parametrization of the torus, the $C^0$-line is obtained by knot insertion at the respective position in the global knot vector.
Figs.~\ref{f:n_c_trs_evlm0_2}--\ref{f:n_c_trs_evlm0_3} illustrate the evolution of the phase field and the mechanical deformation using a mesh consisting of $n_\mathrm{CP}=4128$ control points and the Lagrange multiplier method with constant interpolation to enforce the two constraints of Eqs.~\eqref{e:p_g_cnstrsmpl} and \eqref{e:p_c_cnstr4smpl}. The results show good agreement with the results from an overall $C^1$-continuous discretization with $n_\mathrm{CP}=4096$ control points. This is illustrated in \figref{f:n_c_trs_err}, where the relative error of the Cahn-Hilliard energy is plotted over time.\footnote{The time steps of the two simulations are not coinciding. Therefore, the energy is linearly interpolated between the time steps to compute the relative energy difference.\label{footnote_c_trs_err}} The enforcement of the continuity constraints using the penalty method shows similarly good results. The black dash-dotted line shows the relative error when the continuity constraints are not enforced. As a result of this insufficient discretization, the error is several orders of magnitude larger compared to the results without constraint enforcement.
\begin{figure}[!ht]
	\centering
		\subfloat[Phase field at $t\approx1.5652\,T_0$\label{f:n_c_trs_no1}]{\includegraphics[width=0.3\textwidth]{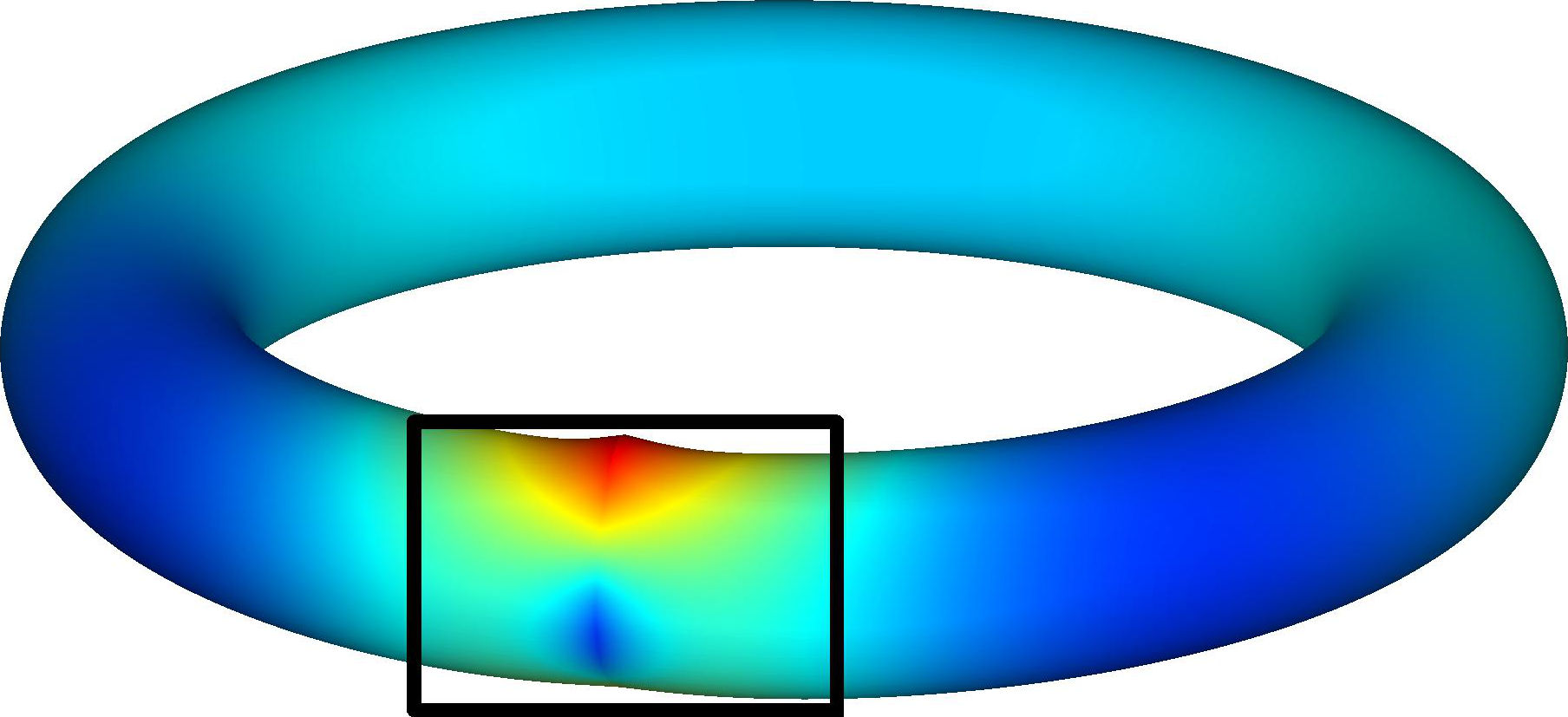}}
	\qquad
		\subfloat[Phase field {$\phi$ $[-]$}\label{f:n_c_trs_no2}]{\includegraphics[width=0.25\textwidth]{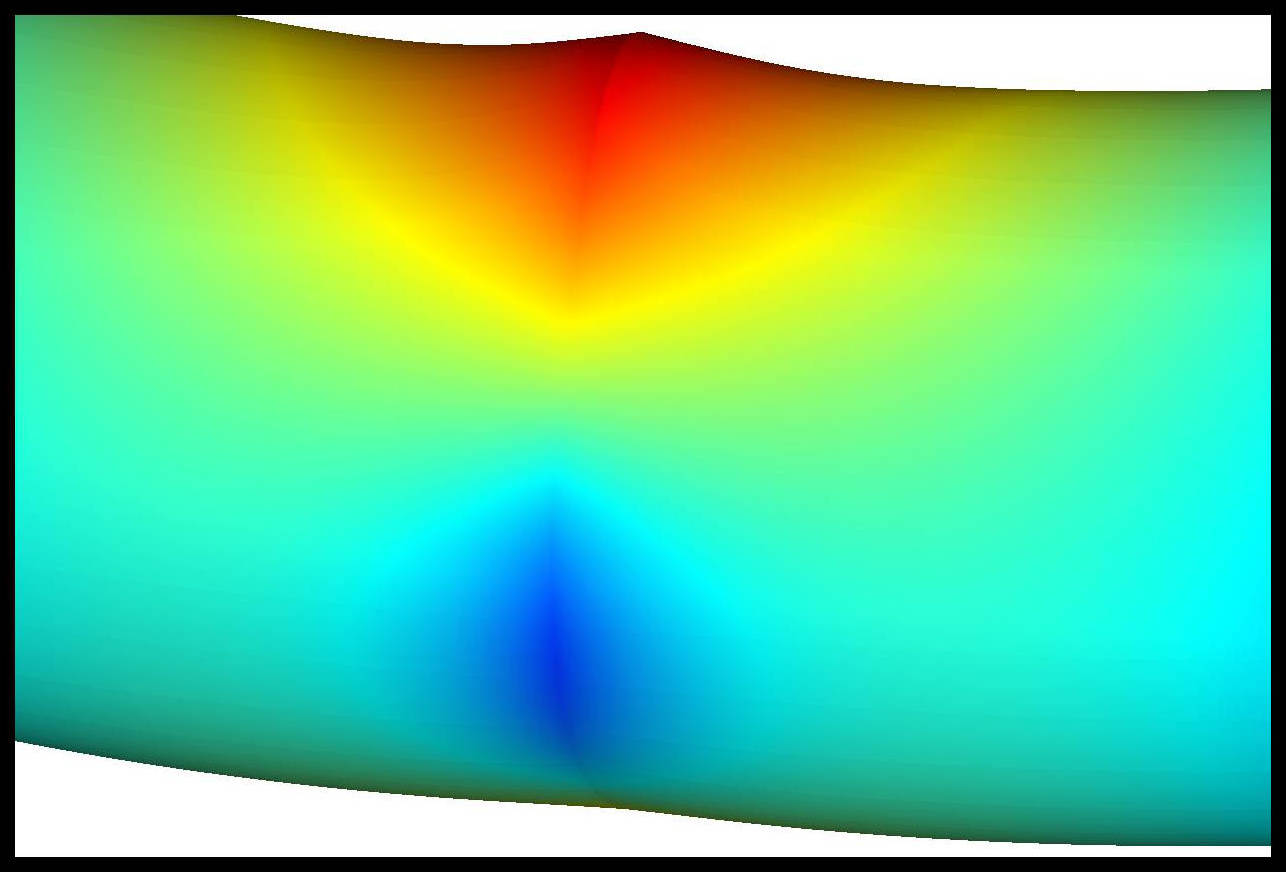}}
	\qquad
		\subfloat[Elastic surface tension {$\gamma_\mathrm{el}$ $[N\omega]$}\label{f:n_c_trs_no3}]{\includegraphics[width=0.25\textwidth]{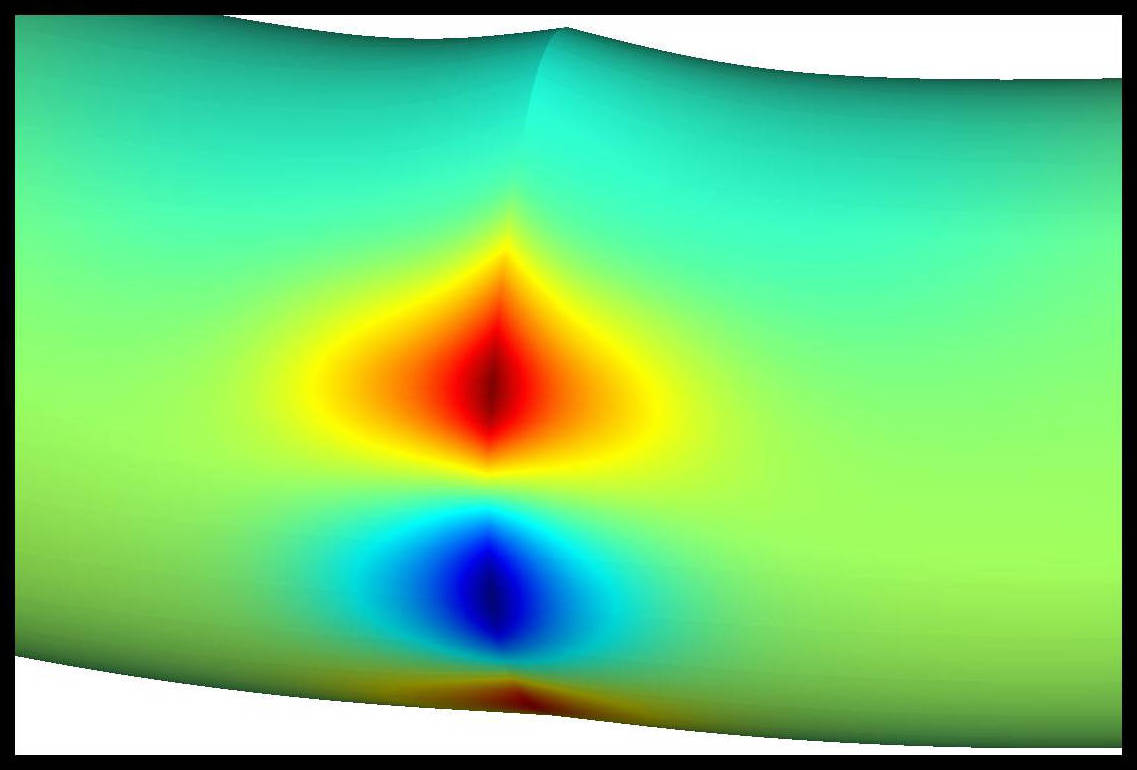}\hspace{2mm}\includegraphics[width=0.0558\textwidth]{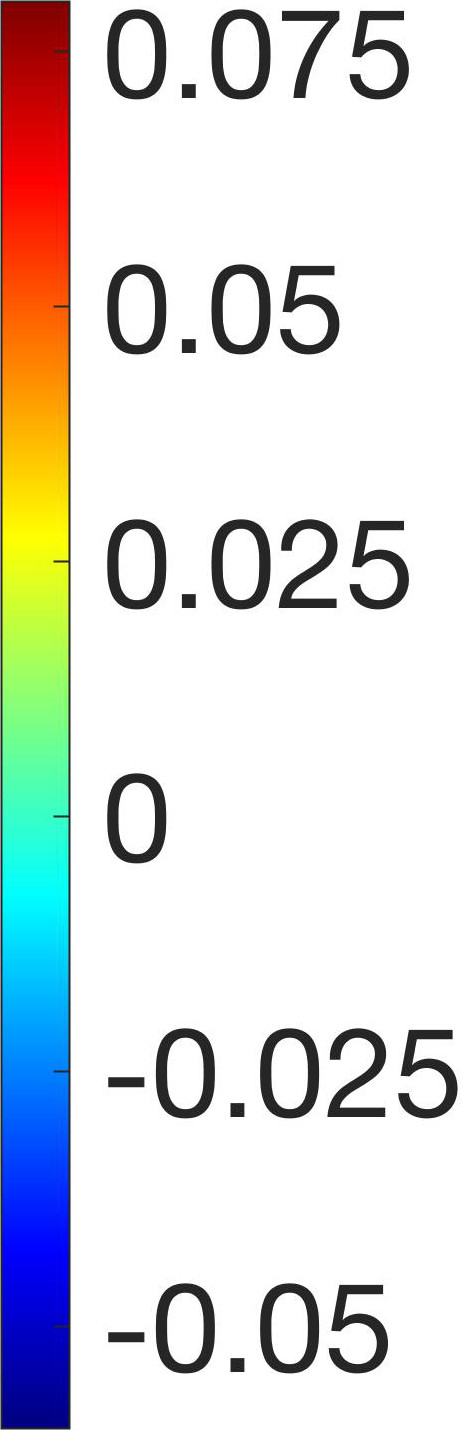}}
	\caption{Pressurized torus: (a) Phase field at early times when no constraints are enforced; Enlargement of (b) phase field and (c) elastic surface tension. The insufficient discretization leads to the formation of kinks, non-physicial phase separation and stress peaks.} \label{f:n_c_trs_no}
\end{figure}
\figref{f:n_c_trs_no} visualizes the deformed configuration for the case that the constraints are not enforced. Already at early times, the insufficient discretization leads to kinks in the geometry that attract phase separation, see \figref{f:n_c_trs_no1} and its enlargement in \figref{f:n_c_trs_no2}. Also, the stresses show non-physical peaks at the $C^0$-line, see \figref{f:n_c_trs_no3}. Here, the elastic surface tension
\eqb{l}
	\gamma_\mathrm{el}=\dfrac{1}{2}N_\mathrm{el}^\ab\,a_\ab \label{e:s_gamel}
\eqe
is shown, where $N^\ab_\mathrm{el}$ denotes the elastic Cauchy stress components \citep{sauer2017a}. The wrong result at early times leads to a different final state. Thus, the relative error of the Cahn-Hilliard energy is much higher for the unconstrained problem, see \figref{f:n_c_trs_err}.

\subsubsection{Phase separation on a deforming sphere} \label{s:n_c_sphr}
\begin{figure}[!ht]
		\subfloat[Six-patch discretization for $m=8$\label{f:n_c_6ptch}]{
			\includegraphics[width=0.2\textwidth,valign=t]{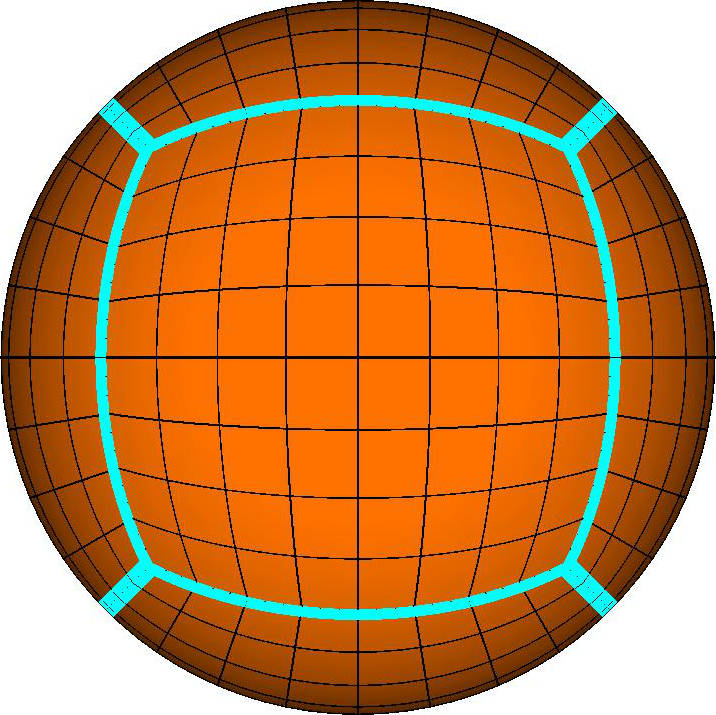}
			\quad
			\begin{tabular}[t]{r r r}
					$m$ & $n_\mathrm{CP}$ & $n_\mathrm{sel}$\\\hline
					$8$ & $488$ & $384$ \\
					$16$ & $1736$ & $1536$ \\
					$32$ & $6536$ & $6144$ \\
					$64$ & $25352$ & $24576$ \\
			\end{tabular}
		}
	\qquad
		\subfloat[Unstructured spline discretization for $r=0$\label{f:n_c_us}]{
			\includegraphics[width=0.2\textwidth,valign=t]{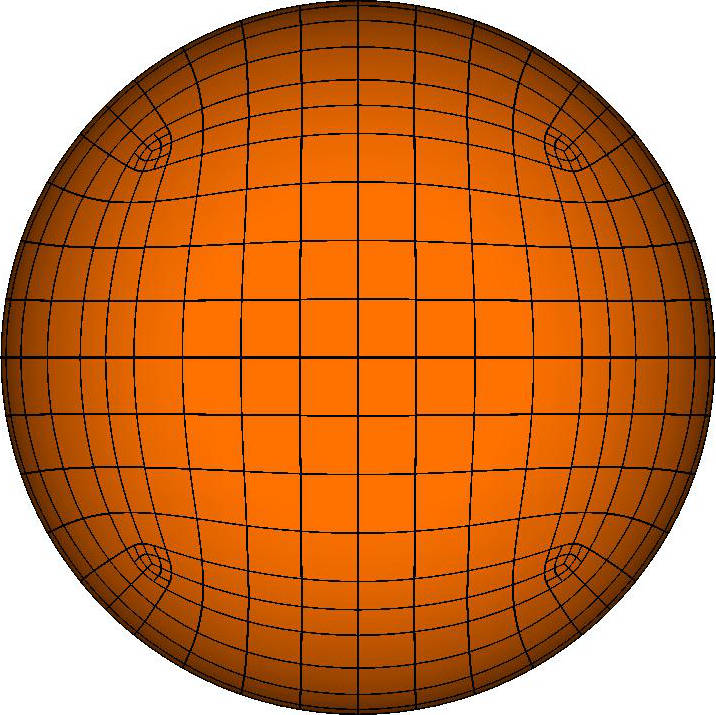}
			\quad
			\begin{tabular}[t]{r r r}
					$r$ & $n_\mathrm{CP}$ & $n_\mathrm{sel}$\\\hline
					$0$ & $690$ & $672$ \\
					$1$ & $2490$ & $2472$ \\
					$2$ & $9690$ & $9672$ \\
					$3$ & $38490$ & $38472$ \\
			\end{tabular}
		}
	\caption{Spline discretizations of a sphere and corresponding mesh properties for different refinement levels for (a) the six-patch discretization (the bold cyan-colored lines mark the patch interfaces) and (b) an unstructrued spline discretization. A comparison between the two discretizations can be found in \appref{s:sphr_tot}.} \label{f:n_c_sphrs}
\end{figure}

This section investigates phase separation on a spherical shell. It is either discretized by six patches, see \figref{f:n_c_6ptch}, or by unstructured cubic splines, see \figref{f:n_c_us}. The latter serves as reference as it provides $C^2$-continuity everywhere except for eight so-called extraordinary points, where only $C^1$-continuity is maintained \citep{toshniwal2017b}. A comparison between the two discretizations is provided in \appref{s:sphr_tot}. Further constructions of multi-patch sphere discretizations with conforming or non-conforming meshes can be found in \cite{dedoncker2018}.

First, this section investigates the phase field evolution on a sphere for the interface parameter $\ell=\sqrt{0.001}\,L_0$. The initial radius of the sphere is $L_0$ and the constant pressure $p=0.04\,E_0\,L_0^{-1}$ is applied to its interior surface. The mobility constant is $D=2.5\,T_0$.
\begin{figure}
	\centering
		\includegraphics[scale=1]{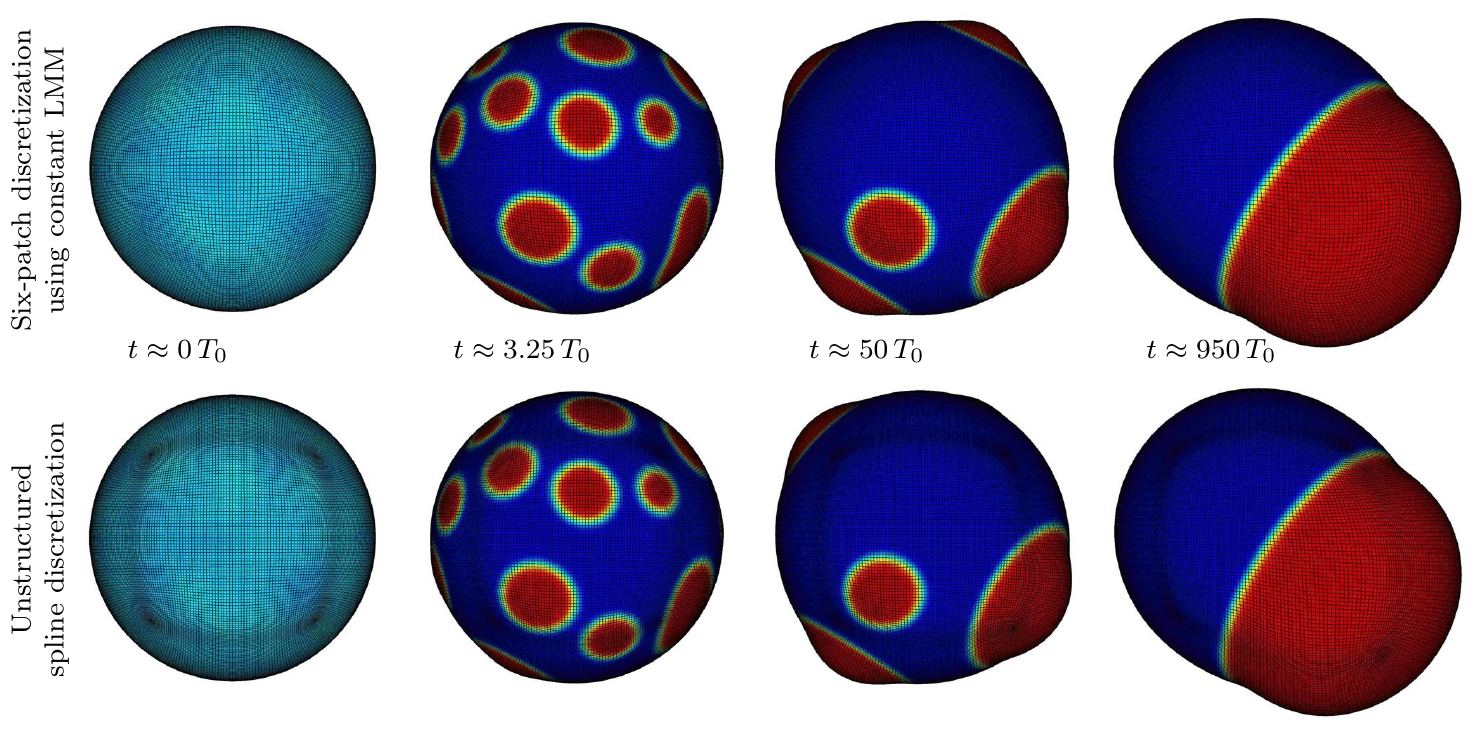}
	\caption{Phase separation on a deforming sphere: The top row shows the phase evolution for the six-patch discretization, the bottom row for the unstructured spline discretization. The constraints for the first one are enforced by the Lagrange multiplier method with constant interpolation. Both cases show excellent agreement.} \label{f:n_c_sphr_evo}
\end{figure}
\figref{f:n_c_sphr_evo} shows several snapshots for the phase separation for the six-patch discretization with $m=64$ in the first row, and the unstructured spline discretization with $r=3$ in the second row. Both show excellent agreement in space and time. Due to the small interface parameter, multiple red phase nuclei appear. As the time progresses, these bulge, merge and evolve. The deformation is larger at locations where the nuclei grow.
\begin{figure}
	\centering
		\includegraphics[scale=1]{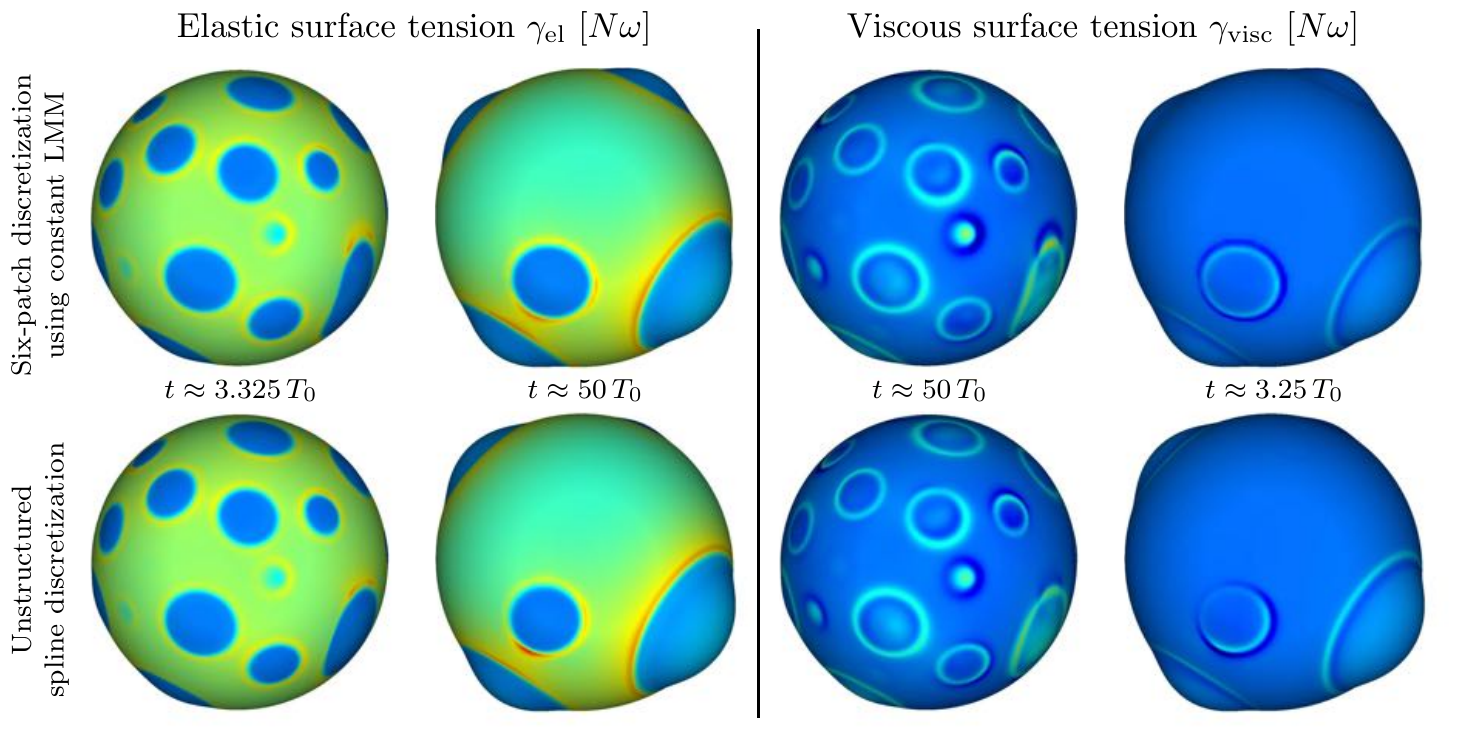}\\
		\includegraphics[scale=0.3]{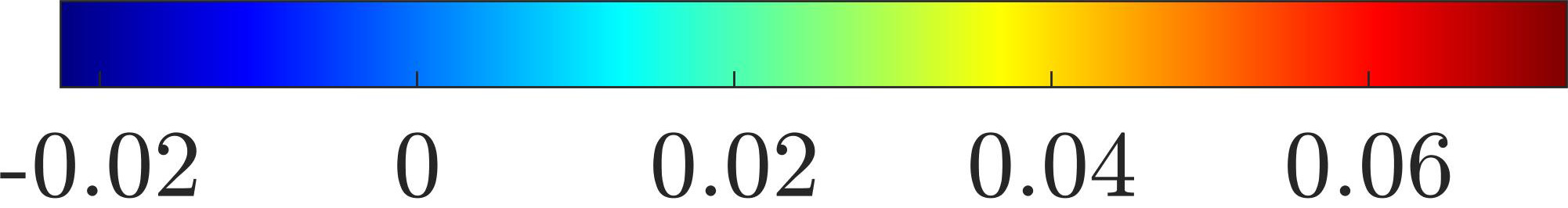}
	\caption{Phase separation on a deforming sphere: The top row shows the elastic and viscous surface tensions for the six-patch discretization, the bottom row for the unstructured spline discretization. The constraints for the first one are enforced by the Lagrange multiplier method with constant interpolation. Both cases show excellent agreement. } \label{f:n_c_sphr_gam}
\end{figure}
In \figref{f:n_c_sphr_gam}, the elastic and viscous surface tensions are shown for the two different discretization techniques. The viscous surface tension $\gamma_\mathrm{visc}$ follows in analogy to \eqsref{e:s_gamel} from the viscous Cauchy stress. The largest values of the elastic surface tension are obtained around the evolving bulges, while the smallest are in their center. Viscous effects are comparatively large at disappearing and around growing bulges. Again, both discretization techniques show excellent agreement.

\begin{figure}[!ht]
\centering
	\begin{tikzpicture}  	
		\def\cdot{\times}
		\pgfplotsset{set layers,every linear axis/.style={scale only axis, xmin=-1.1,xmax=1.6, }
    }
    \begin{axis}[height=0.35\textwidth,width=0.8\textwidth,grid=both,xlabel={Height-coordinate of the illustrated cutting plane $[L_0]$},ylabel={Elastic surface tension $\gamma_\mathrm{el}\:[N\omega]$},axis y line*=left,legend cell align={left},legend style={nodes={scale=1, transform shape},at={(0.1,0.975)},anchor=north},yticklabel style={ /pgf/number format/fixed,/pgf/number format/precision=5},ytick={0.01,0.02,0.03,0.04,0.05,0.06},xtick={-1,-0.5,0,0.5,1,1.5},ymin=1e-2,ymax=6e-2]
			\addplot[violet,solid,line width=1] table [x index = {0}, y index = {2},col sep=comma,]{fig/ch/sphr/sphr01.csv};
			\addlegendentry{$r=0$};
			\addplot[orange,solid,line width=1] table [x index = {0}, y index = {1},col sep=comma,]{fig/ch/sphr/sphr02.csv};
			\addlegendentry{$r=1$};
			\addplot[blue,solid,line width=1] table [x index = {0}, y index = {1},col sep=comma,]{fig/ch/sphr/sphr03.csv};
			\addlegendentry{$r=2$};
			\addplot[red,solid,line width=1] table [x index = {0}, y index = {1},col sep=comma,]{fig/ch/sphr/sphr04.csv};
			\addlegendentry{$r=3$};
			\addplot[violet,dashed,line width=1] table [x index = {0}, y index = {1},col sep=comma,]{fig/ch/sphr/sphr05.csv};
			\addlegendentry{$m=16$};
			\addplot[orange,dashed,line width=1] table [x index = {0}, y index = {1},col sep=comma,]{fig/ch/sphr/sphr06.csv};
			\addlegendentry{$m=21$};
			\addplot[blue,dashed,line width=1] table [x index = {0}, y index = {1},col sep=comma,]{fig/ch/sphr/sphr07.csv};
			\addlegendentry{$m=31$};
			\addplot[red,dashed,line width=1] table [x index = {0}, y index = {1},col sep=comma,]{fig/ch/sphr/sphr08.csv};
			\addlegendentry{$m=64$};
			\node[anchor=center,draw,black,line width=0.5,rectangle,outer sep=0pt,fill=white,align=left] at (1.2,3.9e-2) {\includegraphics[scale=0.135]{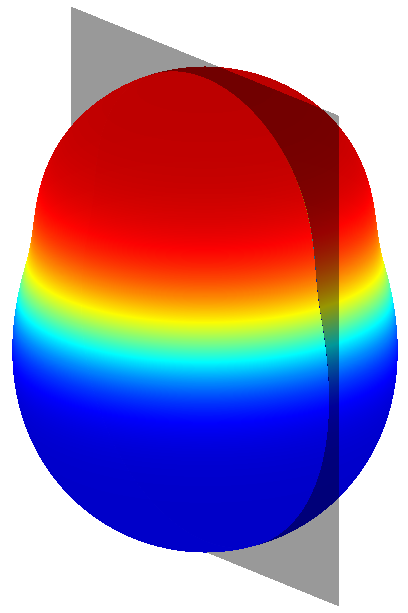}};
    \end{axis}
    \begin{axis}[height=0.35\textwidth,width=0.8\textwidth,axis y line*=right,axis x line=none,ylabel={Phase field $\phi\:[-]$},ymin=0,ymax=1,legend cell align={left},legend style={at={(0.25,0.975)},anchor=north}]      
			\addplot[black,densely dotted,line width=1] table [x index = {0}, y index = {1},col sep=comma,]{fig/ch/sphr/sphr01.csv};
			\addlegendentry{$\phi$};
    \end{axis}
	\end{tikzpicture}
\caption{Phase separation on a deforming sphere: Final deformation and phase distribution for the initial state in \eqsref{e:n_c_sphrphi0}. Surface tension (left axis) and phase field (right axis) over the height-coordinate. The former is shown for the different discretizations (dashed line: six-patch discretization; solid line: unstructured spline discretization). The results of the two discretization techniques converge for an increasing number of control points.} \label{f:n_c_sphr_crsssctn}
\end{figure}
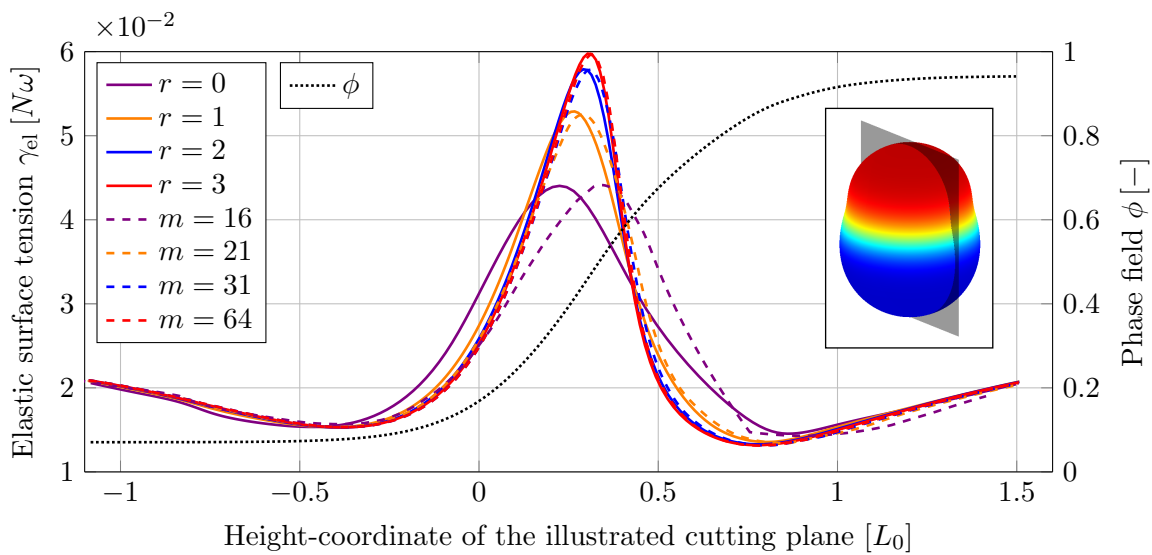

Second, the results from the two discretization approaches are compared for the same pressure loading but the initial concentration field
\eqb{l}
	\phi(\xi^\alpha,0)=\begin{cases}1/2\,,&z\geq0\\1/6\,,&z<0\end{cases}\,. \label{e:n_c_sphrphi0}
\eqe
The interface parameter is now set to $\ell=\sqrt{0.05}\,L_0$. The final deformation and phase distribution is visualized on the inset of \figref{f:n_c_sphr_crsssctn}. Next to it, the black dotted line shows the phase field value over the height-coordinate of the illustrated cutting plane for the unstructured spline discretization with $r=0$ (see \figref{f:n_c_us}). Further, the elastic surface tension for different refinement levels $m$ and $r$ (see \figref{f:n_c_sphrs}) is illustrated in \figref{f:n_c_sphr_crsssctn}.
For an increasing number of control points, the results from the six-patch discretization and the unstructured spline discretization converge. This shows that the proposed patch constraints successfully enforce continuity for the coupled problem.

\subsection{Fracture of deforming shells} \label{s:n_f_tot}
This section illustrates the modeling of crack evolution on multi-patch shells with the constraints from \secref{s:p_tot}. Note that the gradients of $\phi$ are now defined in the reference configuration and the constraint from \eqsref{e:p_c_cnstr2} is considered. The following material parameters are used throughout this section (see Eqs.~\eqref{e:s_psimem}--\eqref{e:s_psibnd})
\eqb{l}
	K=\dfrac{E\,\nu}{\lr{1+\nu}\lr{1-2\nu}}\,,\qquad G=\dfrac{E}{2\lr{1+\nu}}\,,\qquad c=0.1\,E_0\,L_0^2\,, \label{e:n_f_mat}
\eqe
with 2D Young's modulus $E$ and Poisson's ratio $\nu$. All quantities in this section are non-dimensionalized by the introduction of a reference length $L_0$, time $T_0$ and density $\rho_0$. From this, the reference modulus $E_0:=\rho_0\,L_0^2\,T_0^{-2}$ with units $[\mathrm{N}/\mathrm{m}]$ follows \citep{paul2020}. The color coding for the fracture field ranges from the fully fractured state ($\phi=0$, red color) to the undamaged state ($\phi=1$, blue color), see \figref{f:s_mtvtn2}. The lines of reduced continuity are again referred to as $C^0$-lines. The present brittle fracture model does not incorporate viscosity, such that $\eta=0$ in Eq.~(\ref{e:s_sigM}.1).

\subsubsection{Verification of the continuity constraint on the fracture field} \label{s:n_f_vld}
\begin{figure}[!ht]
	\centering
		\subfloat[Setup with induced crack and $C^0$-lines\label{f:n_f_vld_setup1}]{\includegraphics[scale=1,valign=t]{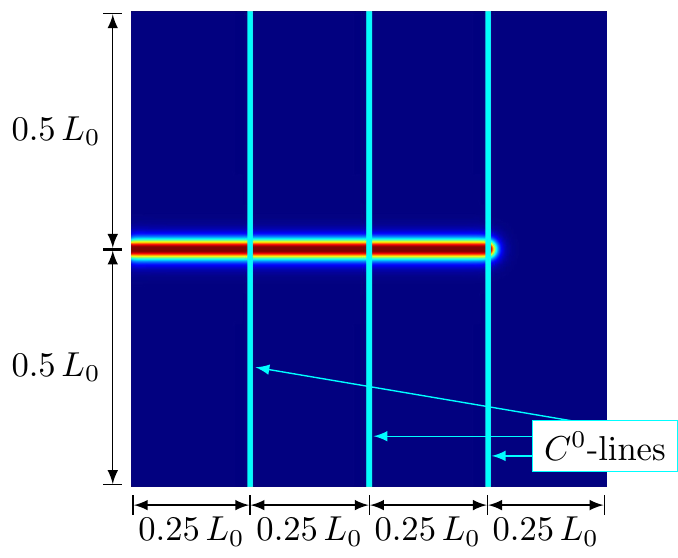}}
	\qquad
		\subfloat[LR mesh\label{f:n_f_vld_setup2}]{\vspace{4mm}\includegraphics[scale=1,valign=t]{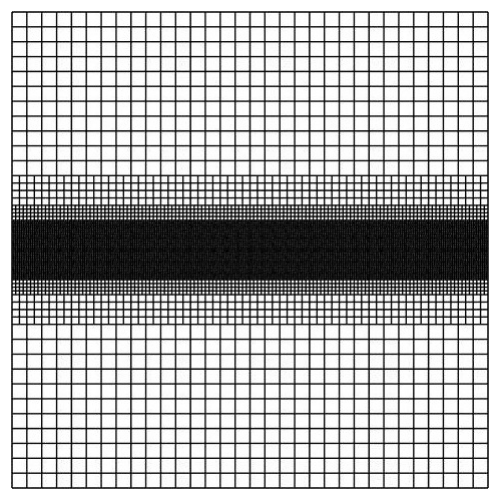}%
			\vphantom{\includegraphics[scale=1,valign=t]{fig/frc/vld/vldSetup}}}
	\caption{Verification of the continuity constraint on the fracture field. (a) Setup showing the induced initial crack and three $C^0$-lines treated with the Lagrange multiplier method using linear interpolation. (b) LR mesh used for all computations.} \label{f:n_f_vld_setup}
\end{figure}

This section verifies the constraint from \eqsref{e:p_c_cnstr2} in a two-dimensional setup without mechanical loading. The flat sheet shown in \figref{f:n_f_vld_setup1} is considered. The length scale parameter is $\ell_0=0.008\,L_0$ and the fracture toughness is $\sG_\mrc=0.0005\,E_0\,L_0$. The used LR mesh \citep{dokken2013,zimmermann2017} is shown in \figref{f:n_f_vld_setup2}. The three $C^0$-lines are inserted by knot insertion. The crack in \figref{f:n_f_vld_setup1} is computed by manually imposing values for the history field, see \eqsref{e:f_hlmhltz1}, and solving the phase field equation.
\begin{figure}[!ht]
	\centering
		\subfloat[$|g_\nabla|$ for the proposed penalty parameter\label{f:n_f_vld_errconv1}]{
			\begin{tikzpicture}
				\def\cdot{\times}
				\begin{axis}[ymode=log,grid=both,xlabel={$y$-position $[L_0]$ ($x=0.75\,L_0$)},ylabel={Constraint value $|g_\nabla|\:[L_0^{-1}]$},width=0.45\textwidth,height=0.4\textwidth,xmin=0,xmax=1,ymin=1e-16,ymax=1e-7,xtick={0,0.1,0.2,0.3,0.4,0.5,0.6,0.7,0.8,0.9,1},
					xticklabels={$0$,,$0.2$,,$0.4$,,$0.6$,,$0.8$,,$1$},
					ytick={1e-16,1e-15,1e-14,1e-13,1e-12,1e-11,1e-10,1e-9,1e-8,1e-7,1e-6,1e-5},
					yticklabels={,$10^{-15}$,,$10^{-13}$,,$10^{-11}$,,$10^{-9}$,,$10^{-7}$},
					tick label style={font=\footnotesize},legend cell align={left},legend style={at={(0.5,1.05)},anchor=south}]
					\addplot[blue,solid,line width=1]table [x index = {0}, y index = {1},col sep=comma]{fig/frc/vld/epsOpt.csv};
					\addlegendentry{$\eps_\phi$ scaled with $\Delta x^{\Gamma}_{\max}/\Delta x^{\Gamma}$};
					\addplot[red,densely dashed,line width=1]table [x index = {0}, y index = {2},col sep=comma]{fig/frc/vld/epsOpt.csv};
					\addlegendentry{$\eps_\phi$ not scaled with $\Delta x^{\Gamma}_{\max}/\Delta x^{\Gamma}$};
				\end{axis}
			\end{tikzpicture}
		}
	\qquad
		\subfloat[Error plot for increasing penalty parameter\label{f:n_f_vld_errconv2}]{
				\begin{tikzpicture}
					\def\cdot{\times}
					\begin{axis}[xmode=log,ymode=log,grid=major,xlabel={$\eps_{\phi}^0/\Delta x^{\Gamma}_{\min}\:[E_0\,L_0^3]$},ylabel={Relative error in fracture energy $[-]$},width=0.4\textwidth,height=0.4\textwidth,xmin=1e-12,xmax=2560,ymin=1e-17,ymax=1e-2,legend cell align={left},legend style={nodes={scale=1, transform shape},at={(0.05,0.34)},anchor=west,cells={align=left}},tick label style={font=\footnotesize},
					xtick={1e-12,1e-11,1e-10,1e-9,1e-8,1e-7,1e-6,1e-5,1e-4,1e-3,1e-2,1e-1,1e0,1e1,1e2,1e3,1e4},
					xticklabels={$10^{-12}$,,,$10^{-9}$,,,$10^{-6}$,,,$10^{-3}$,,,$10^{0}$,,,$ 10^{3}$,},
					ytick={1e-16,1e-15,1e-14,1e-13,1e-12,1e-11,1e-10,1e-9,1e-8,1e-7,1e-6,1e-5,1e-4,1e-3,1e-2},
					yticklabels={$10^{-16}$,,$10^{-14}$,,$10^{-12}$,,$10^{-10}$,,$10^{-8}$,,$10^{-6}$,,$10^{-4}$,,$10^{-2}$},
					x label style={yshift=5.3pt},]
						\fill[gray,opacity=0.6,draw=black] (25.6,1e-17) rectangle(204.8,1e-2);
						\addplot[blue,densely dashed,line width=1,log basis x=10]table [x index = {0}, y index = {1},col sep=comma,restrict expr to domain={\coordindex}{16:17}]{fig/frc/vld/errPsiFracThreeC0lines.csv};
						\addlegendentry{$g_\nabla$ not\\enforced};
						\addplot[red,line width=1]table [x index = {0}, y index = {1},col sep=comma,restrict expr to domain={\coordindex}{0:15}]{fig/frc/vld/errPsiFracThreeC0lines.csv};
						\addlegendentry{$\mathrm{PM}$};
						\addplot[black,densely dotted,line width=1]table [x index = {0}, y index = {1},col sep=comma,restrict expr to domain={\coordindex}{18:19}]{fig/frc/vld/errPsiFracThreeC0lines.csv};
						\addlegendentry{$\mathrm{LMM}$};
							\addplot[black,solid,line width=.5,mark=none,domain=1e-3:1e0,samples=2]{1.027323441113802e-10*x^-0.996097600730106};
							\draw[black,solid,line width=.5] (1e0,1e-7) -- (1e0,1.027323441113802e-10); \node[anchor=west] at (.6,9e-9) {\scriptsize $\boldsymbol{1}$};
							\draw[black,solid,line width=.5] (1e-3,1e-7) -- ( 1e0,1e-7); \node[anchor=south] at (1.2e-1,7e-8) {\scriptsize $\boldsymbol{1}$};
					\end{axis}
				\end{tikzpicture}
		}
	\caption[]{Verification of the continuity constraint on the fracture field. (a) The absolute value of the phase field constraint along the third $C^0$-line (at $x=0.75\,L_0$) for the penalty parameter $\eps_\phi$ according to \eqsref{e:p_c_epsf}. (b) Relative error in the fracture energy for different enforcement techniques and increasing penalty parameter. The gray marked area shows the range of the proposed penalty paramter according to \eqsref{e:p_c_epsf}.} \label{f:n_f_vld_errconv}
\end{figure}
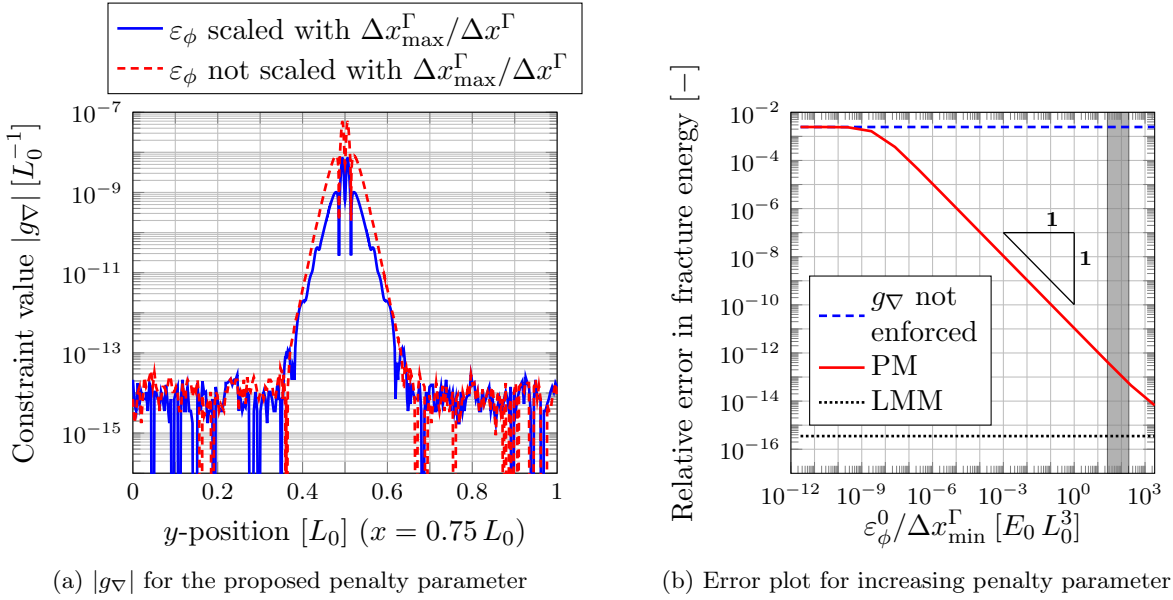

\figref{f:n_f_vld_errconv1} illustrates the absolute value of the constraint $g_\nabla$ along the third $C^0$-line (at $x=0.75\,L_0$) over the $y$-position using the penalty method. It shows the two cases when the scaling of the penalty parameter with $\Delta x_{\max}^{\Gamma}/\Delta x^{\Gamma}$ is used or not, see \eqsref{e:p_c_epsf}. For both cases, the maximum values are obtained at the center where the crack is located, but the scaling factor reduces the constraint value by one order of magnitude. Thus, the constraint is fulfilled more accurately in regions of fracture, which are the regions of primary interest. The Lagrange multiplier method with constant or linear interpolation leads to values of order $10^{-12}$ for $|g_\nabla|$. \figref{f:n_f_vld_errconv2} shows the relative error in the fracture energy for an increasing penalty parameter. The fracture energy of the sheet without manually inserted $C^0$-lines is taken as a reference solution. As the penalty parameter increases, the relative error decreases approximately linear until it is close to machine precision. Non-convergence is encountered for $\eps_\phi^0/\Delta x^{\Gamma}_{\min}\geq10^{4}\,E_0\,L_0^3$ due to ill-conditioning. The gray marked area reflects the range of the proposed penalty parameter from \eqsref{e:p_c_epsf} when the scaling factor $\Delta x_{\max}^{\Gamma}/\Delta x^{\Gamma}$ is used. Its largest values are obtained at regions of the highest resolved mesh. It ensures sufficiently accurate results while providing good convergence of the Newton-Raphson scheme. Machine precision is directly reached when the Lagrange multiplier method is used. The blue dashed line marks the error when no constraints are enforced. In this case, the insufficient discretization yields comparatively high errors that can have a huge influence on the solution as loads are applied. Note that also computations with two and four refinement levels and length scale parameters of $\ell_0=0.02\,L_0$ and $\ell_0=0.004\,L_0$, respectively, were performed. The results are similar to the ones shown in  \figref{f:n_f_vld_errconv2}.

\subsubsection{Fracturing sphere} \label{s:n_f_sphr}
This section investigates crack evolution on a spherical shell with radius $L_0$. Two initial cracks with perpendicular orientation are placed on two opposite sides of the sphere, as shown in \figref{f:n_f_sphr_evo}. The sphere is subjected to the internal pressure $p$.\footnote{Note that here, the pressure is not depending on the phase field as in \cite{paul2020}.} \figref{f:n_f_sphr_evo} also shows the positions of the patch interfaces. The parameters $m\in\{32,64\}$ are used for the six-patch discretization (see \figref{f:n_c_6ptch}),  and the parameters $r\in\{2,3\}$ are used for the unstructured spline discretization (see \figref{f:n_c_us}). In this section, a constant penalty parameter $\eps_\phi$, which does not depend on mesh or time step sizes, is used. The material parameters are listed in Table~\ref{t:n_f_sphr_mat}. The parameter $\ell_0$ is chosen in a way, such that the interface is properly resolved for both kind of discretizations. For the coarse mesh ($m=32$, $r=2$), it is set to $0.05\,L_0$, whereas it is set to $0.03\,L_0$ for the fine mesh ($m=64$, $r=3$).
\begin{table}[!ht]
	\centering
	 \caption{Fracturing sphere: Material parameters and imposed internal pressure.}
	\setlength{\tabcolsep}{8pt}
	\renewcommand{\arraystretch}{1.25}
  	\begin{tabular}{c c c c c c c }
  		$E$ $[E_0]$ & $\nu$ $[-]$ & $\sG_c$ $[E_0\,L_0]$ & $T$ $[L_0]$  & $p$ $[E_0\,L_0^{-1}]$
  		\\ \hline
  		$10$ & $0.3$ & $0.0005$ & $0.0125$ & $0.1$
  \end{tabular}
  \label{t:n_f_sphr_mat}
\end{table}

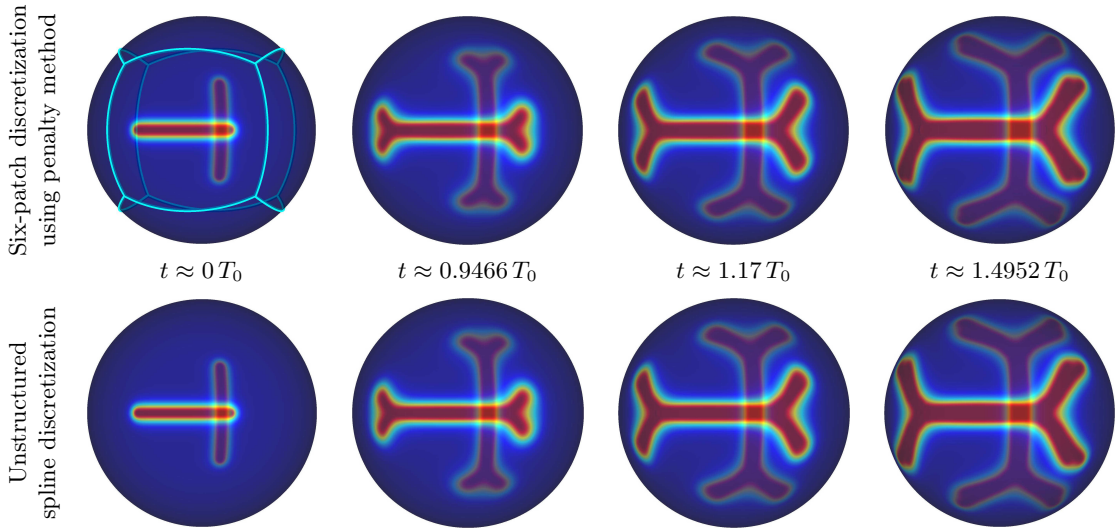
\begin{figure}[!ht]
	\centering
		\begin{tikzpicture}	
			\node at (0,0) {\includegraphics[scale=0.3]{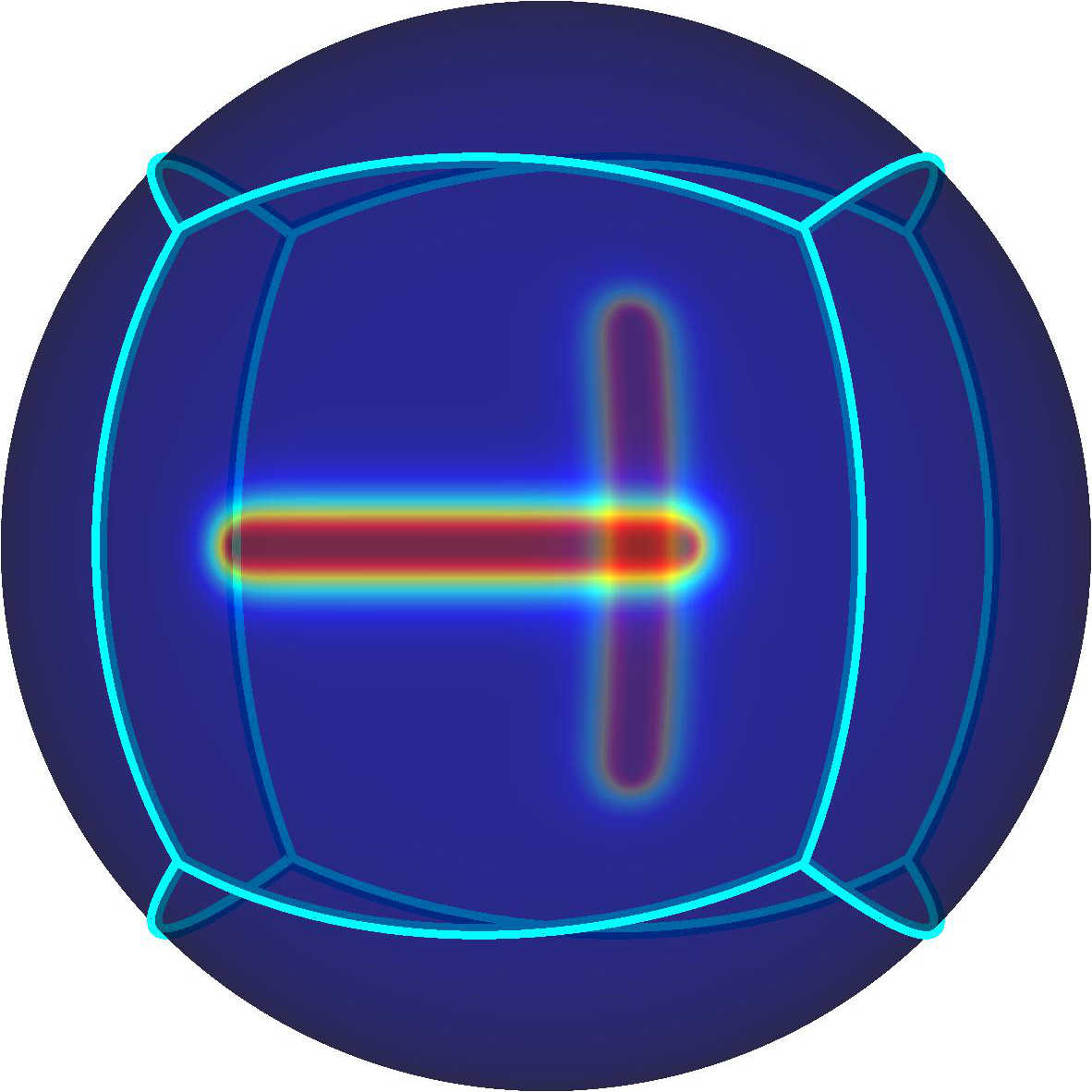}};	
			\node at (3.5,0) {\includegraphics[scale=0.3]{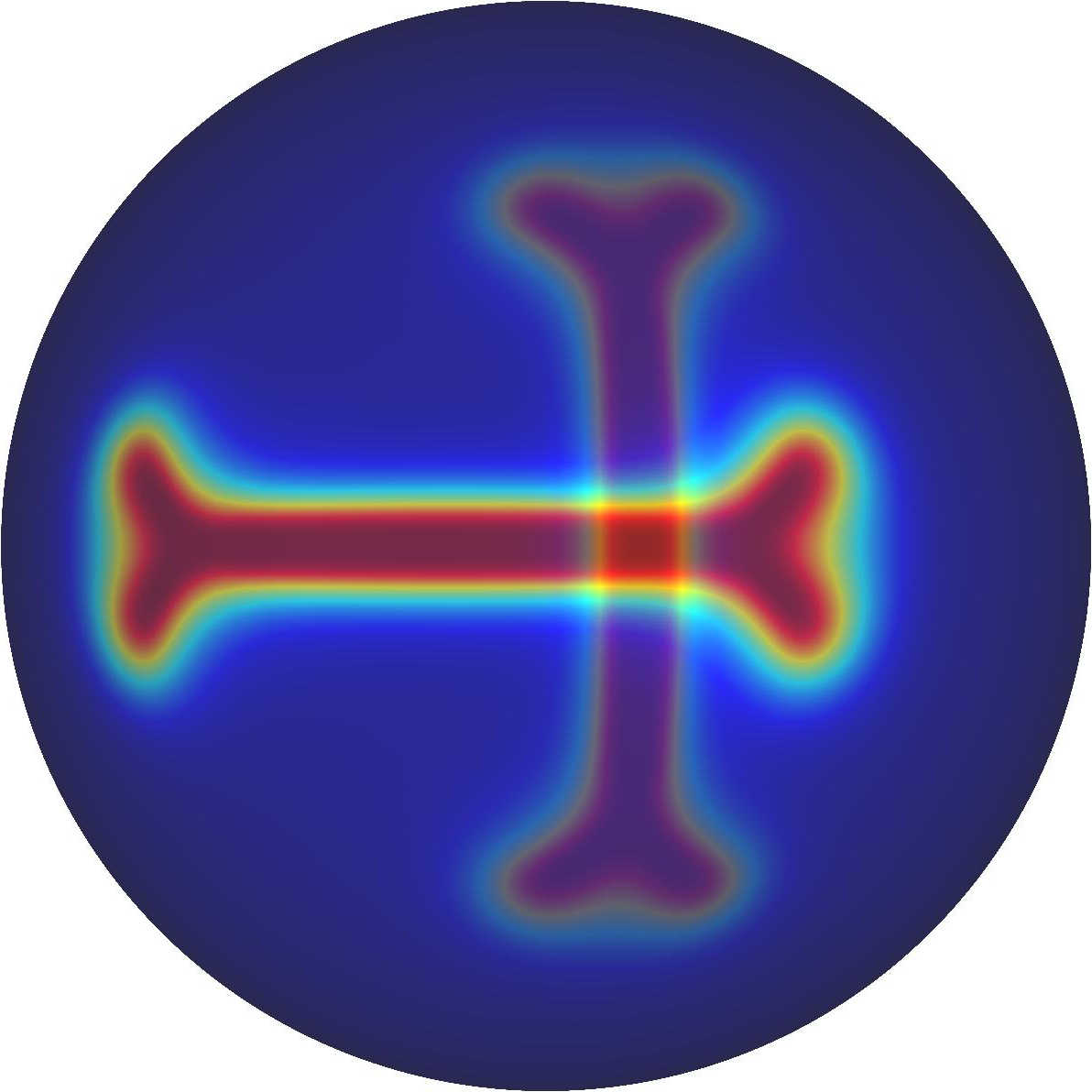}};	
			\node at (7,0) {\includegraphics[scale=0.3]{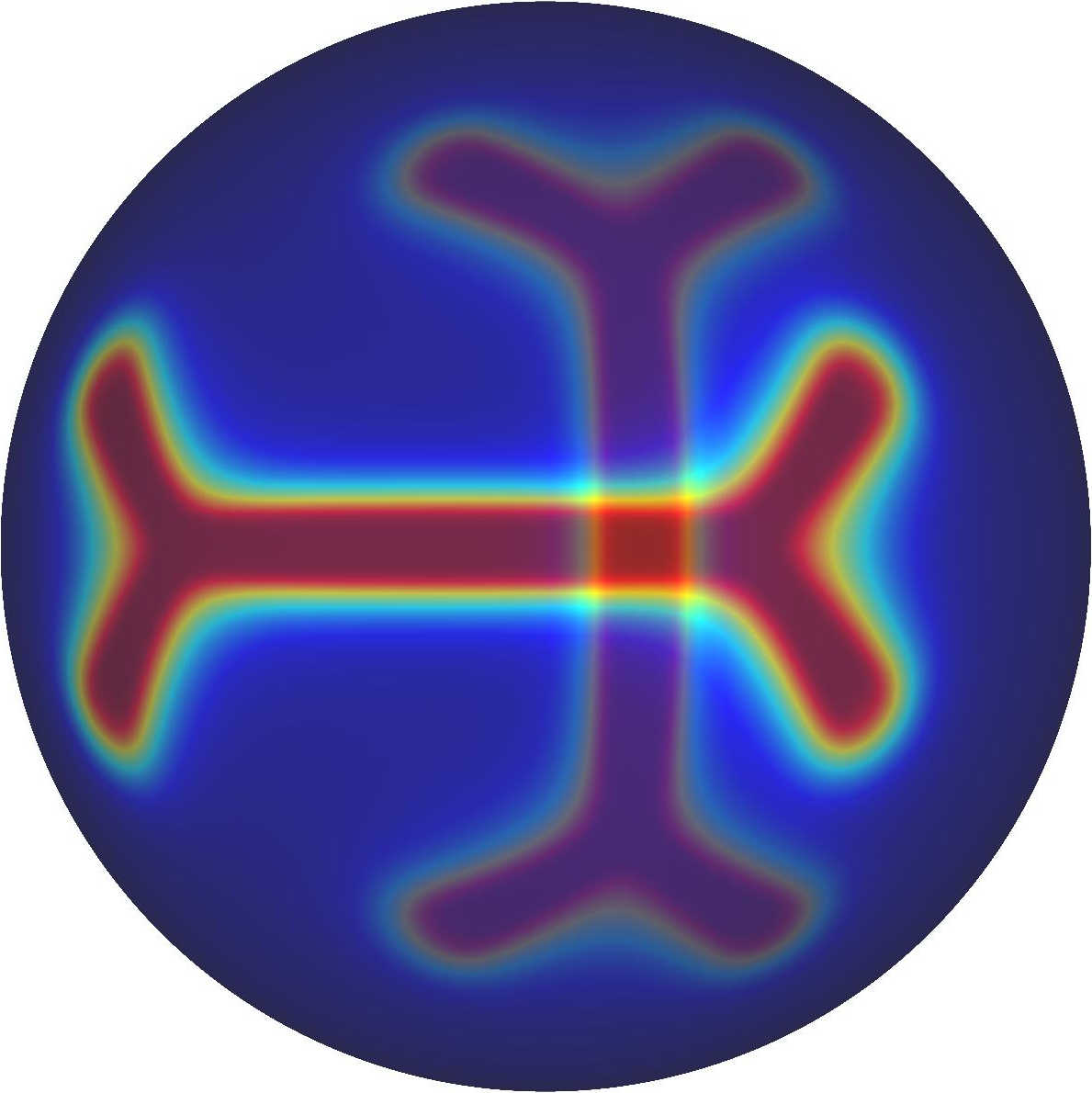}};	
			\node at (10.5,0) {\includegraphics[scale=0.3]{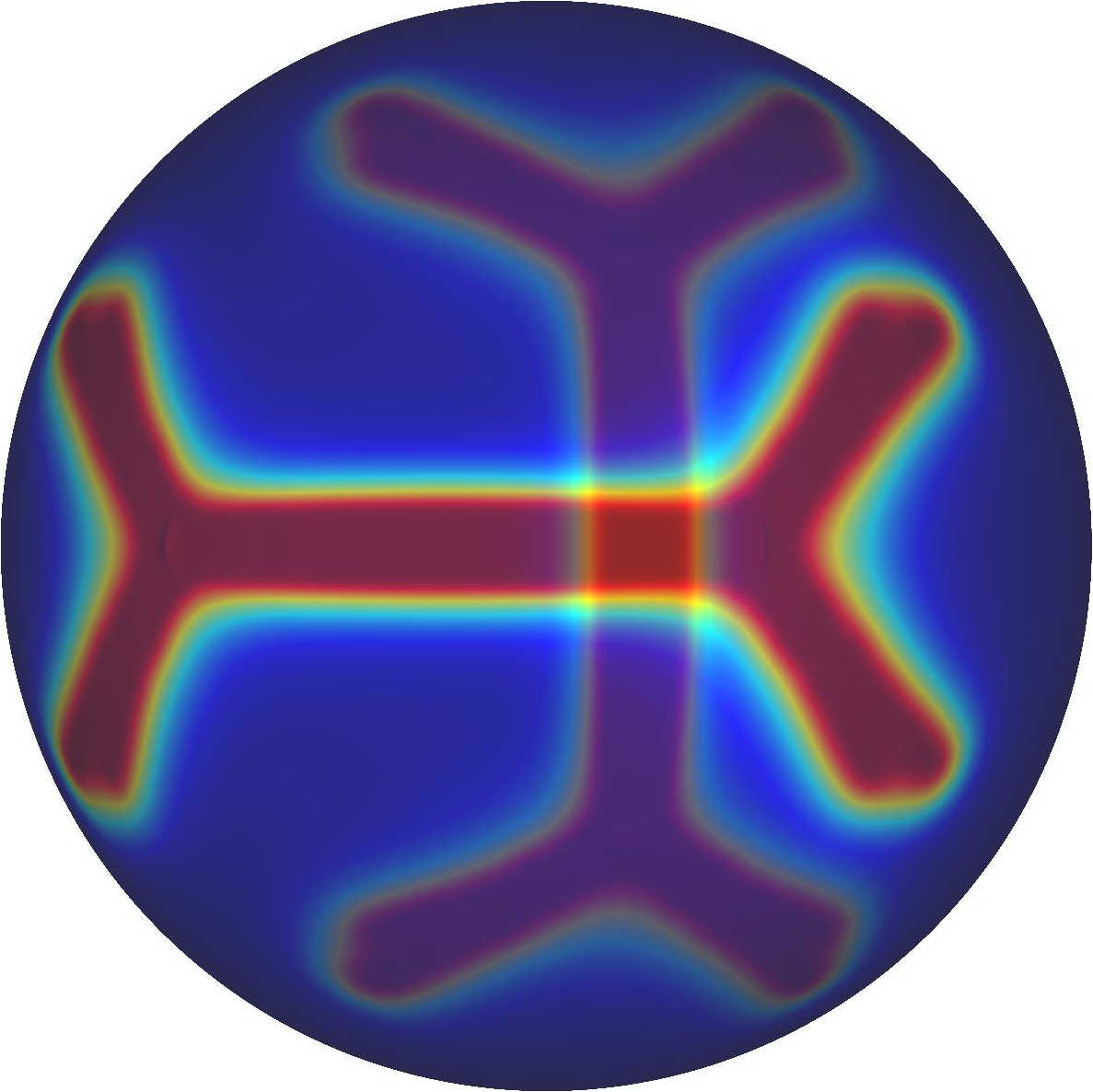}};	
			\node at (0,-3.75) {\includegraphics[scale=0.3]{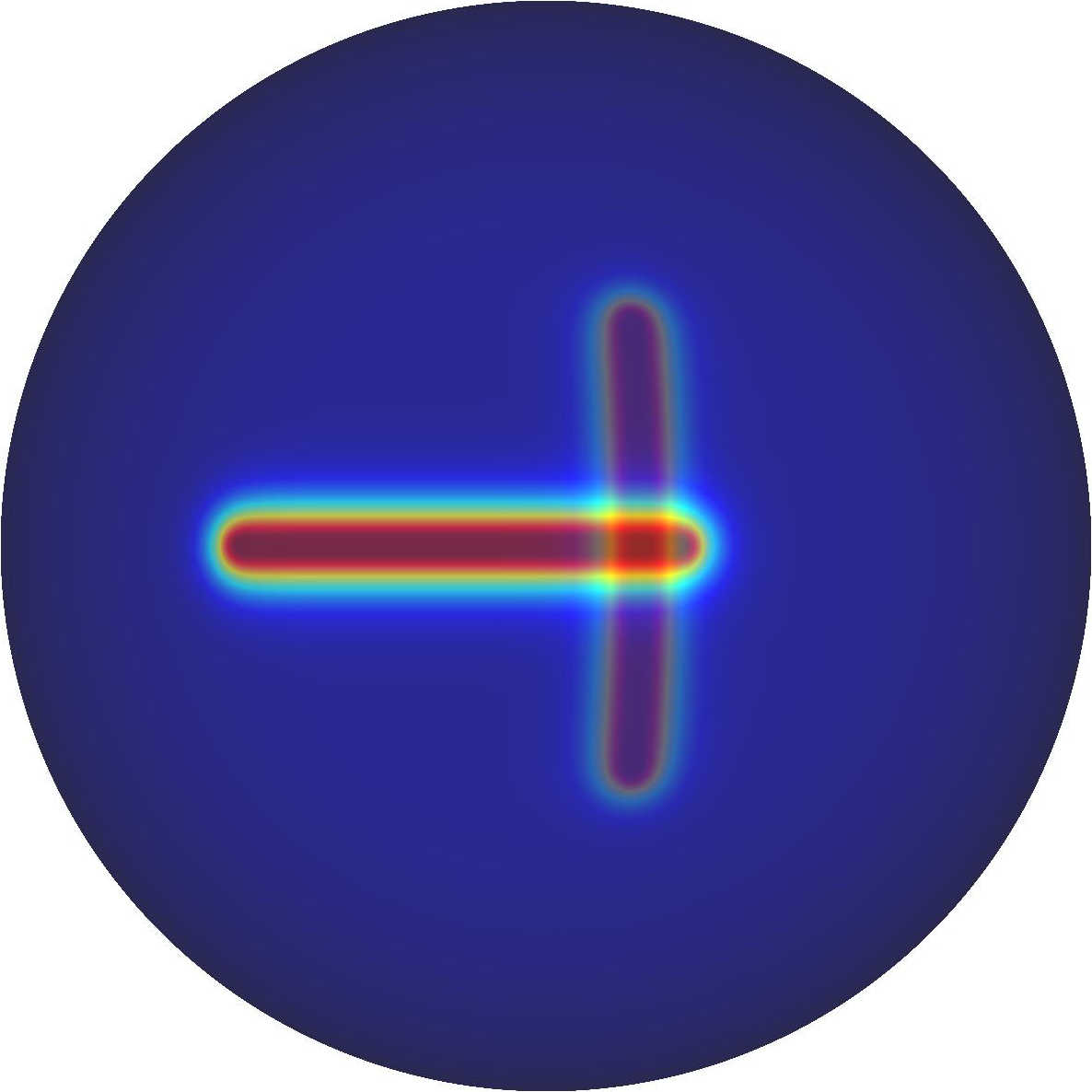}};	
			\node at (3.5,-3.75) {\includegraphics[scale=0.3]{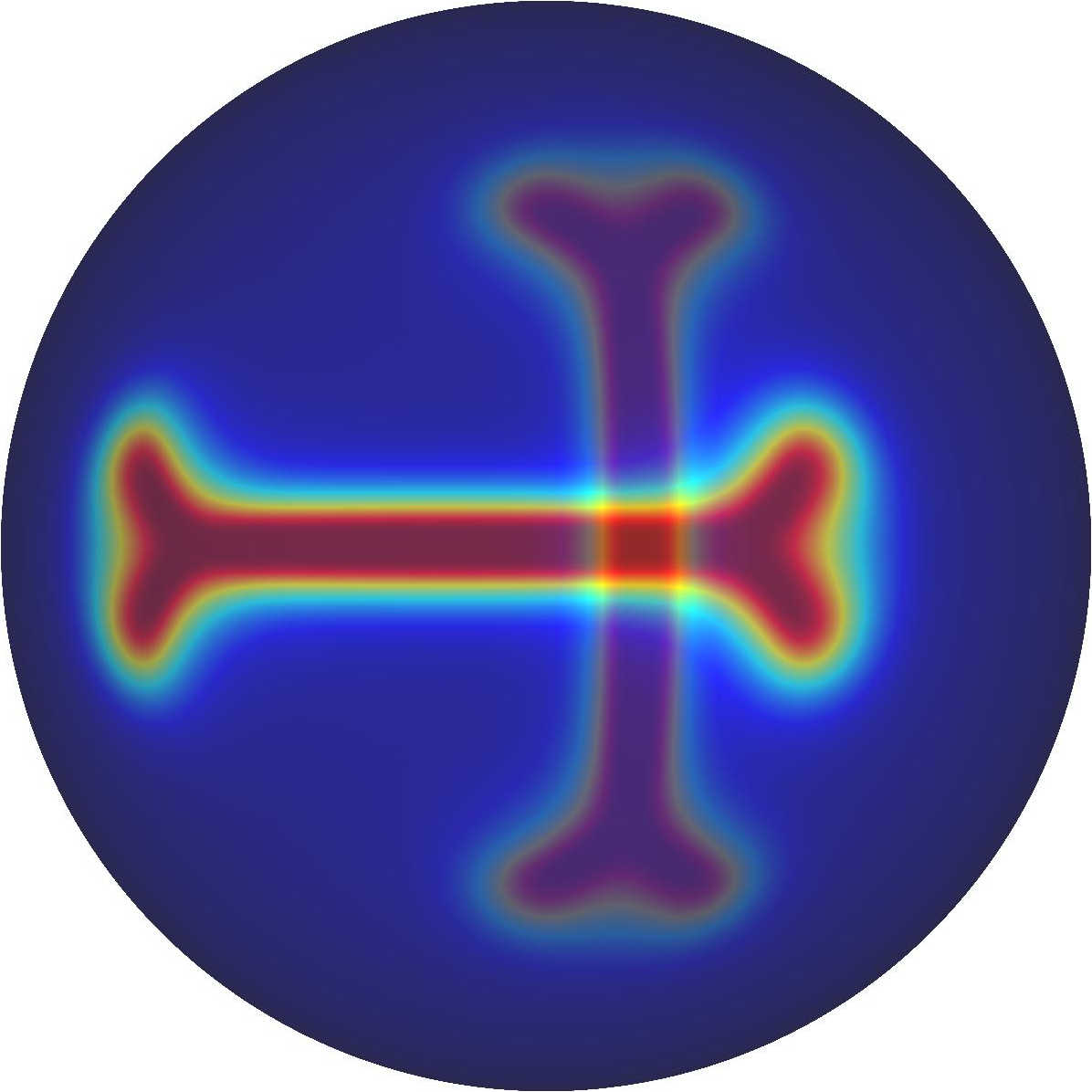}};	
			\node at (7,-3.75) {\includegraphics[scale=0.3]{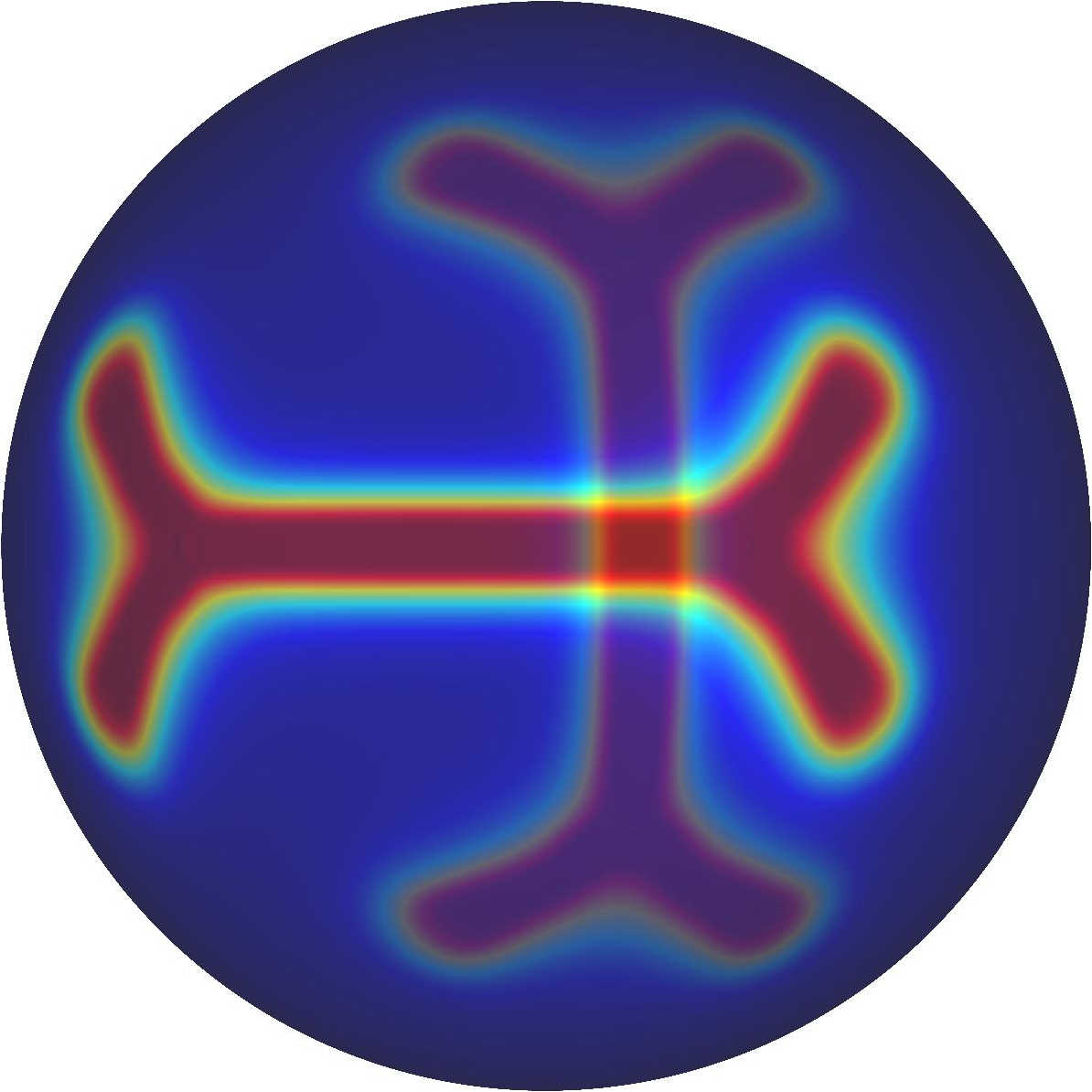}};	
			\node at (10.5,-3.75) {\includegraphics[scale=0.3]{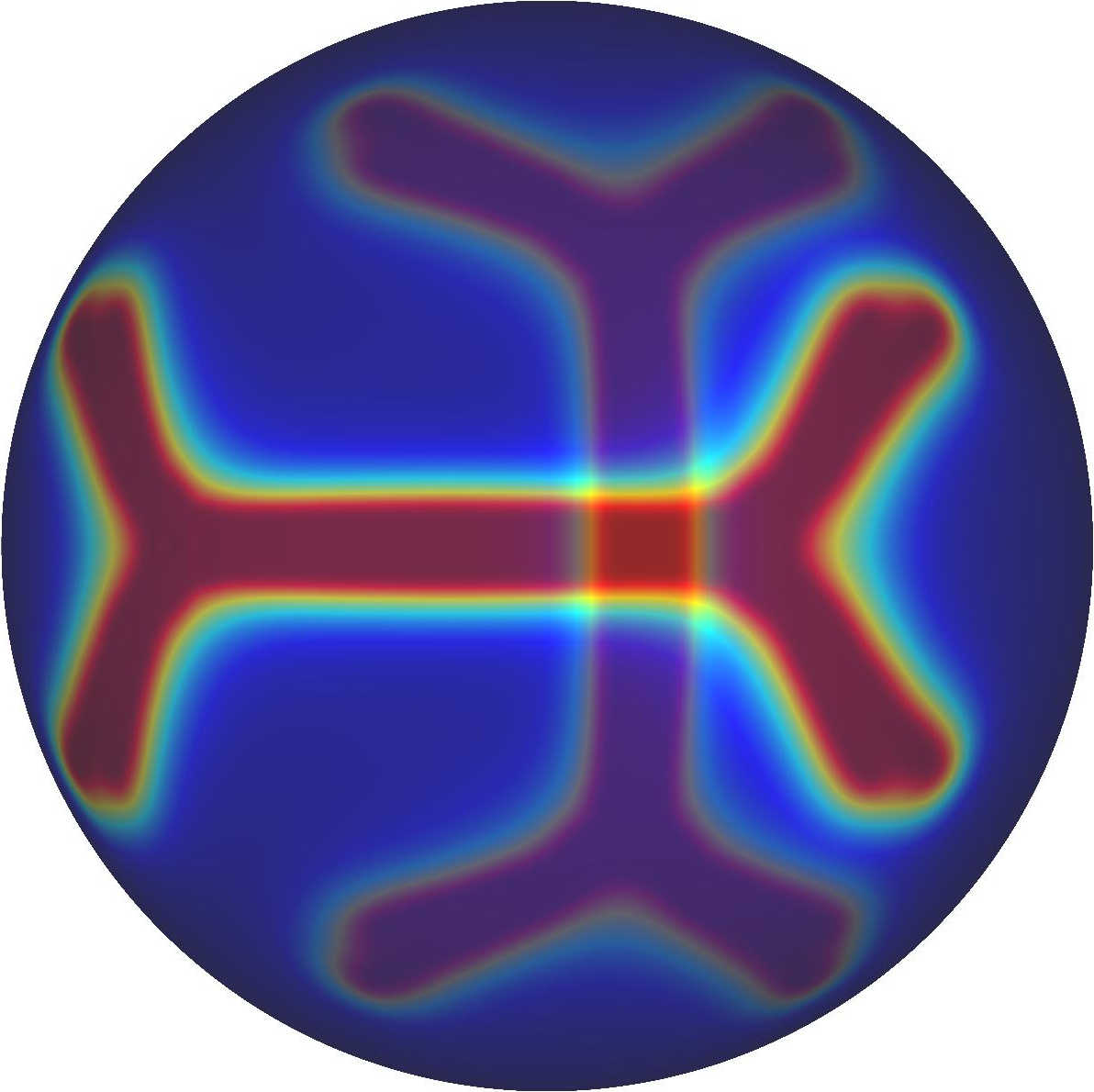}};	
			\node[align=center,font=\footnotesize] at (0,-1.875) {$t\approx0\,T_0$};
			\node[align=center,font=\footnotesize] at (3.5,-1.875) {$t\approx0.9466\,T_0$};
			\node[align=center,font=\footnotesize] at (7,-1.875) {$t\approx1.17\,T_0$};
			\node[align=center,font=\footnotesize] at (10.5,-1.875) {$t\approx1.4952\,T_0$};
			\node[align=center,rotate=90,font=\footnotesize] at (-2.2,0) {Six-patch discretization\\using penalty method};
			\node[align=center,rotate=90,font=\footnotesize] at (-2.2,-3.75) {Unstructured\\spline discretization};
		\end{tikzpicture}
	\caption{Fracturing sphere: Crack evolution for $\ell_0=0.03\,L_0$ using the six-patch discretization ($m=64$) and the penalty method with penalty parameters $\eps_\phi=1000\,E_0\,L_0^3$ and $\eps_\mrn=1000\,E_0\,L_0$ in the top row, and the unstructured spline discretization ($r=3$) in the bottom row. Both cases show excellent agreement. The spheres are visualized transparently.} \label{f:n_f_sphr_evo}
\end{figure}
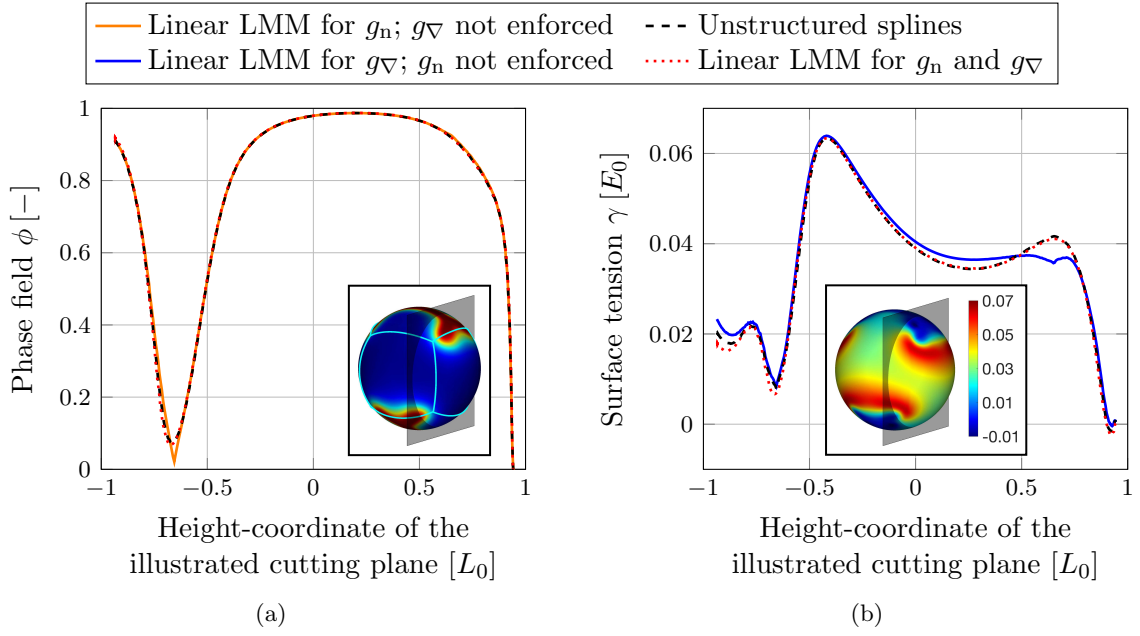
\begin{figure}[!ht]
	\centering
		\begin{tikzpicture}
		\pgfplotsset{width=1\textwidth,height=0.2\textwidth,compat=newest,}
		\begin{axis}[hide axis,xmin=0,xmax=0.00001,ymin=0,ymax=0.00001,legend cell align={left},
					 legend columns=2,legend style={/tikz/every even column/.append style={column sep=2ex}}]
  			  	\addlegendimage{orange,line width=1,}
  			  	\addlegendentry{Linear $\mathrm{LMM}$ for $g_\mrn$; $g_\nabla$ not enforced};
  			  	\addlegendimage{black,dashed,line width=1,}
 			   	\addlegendentry{Unstructured splines};
  			  	\addlegendimage{blue,line width=1,}
  			  	\addlegendentry{Linear $\mathrm{LMM}$ for $g_\nabla$; $g_\mrn$ not enforced};
  			  	\addlegendimage{red,dotted,line width=1,}
  			  	\addlegendentry{Linear $\mathrm{LMM}$ for $g_\mrn$ and $g_\nabla$};
  			 \end{axis}
	\end{tikzpicture}
	\\ \vspace{-3mm}
		\subfloat[\label{f:frc_n_sphr_plnphi}]{
			\begin{tikzpicture}
				\def\cdot{\times}
				\begin{axis}[grid=major,xlabel style={align=center},xlabel={Height-coordinate of the\\illustrated cutting plane $[L_0]$},ylabel={Phase field $\phi\:[-]$},width=0.45\textwidth,height=0.4\textwidth,xmin=-1,xmax=1,ymin=0,ymax=1,legend cell align={left},legend style={nodes={scale=1, transform shape},at={(0.,1.15)},anchor=west},tick label style={font=\footnotesize},legend columns=1,legend style={/tikz/every even column/.append style={column sep=2ex}}]
					\addplot[orange,line width=1,]table [x index = {0}, y index = {1},col sep=comma]{fig/frc/sphr/nqp2_14r1_planeValues.csv};
					\addplot[black,dashed,line width=1,]table [x index = {0}, y index = {1},col sep=comma]{fig/frc/sphr/nqp2_14_planeValues.csv};
					\addplot[red,dotted,line width=1,]table [x index = {0}, y index = {1},col sep=comma]{fig/frc/sphr/us01planeValues.csv};
					\node[anchor=center,draw,black,line width=0.75,rectangle,outer sep=0pt,fill=white,align=left] at (0.5,0.274) {\includegraphics[height=0.125\textwidth]{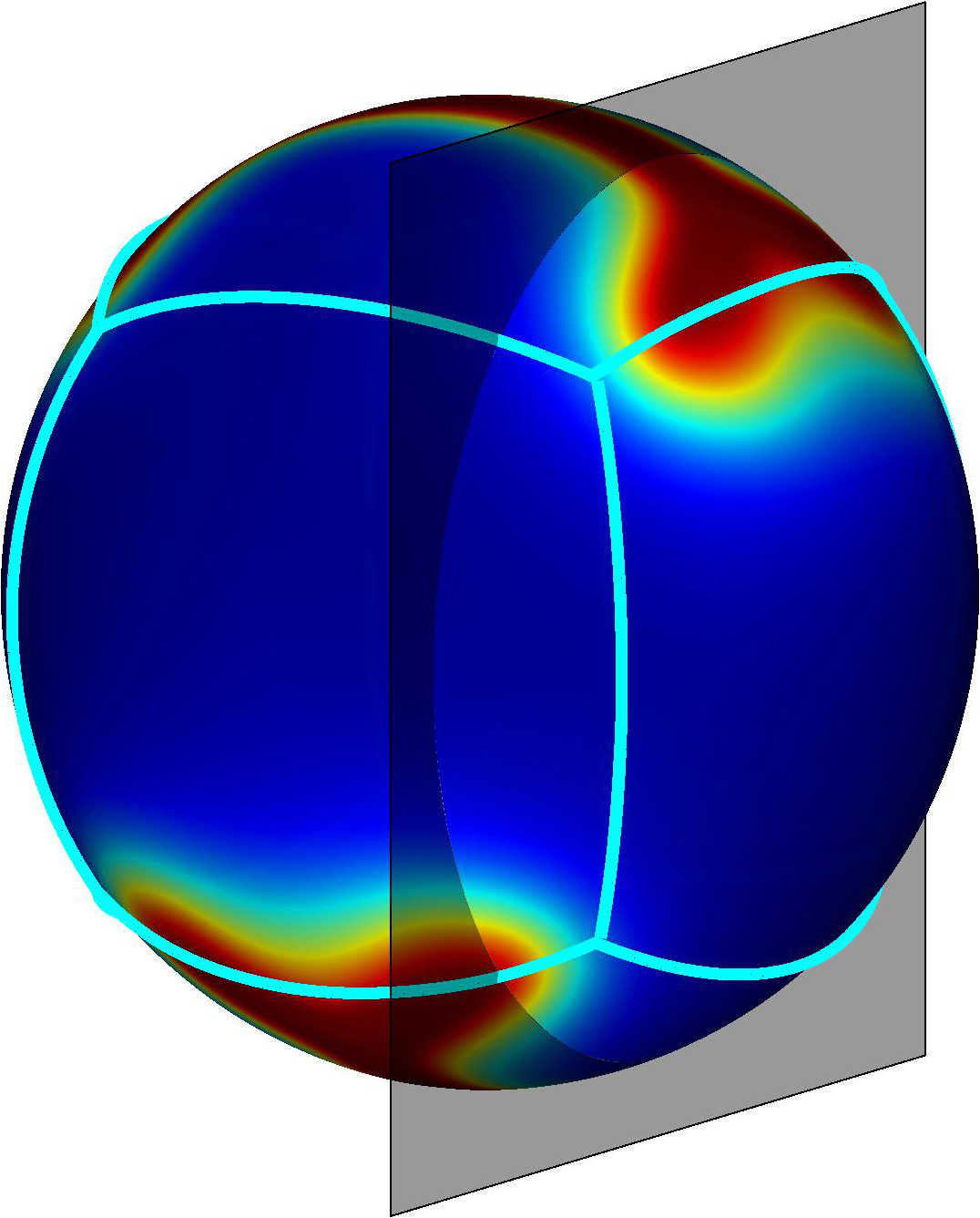}};
				\end{axis}
			\end{tikzpicture}		
		}
	\quad
		\subfloat[\label{f:frc_n_sphr_plngam}]{
			\begin{tikzpicture}
				\def\cdot{\times}
				\begin{axis}[grid=major,xlabel style={align=center},xlabel={Height-coordinate of the\\illustrated cutting plane $[L_0]$},ylabel={Surface tension $\gamma\:[E_0]$},width=0.45\textwidth,height=0.4\textwidth,xmin=-1,xmax=1,ymin=-0.01,ymax=0.07,legend cell align={left},legend style={nodes={scale=1, transform shape},at={(0.275,0.2)},anchor=west},tick label style={font=\footnotesize},yticklabel style={/pgf/number format/fixed,/pgf/number format/precision=5},scaled y ticks=false,legend cell align={left},legend style={nodes={scale=1, transform shape},at={(0.,1.15)},anchor=west},tick label style={font=\footnotesize},legend columns=2,legend style={/tikz/every even column/.append style={column sep=2ex}}]
					\addplot[blue,line width=1,]table [x index = {0}, y index = {2},col sep=comma]{fig/frc/sphr/nqp2_14r2_planeValues.csv};
					\addplot[black,dashed,line width=1,]table [x index = {0}, y index = {2},col sep=comma]{fig/frc/sphr/nqp2_14_planeValues.csv};
					\addplot[red,dotted,line width=1,]table [x index = {0}, y index = {2},col sep=comma]{fig/frc/sphr/us01planeValues.csv};
					\node[anchor=center,draw,black,line width=0.75,rectangle,outer sep=0pt,fill=white,align=left] at (0.05,0.012) {\includegraphics[height=0.125\textwidth]{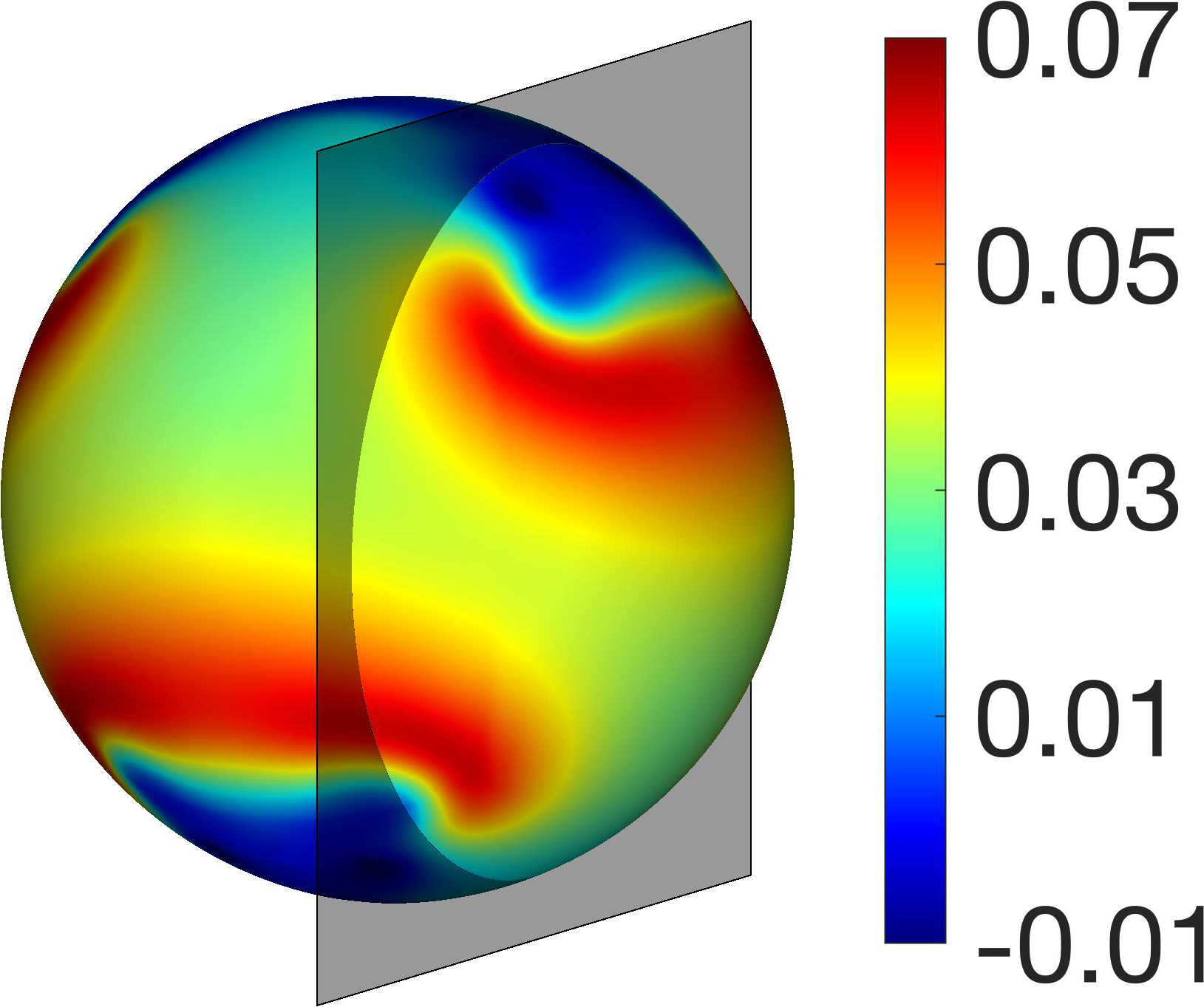}};
				\end{axis}
			\end{tikzpicture}		
		}
	\caption[]{Fracturing sphere: (a) Phase field value and (b) surface tension along the height-coordinate of the illustrated cutting plane at $t\approx1.13\,T_0$ using the Lagrange multiplier method.{\footnotemark} Neglecting one constraint leads to significant offsets in the corrsponding curves.} \label{f:frc_sphr_pln}
\end{figure}
\footnotetext{Note that the time steps for the shown profiles and snapshots slightly differ due to the adaptive time stepping scheme. \label{footnote_n_f_tmstp}}
\begin{figure}[!ht]
	\centering
		\begin{tikzpicture}
		\pgfplotsset{width=1\textwidth,height=0.2\textwidth,compat=newest,}
		\begin{axis}[hide axis,xmin=0,xmax=0.00001,ymin=0,ymax=0.00001,legend cell align={left},
					 legend columns=1,legend style={/tikz/every even column/.append style={column sep=2ex}}]
  			  	\addlegendimage{black,line width=1,}	\addlegendentry{Unstructured splines};
  			  	\addlegendimage{red,densely dotted,line width=1,}	\addlegendentry{$\mathrm{PM}$: $\eps_\mrn=0.1\,E_0\,L_0$, $\eps_\phi=0.1\,E_0\,L_0^3$, $n_\mathrm{qp}=1$};
  			  	\addlegendimage{blue,densely dotted,line width=1,}	\addlegendentry{$\mathrm{PM}$: $\eps_\mrn=0.1\,E_0\,L_0$, $\eps_\phi=0.1\,E_0\,L_0^3$, $n_\mathrm{qp}=2$, Trapezoidal};
  			  	\addlegendimage{orange,dashed,line width=1,}	\addlegendentry{$\mathrm{PM}$: $\eps_\mrn=10\,E_0\,L_0$, $\eps_\phi=10\,E_0\,L_0^3$, $n_\mathrm{qp}=1$};
  			  	\addlegendimage{cyan,dashed,line width=1,}	\addlegendentry{$\mathrm{PM}$: $\eps_\mrn=10\,E_0\,L_0$, $\eps_\phi=10\,E_0\,L_0^3$, $n_\mathrm{qp}=2$, Trapezoidal};
  			  	\addlegendimage{red,dash dot,line width=1,}	\addlegendentry{$\mathrm{PM}$: $\eps_\mrn=100\,E_0\,L_0$, $\eps_\phi=100\,E_0\,L_0^3$, $n_\mathrm{qp}=1$};
  			  	\addlegendimage{blue,dash dot,line width=1,}	\addlegendentry{$\mathrm{PM}$: $\eps_\mrn=100\,E_0\,L_0$, $\eps_\phi=100\,E_0\,L_0^3$, $n_\mathrm{qp}=2$, Trapezoidal};
  			 \end{axis}
	\end{tikzpicture}
	\\ \vspace{-3mm}
		\subfloat[\label{f:frc_sphr_pen2}]{
			\begin{tikzpicture}
				\def\cdot{\times}
				\begin{axis}[grid=major,xlabel style={align=center},xlabel={Height-coordinate $[L_0]$},ylabel={Phase field $\phi\:[-]$},width=0.45\textwidth,height=0.4\textwidth,xmin=-1,xmax=1,ymin=0,ymax=1,legend cell align={left},legend style={nodes={scale=1, transform shape},at={(0.,1.15)},anchor=west},tick label style={font=\footnotesize},legend columns=1,legend style={/tikz/every even column/.append style={column sep=2ex}}]
					\addplot[black,line width=1,]table [x index = {0}, y index = {1},col sep=comma]{fig/frc/sphr/us01planeValues.csv};
					\addplot[red,densely dotted,line width=1,]table [x index = {0}, y index = {1},col sep=comma]{fig/frc/sphr/nqp1_02_planeValues.csv};
					\addplot[orange,dashed,line width=1,]table [x index = {0}, y index = {1},col sep=comma]{fig/frc/sphr/nqp1_04_planeValues.csv};
					\addplot[red,dash dot,line width=1,]table [x index = {0}, y index = {1},col sep=comma]{fig/frc/sphr/nqp1_06_planeValues.csv};
					\addplot[blue,densely dotted,line width=1,]table [x index = {0}, y index = {1},col sep=comma]{fig/frc/sphr/nqp2_12_planeValues.csv};
					\addplot[cyan,dashed,line width=1,]table [x index = {0}, y index = {1},col sep=comma]{fig/frc/sphr/nqp2_10_planeValues.csv};
					\addplot[blue,dash dot,line width=1,]table [x index = {0}, y index = {1},col sep=comma]{fig/frc/sphr/nqp2_08_planeValues.csv}; 
					\coordinate (insertPosition) at (rel axis cs:0.65,0.6);
					\draw[thick,black] (0.72-0.025,0.872-0.025) rectangle (0.73+0.025,0.88+0.025);
					\draw[thick,black,densely dashed] (-0.12,0.6) -- (0.72-0.025,0.872-0.025);
					\draw[thick,black,densely dashed] (0.73,0.6) -- (0.73+0.025,0.872-0.025);
				\end{axis}
				\begin{axis}[at={(insertPosition)},anchor={north},axis background/.style={fill=white},xmin=0.72,xmax=0.73,ymin=0.872,ymax=0.88,width=0.25\textwidth,height=0.25\textwidth,tick label style={font=\footnotesize},grid=both,yticklabel style={/pgf/number format/fixed,/pgf/number format/precision=5},scaled y ticks=false,xtick={0.72,0.725,0.73},ytick={0.875,0.88},axis line style=thick,x tick label style={/pgf/number format/.cd,precision=3,/tikz/.cd}]
					\addplot[black,line width=1,]table [x index = {0}, y index = {1},col sep=comma]{fig/frc/sphr/us01planeValues.csv};
					\addplot[red,densely dotted,line width=1,]table [x index = {0}, y index = {1},col sep=comma]{fig/frc/sphr/nqp1_02_planeValues.csv};
					\addplot[orange,dashed,line width=1,]table [x index = {0}, y index = {1},col sep=comma]{fig/frc/sphr/nqp1_04_planeValues.csv};
					\addplot[red,dash dot,line width=1,]table [x index = {0}, y index = {1},col sep=comma]{fig/frc/sphr/nqp1_06_planeValues.csv};
					\addplot[blue,densely dotted,line width=1,]table [x index = {0}, y index = {1},col sep=comma]{fig/frc/sphr/nqp2_12_planeValues.csv};
					\addplot[cyan,dashed,line width=1,]table [x index = {0}, y index = {1},col sep=comma]{fig/frc/sphr/nqp2_10_planeValues.csv};
					\addplot[blue,dash dot,line width=1,]table [x index = {0}, y index = {1},col sep=comma]{fig/frc/sphr/nqp2_08_planeValues.csv};
				\end{axis}
			\end{tikzpicture}		
		}
	\quad
		\subfloat[\label{f:frc_sphr_pen1}]{
			\begin{tikzpicture}
				\def\cdot{\times}
				\begin{axis}[grid=major,xlabel style={align=center},xlabel={Height-coordinate $[L_0]$},ylabel={Surface tension $\gamma\:[E_0]$},width=0.45\textwidth,height=0.4\textwidth,xmin=-1,xmax=1,ymin=-0.01,ymax=0.07,legend cell align={left},legend style={nodes={scale=1, transform shape},at={(0.275,0.2)},anchor=west},tick label style={font=\footnotesize},yticklabel style={/pgf/number format/fixed,/pgf/number format/precision=5},scaled y ticks=false,legend cell align={left},legend style={nodes={scale=1, transform shape},at={(0.,1.15)},anchor=west},tick label style={font=\footnotesize},legend columns=2,legend style={/tikz/every even column/.append style={column sep=2ex}}]
					\addplot[black,line width=1,]table [x index = {0}, y index = {2},col sep=comma]{fig/frc/sphr/us01planeValues.csv};
					\addplot[red,densely dotted,line width=1,]table [x index = {0}, y index = {2},col sep=comma]{fig/frc/sphr/nqp1_02_planeValues.csv};
					\addplot[orange,dashed,line width=1,]table [x index = {0}, y index = {2},col sep=comma]{fig/frc/sphr/nqp1_04_planeValues.csv};
					\addplot[red,dash dot,line width=1,]table [x index = {0}, y index = {2},col sep=comma]{fig/frc/sphr/nqp1_06_planeValues.csv};
					\addplot[blue,densely dotted,line width=1,]table [x index = {0}, y index = {2},col sep=comma]{fig/frc/sphr/nqp2_12_planeValues.csv};
					\addplot[cyan,dashed,line width=1,]table [x index = {0}, y index = {2},col sep=comma]{fig/frc/sphr/nqp2_10_planeValues.csv};
					\addplot[blue,dash dot,line width=1,]table [x index = {0}, y index = {2},col sep=comma]{fig/frc/sphr/nqp2_08_planeValues.csv};
					\coordinate (insertPosition) at (rel axis cs:0.6,0.45);
					\draw[thick,black] (-0.22-0.02,0.048) rectangle (-0.18+0.02,0.051);
					\draw[thick,black,densely dashed] (-0.235,0.026) -- (-0.22-0.02,0.048);
					\draw[thick,black,densely dashed] (0.625,0.026) -- (-0.18+0.02,0.048);
				\end{axis}
				\begin{axis}[at={(insertPosition)},anchor={north},axis background/.style={fill=white},xmin=-0.22,xmax=-0.18,ymin=0.048,ymax=0.051,width=0.25\textwidth,height=0.21\textwidth,tick label style={font=\footnotesize},grid=both,yticklabel style={/pgf/number format/fixed,/pgf/number format/precision=5},scaled y ticks=false,xtick={-0.22,-0.2,-0.18},ytick={0.049,0.051},axis line style= thick]
					\addplot[black,line width=1,]table [x index = {0}, y index = {2},col sep=comma]{fig/frc/sphr/us01planeValues.csv};
					\addplot[red,densely dotted,line width=1,]table [x index = {0}, y index = {2},col sep=comma]{fig/frc/sphr/nqp1_02_planeValues.csv};
					\addplot[orange,dashed,line width=1,]table [x index = {0}, y index = {2},col sep=comma]{fig/frc/sphr/nqp1_04_planeValues.csv};
					\addplot[red,dash dot,line width=1,]table [x index = {0}, y index = {2},col sep=comma]{fig/frc/sphr/nqp1_06_planeValues.csv};
					\addplot[blue,densely dotted,line width=1,]table [x index = {0}, y index = {2},col sep=comma]{fig/frc/sphr/nqp2_12_planeValues.csv};
					\addplot[cyan,dashed,line width=1,]table [x index = {0}, y index = {2},col sep=comma]{fig/frc/sphr/nqp2_10_planeValues.csv};
					\addplot[blue,dash dot,line width=1,]table [x index = {0}, y index = {2},col sep=comma]{fig/frc/sphr/nqp2_08_planeValues.csv};
				\end{axis}
			\end{tikzpicture}		
		}
	\caption{Fracturing sphere: (a) Phase field value and (b) surface tension along the height-coordinate of the cutting plane from \figref{f:frc_sphr_pln} at $t\approx1.13\,T_0$ using the penalty method.\textsuperscript{\ref{footnote_n_f_tmstp}} The penalty parameters and the quadrature rule are varied. As the penalty parameters are increased, good agreement with the reference solution is obtained.} \label{f:frc_sphr_pen}
\end{figure}
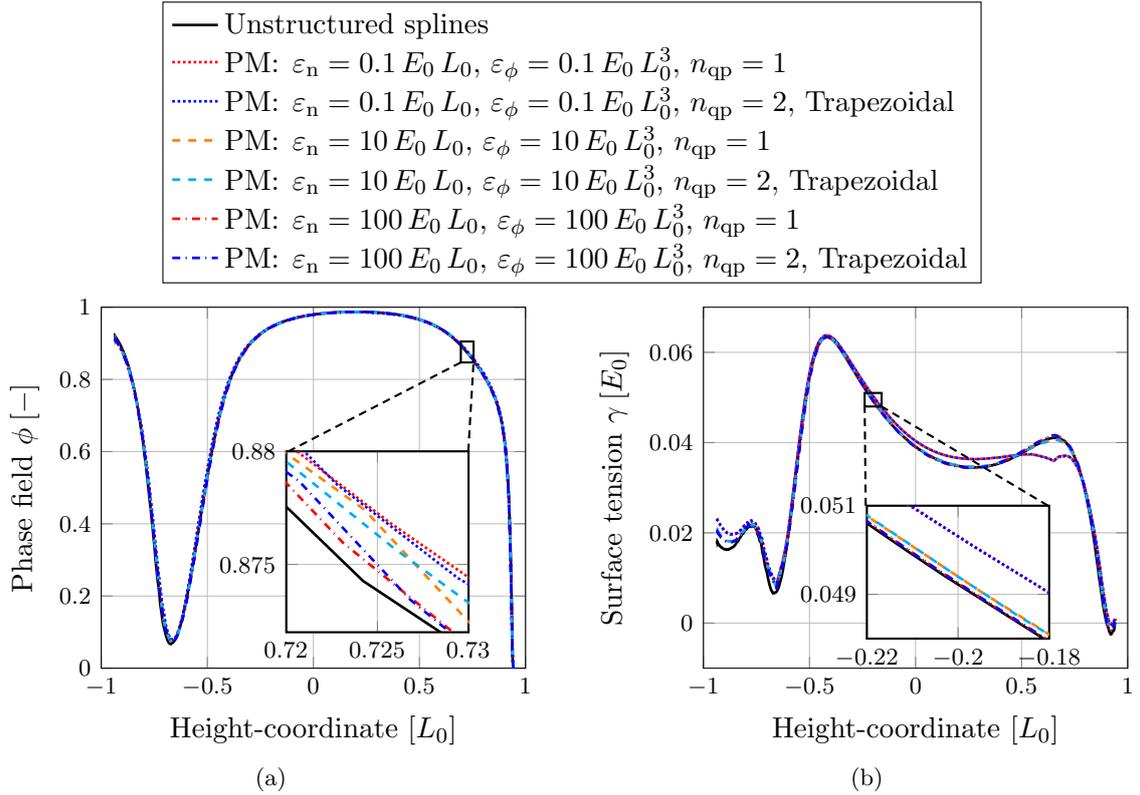

The evolution of the crack is illustrated in \figref{f:n_f_sphr_evo}. At first, the cracks propagate along their initial direction, but then start branching. In the process, they cross several patch interfaces. For the remainder of this section, the coarse meshes ($m=32$, $r=2$) and the larger length scale parameter $\ell_0=0.05\,L_0$ are used.

\begin{figure}[!ht]
	\centering	
		\subfloat[Lagrange multiplier solution\label{f:n_frc_sphr_pmref}]{
			\begin{tikzpicture}
				\node at (-3.1,.53) {\includegraphics[width=0.12\textwidth]{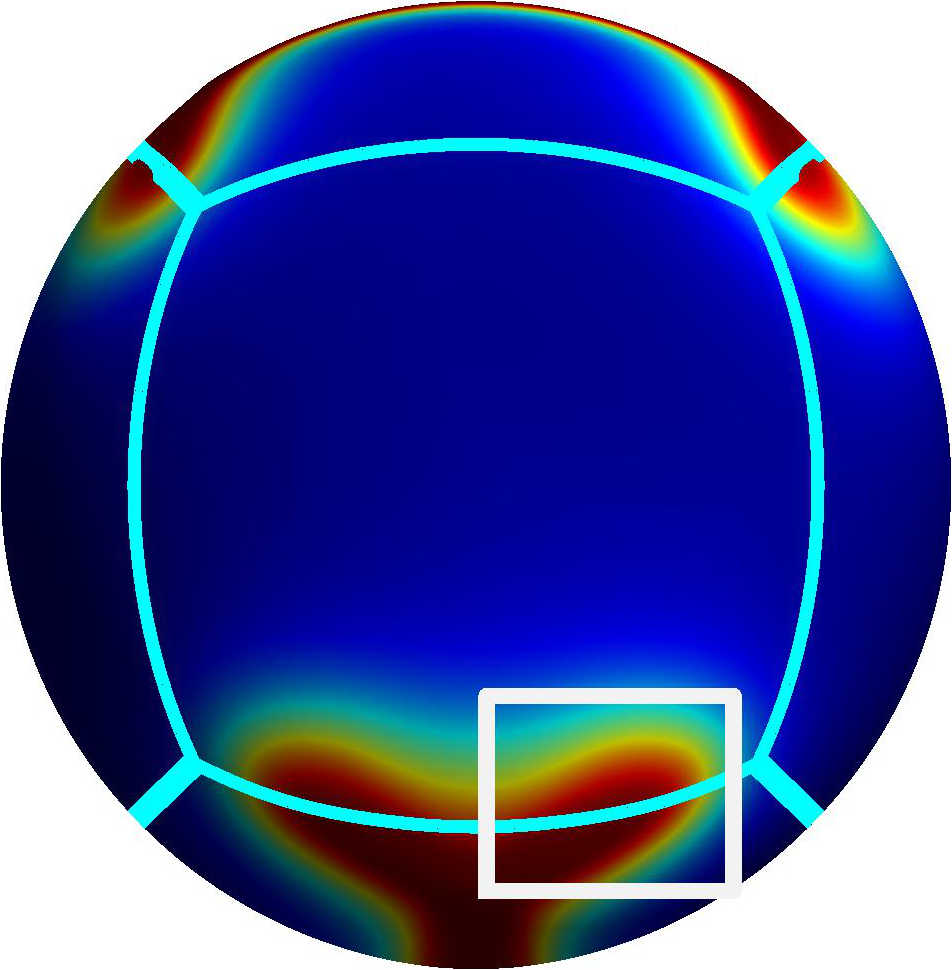}};
				\node at (0,0) {\includegraphics[width=0.24\textwidth]{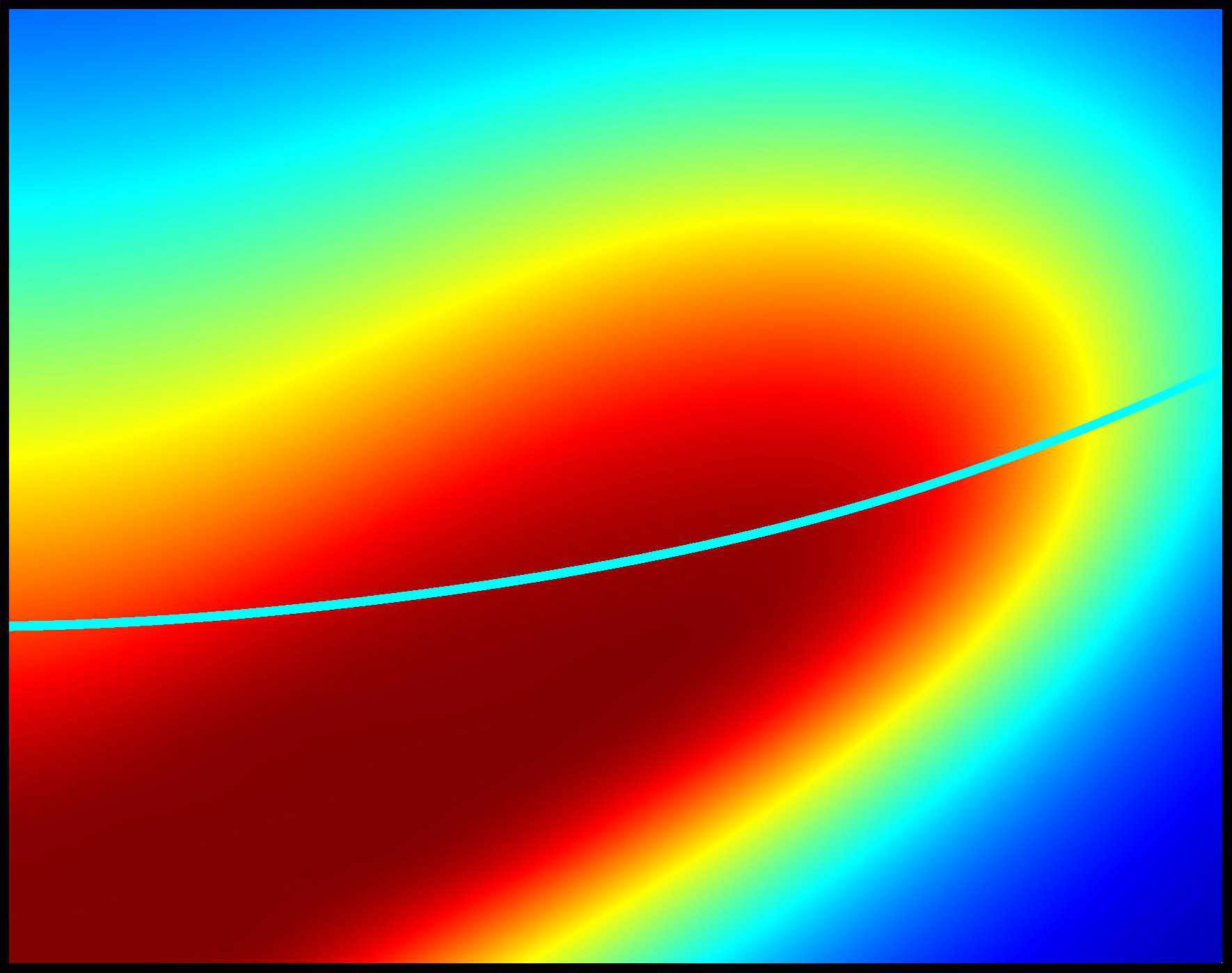}};
			\end{tikzpicture}
		}
	\quad
		\subfloat[Underconstraining\label{f:n_frc_sphr_pmunder}]{
			\begin{tikzpicture}
				\node at (0,0) {\includegraphics[width=0.24\textwidth]{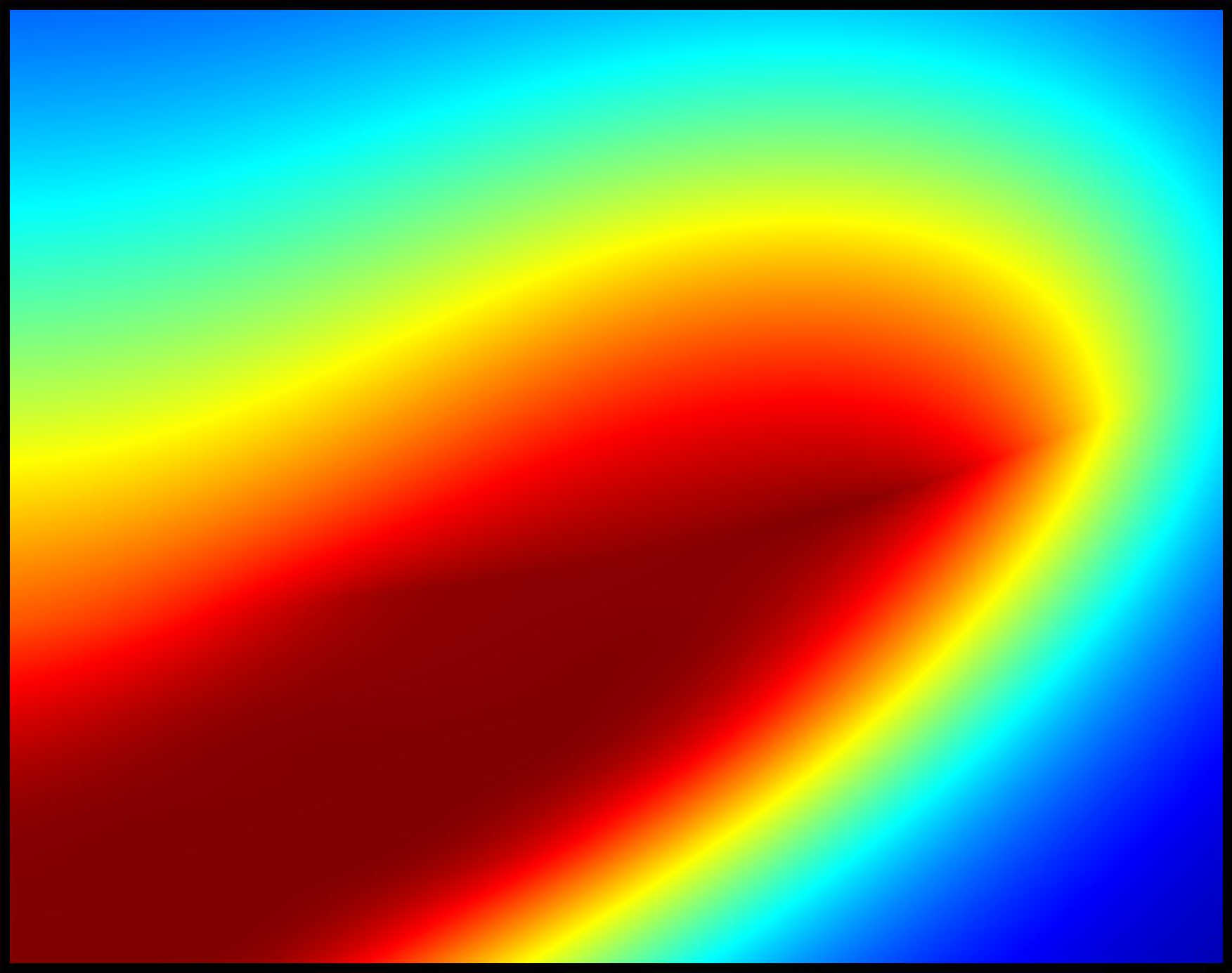}};
			\end{tikzpicture}
		}
	\quad
		\subfloat[Overconstraining\label{f:n_frc_sphr_pmover}]{
			\begin{tikzpicture}
				\node at (0,0) {\includegraphics[width=0.24\textwidth]{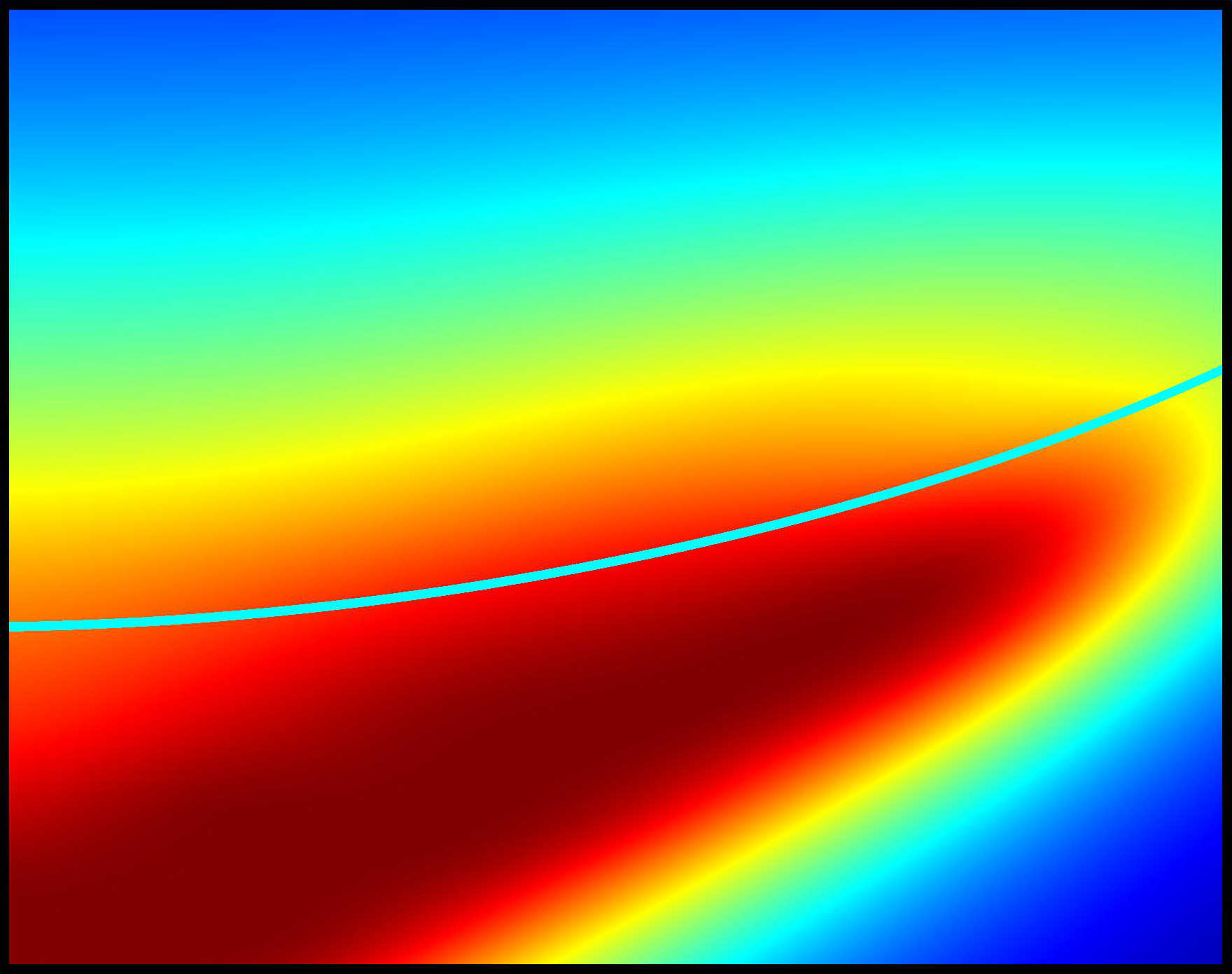}};
			\end{tikzpicture}
		}
	\caption{Fracturing sphere: (a) Phase field at $t\approx1.13\,T_0$ for the Lagrange multiplier method with linear interpolation ($n_\mathrm{qp}=2$, Gaussian quadrature). The enlargement is shown next to it. (b) The effect of a too small penalty parameter ($\eps_\phi=0$, $\eps_\mrn=100\,E_0\,L_0$, $n_\mathrm{qp}=2$, Trapezoidal) on the phase field (the patch line is omitted for better visualization). (c) The effect of overconstraining the penalty method ($\eps_\phi=100\,E_0\,L_0^3$, $\eps_\mrn=100\,E_0\,L_0$) using $n_\mathrm{qp}=3$ Gaussian quadrature points.} \label{f:n_frc_sphr_pmerr}
\end{figure}

\figref{f:frc_sphr_pln} shows the phase field value and surface tension (see \eqsref{e:s_gamel}) over the height-coordinate of the illustrated cutting planes. The curves for the Lagrange multiplier method with linear interpolation and the reference solution based on the unstructured spline discretization show good agreement. The solid line in both figures shows the profile in case one of the constraints is neglected. If the phase field constraint is neglected, a kink will appear in $\phi$ at position $\approx\!-0.65\,L_0$, as the figure shows. This is further illustrated in \figref{f:n_frc_sphr_pmunder}. These kinks are avoided if the constraint is fulfilled. Further inaccuracies for $\gamma$ can be observed at $\approx\!\{-0.93,-0.65,0.2,0.65,0.94\}\,L_0$.

\figref{f:frc_sphr_pen} illustrates the phase field and surface tension over the height-coordinate of the same cutting plane, but now, the constraints are enforced by the penalty method. As the penalty parameters increases, good agreement with the reference solution is obtained. There are no significant differences between the results from the midpoint or trapezoidal rule (note that the quadrature points for these quadrature rules coincide with the locations of the constant or linear Lagrange multipliers, respectively, see \figref{f:p_css}). Note that Gaussian quadrature with two or more quadrature points leads to overconstraining of the solution. The constraints cannot be appropriately satisfied at too many locations and thus, the fracture field shows unphysical behavior, see \figref{f:n_frc_sphr_pmover}.

\begin{figure}[!ht]
	\centering
		\subfloat[Constant $\mathrm{LMM}$ for $g_\nabla$, $g_\mrn$ not enforced\label{f:n_frc_sphr_gam1}]{\includegraphics[scale=0.42]{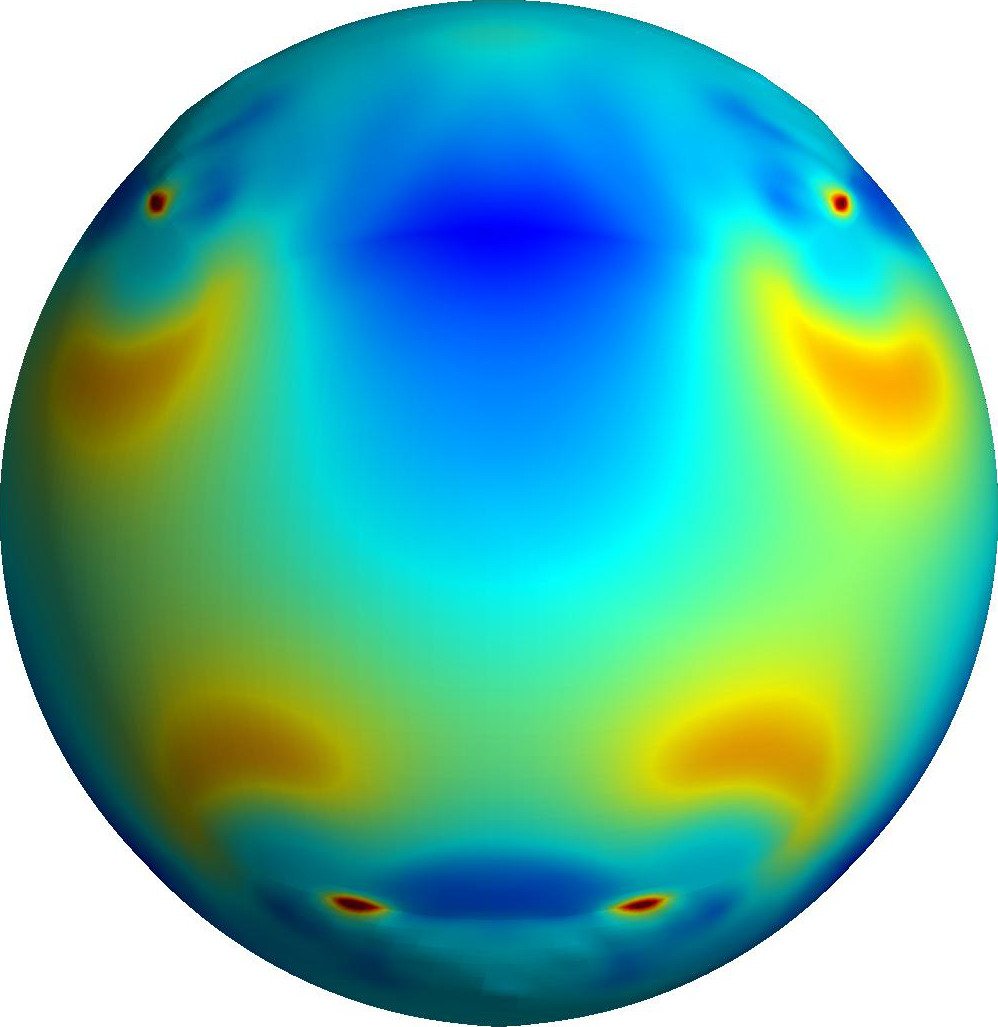}}
	\quad
		\subfloat[$\mathrm{PM}$: $\eps_\phi=100\,E_0\,L_0^3$,\newline$\eps_\mrn=100\,E_0\,L_0$, $n_\mathrm{qp}=2$, Trapezoidal\label{f:n_frc_sphr_gam2}]{\includegraphics[scale=0.42]{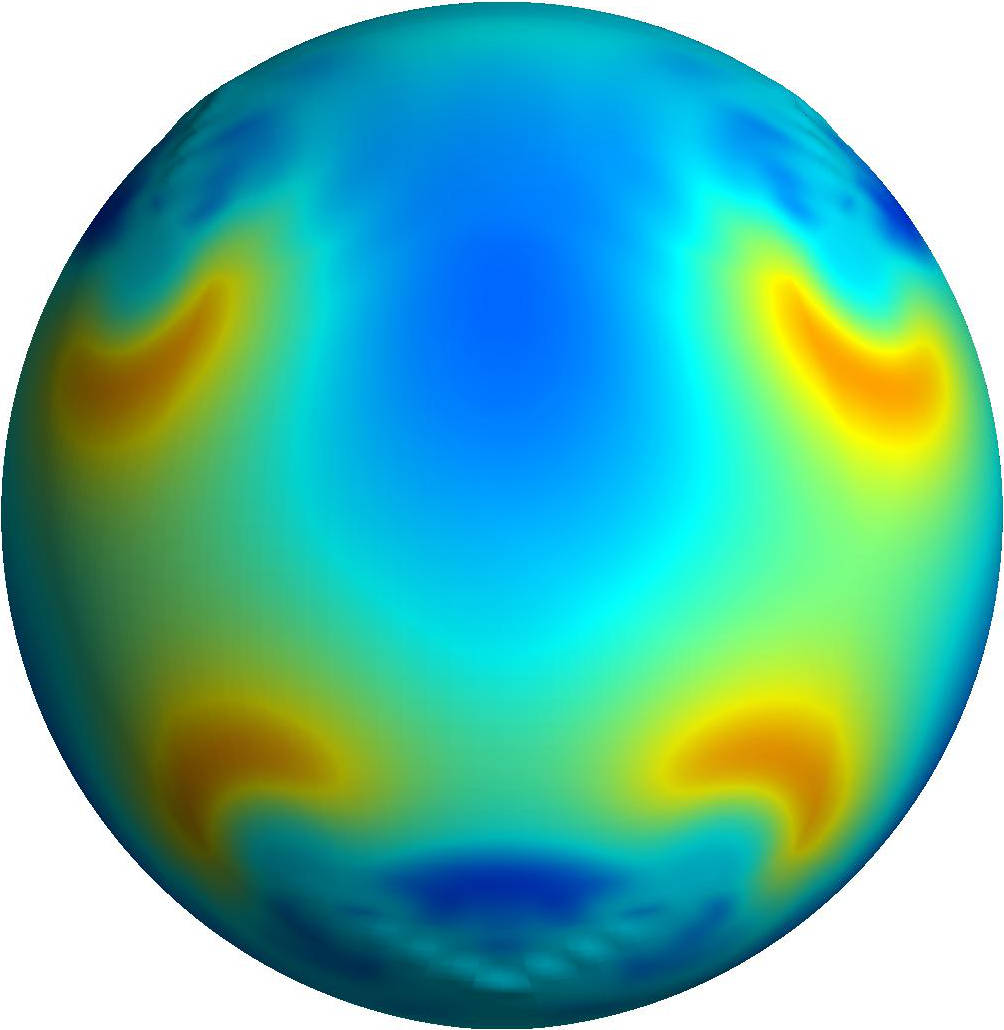}}
	\quad
		\subfloat[Linear $\mathrm{LMM}$ for $g_\mrn$ and $g_\nabla$\label{f:n_frc_sphr_gam3}]{\includegraphics[scale=0.42]{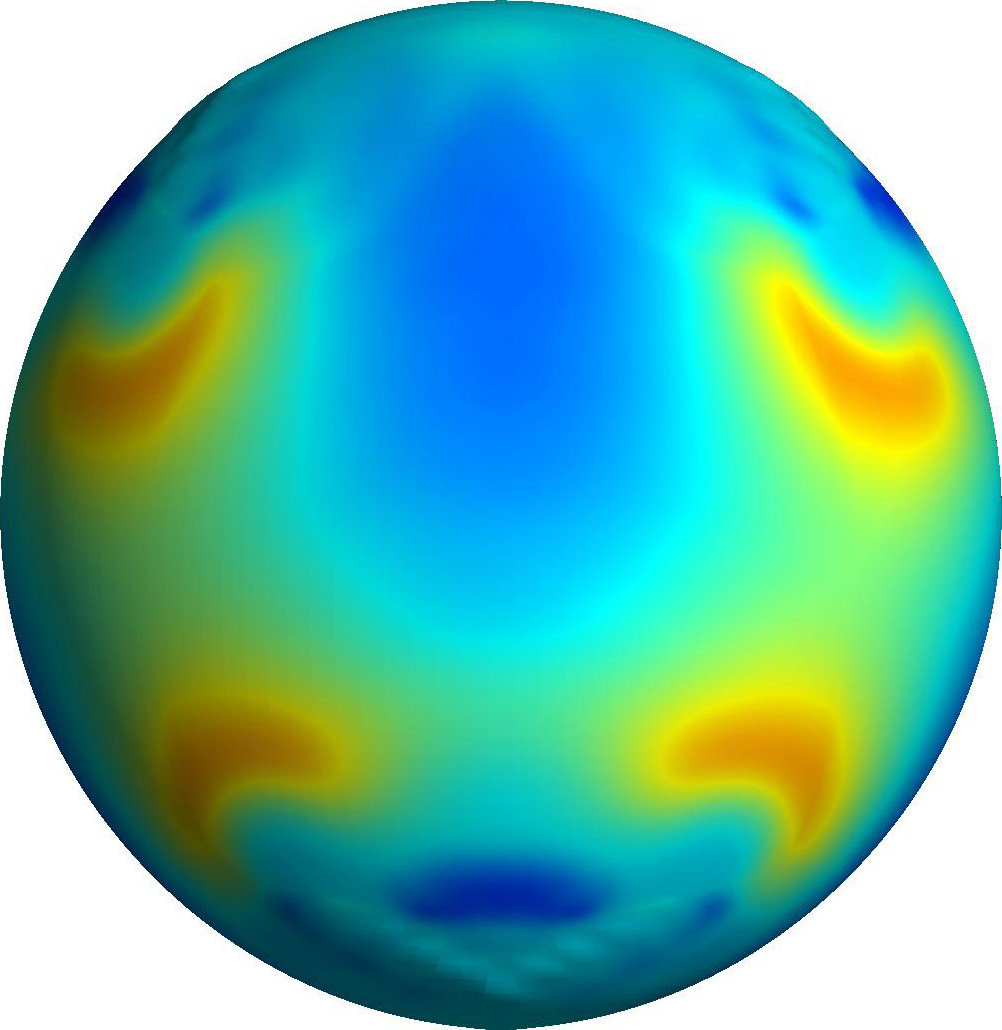}}
	\quad
		\subfloat[Unstructured splines (no patch interfaces)\label{f:n_frc_sphr_gam4}]{\includegraphics[scale=0.42]{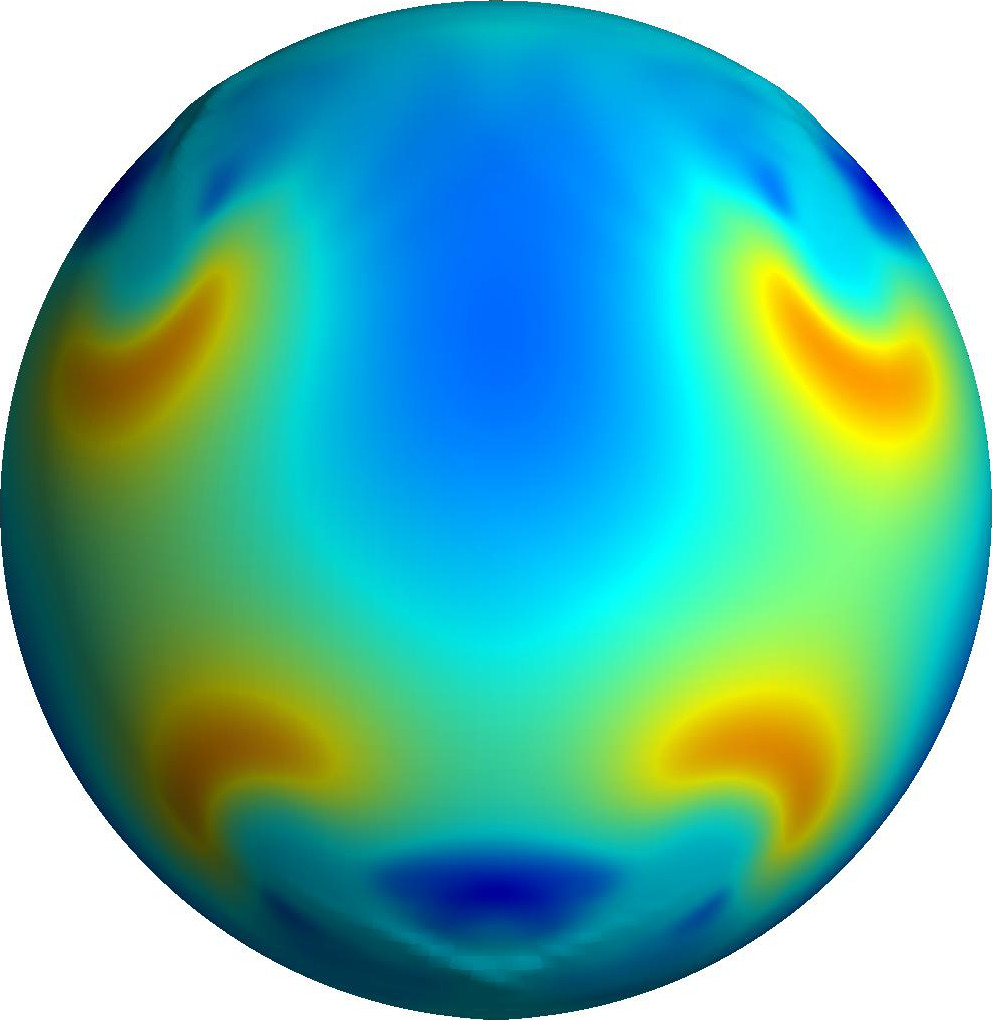}}
	\\
		\includegraphics[width=0.8\textwidth]{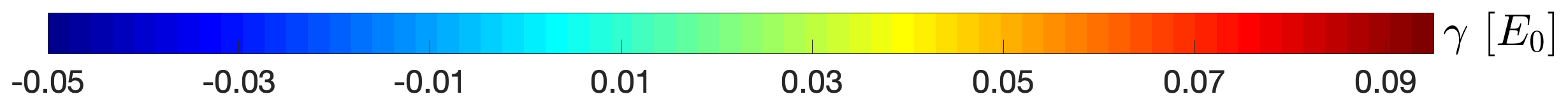}
	\caption{Fracturing sphere: Surface tension at $t\approx1.3263\,T_0$ for different patch constraint enforcement techniques.\textsuperscript{\ref{footnote_n_f_tmstp}}} \label{f:n_frc_sphr_gam}
\end{figure}

The effect on the phase field and surface tension, when either $g_\nabla$ or $g_\mrn$ are ignored, are illustrated in Figs.~\ref{f:n_frc_sphr_pmerr}--\ref{f:n_frc_sphr_gam}.
If the phase field is underconstrained, it will be discontinuous at the interface, see \figref{f:n_frc_sphr_pmunder}. The choice of an inappropriate quadrature rule for the penalty method leads to overconstraining of the solution and an unphysical deflection of the phase field at the patch interfaces will occur, see \figref{f:n_frc_sphr_pmover}.
Finally, the effect of an insufficient enforcement of the $G^1$-continuity constraint is shown in \figref{f:n_frc_sphr_gam}. If this constraint is not appropriately fulfilled, stress peaks will appear at the patch interfaces, see \figref{f:n_frc_sphr_gam1}. Figs.~\ref{f:n_frc_sphr_gam2}--\ref{f:n_frc_sphr_gam3} show the surface tension for different enforcement techniques in comparison to the fully $C^1$-continuous discretization in \figref{f:n_frc_sphr_gam4} at $t\approx1.3263\,T_0$. Again, the results from the different discretization techniques show excellent agreement.

\subsubsection{Crack propagation across kinks} \label{s:n_f_knk}
\begin{figure}[!ht]
	\centering
		\subfloat[Top view\label{f:n_f_knk_geom1}]{\includegraphics[scale=1]{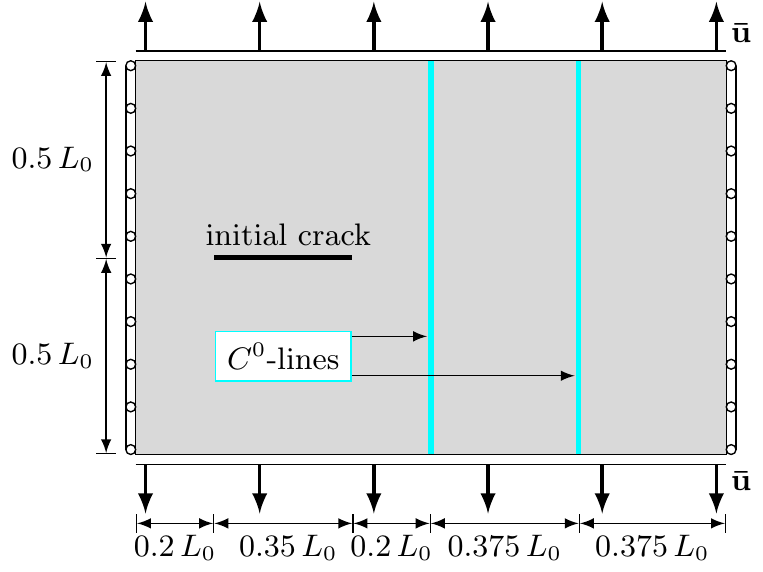}}
	\qquad
		\subfloat[Side view\label{f:n_f_knk_geom2}]{\includegraphics[scale=1]{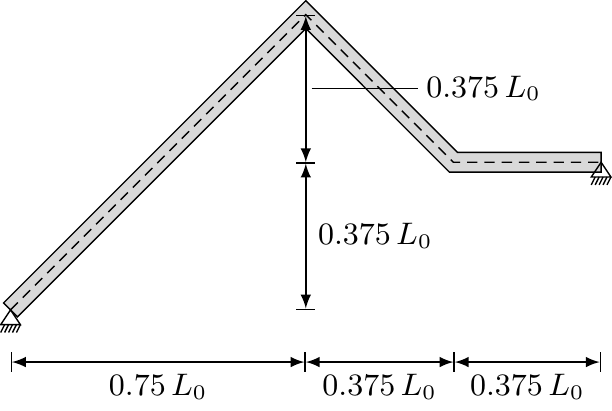}}
	\caption{Crack propagation across kinks: Geometry with initial crack. The two $C^0$-lines are marked by bold cyan-colored lines.} \label{f:n_f_knk_geom}
\end{figure}

\begin{figure}[!ht]
	\centering
		\subfloat[Initial phase field\label{f:n_f_knk_geom3}]{\includegraphics[scale=1,valign=t]{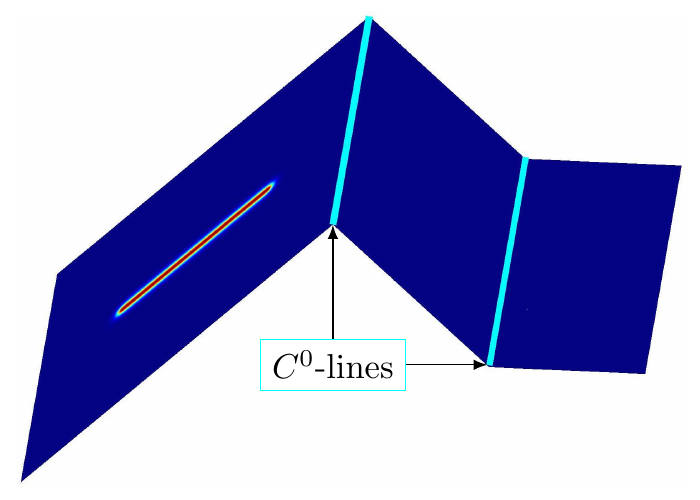}}
	\qquad\qquad
		\subfloat[LR mesh\label{f:n_f_knk_geom4}]{\includegraphics[scale=1,valign=t]{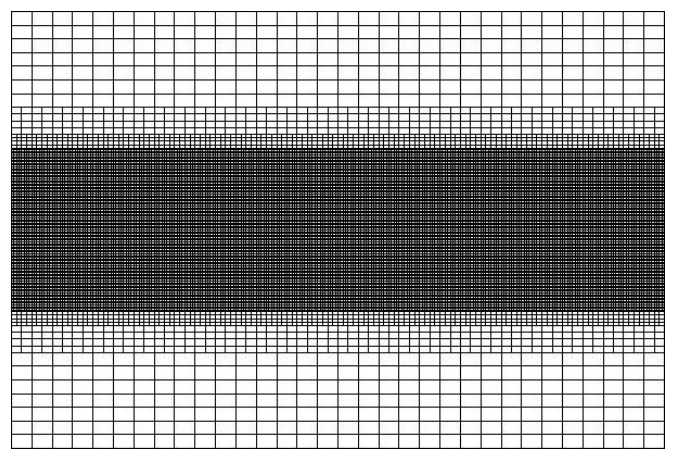}
			\vphantom{\includegraphics[scale=1,valign=t]{fig/frc/knk/knkPhi0}}
		}
	\caption{Crack propagation across kinks: (a) Initial phase field and (b) LR mesh that is locally refined a priori based on the expected crack path.}
\end{figure}

\begin{figure}[!ht]
	\centering
	\begin{tikzpicture}
		\node at (0,0) {\includegraphics[width=0.9\textwidth]{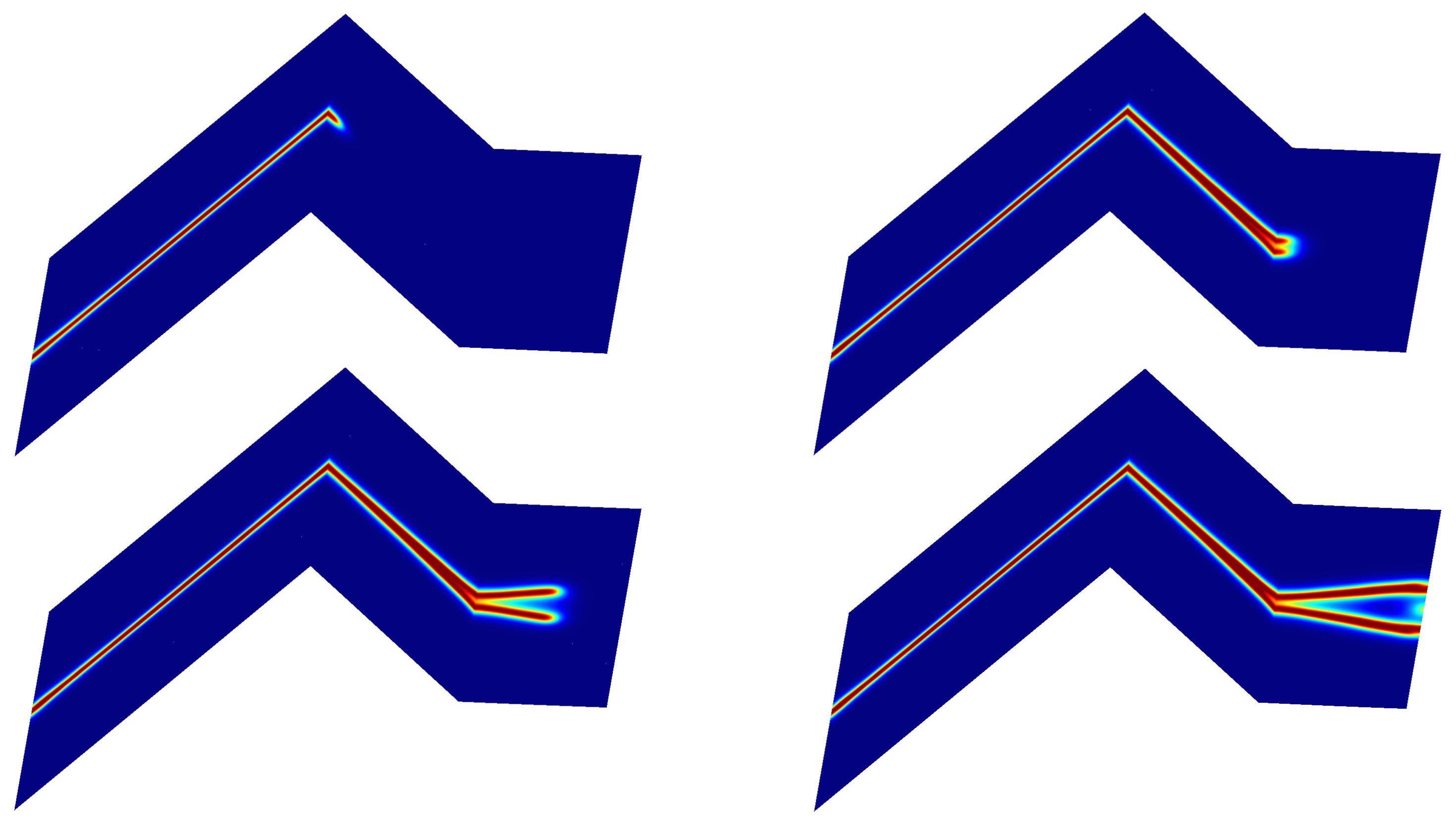}};
		\node[align=center,font=\footnotesize] at (-1.9,0.3) {$t\approx0.5893\,T_0$};
		\node[align=center,font=\footnotesize] at (6,0.3) {$t\approx0.7584\,T_0$};
		\node[align=center,font=\footnotesize] at (-1.9,-3.2) {$t\approx0.8092\,T_0$};
		\node[align=center,font=\footnotesize] at (6,-3.2) {$t\approx0.8646\,T_0$};
	\end{tikzpicture}
	\caption{Crack propagation across kinks: Crack evolution where both constraints are fulfilled via the Lagrange multiplier method with constant interpolation.} \label{f:n_f_knk_evo}
\end{figure}
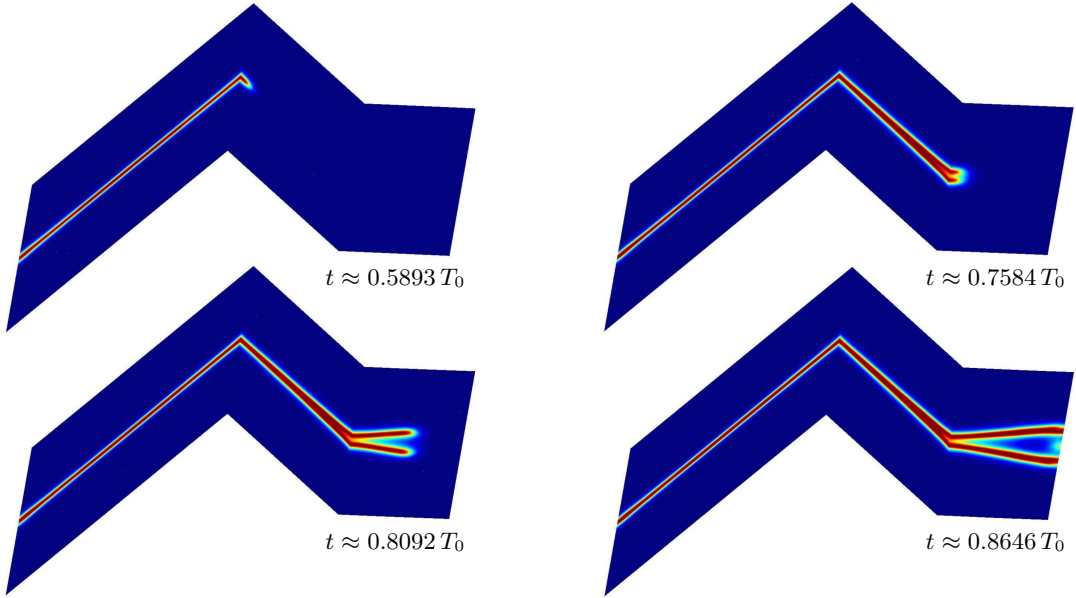

This section focuses on the general constraints for non-smooth patch connections. The initial setup with an existing initial crack is shown in \figref{f:n_f_knk_geom}--\ref{f:n_f_knk_geom3}. The mesh is locally refined a priori based on the expected crack path, see \figref{f:n_f_knk_geom4}. The refinement depth is $d=3$ \citep{paul2020}.
The material parameters are listed in Table~\ref{t:n_f_knk_mat}. A displacement is applied as shown in \figref{f:n_f_knk_geom1}. At each time step, the displacement increment $\Delta\bar{u}=\bar{v}\,\Delta t$ is imposed to obtain a constant loading velocity $\bar{v}$.
\begin{table}[!ht]
  	\caption{Crack propagation across kinks: Material parameters and loading velocity $\bar{v}$.}
	\centering
	\setlength{\tabcolsep}{8pt}
	\renewcommand{\arraystretch}{1.25}
  	\begin{tabular}{c c c c c c c }
  		$E$ $[E_0]$ & $\nu$ $[-]$ & $\sG_c$ $[E_0\,L_0]$  & $\ell_0$ $[L_0]$ & $T$ $[L_0]$  & $\bar{v}$ $[L_0\,T_0^{-1}]$
  		\\ \hline
  		$100$ & $0.3$ & $0.0005$ & $0.008$ & $0.0125$ & $0.004$
  \end{tabular}
  \label{t:n_f_knk_mat}
\end{table}
The crack evolution is illustrated in \figref{f:n_f_knk_evo} for four different time steps.
The crack starts propagating in both directions. While one end reaches the end of the sheet on the left, the right end of the crack propagates across the kinks. Crack branching occurs shortly before the crack reaches the second kink. A very similar fracture pattern is observed when the two surface kinks are smoothed and a fully $C^1$-continuous parametrization is used. If the phase field constraint is neglected, the phase field will smear at the two kinks and will slightly evolve along them, similar as in \figref{f:n_frc_sphr_pmunder}. It can be concluded that crack evolution including branching works well in the proposed framework.

\subsection{Evaluation of the constraint enforcement techniques} \label{s:n_evl}
This section briefly discusses and compares the proposed enforcement techniques. For pure shells, no significant differences between constant and linear interpolation of the Lagrange multiplier were found (using bi-quadratic NURBS for the surface representation). While requiring more implementation effort than the penalty method, the Lagrange multiplier method with element-wise constant interpolation is the safest choice, as it guarantees a stable and well conditioned system. The penalty method on the other hand is easier to implement but requires a careful choice of the penalty parameter.

\begin{figure}[!ht]
	\centering
	\begin{tikzpicture}
		\pgfplotsset{width=1\textwidth,height=0.2\textwidth,compat=newest,}
		\begin{axis}[hide axis,xmin=0,xmax=0.00001,ymin=0,ymax=0.00001,legend cell align={left},
					 legend columns=2,legend style={/tikz/every even column/.append style={column sep=2ex}}]
    			\addlegendimage{red,densely dashed,line width=1}
   			 	\addlegendentry{$n_\mathrm{qp}=1$, $\mathrm{PM}$};
  			  	\addlegendimage{red,line width=1}
  			  	\addlegendentry{$n_\mathrm{qp}=2$, Trapezoidal, $\mathrm{PM}$};
  			  	\addlegendimage{blue,densely dashed,line width=1}
 			   	\addlegendentry{$n_\mathrm{qp}=1$, constant $\mathrm{LMM}$};
  			  	\addlegendimage{blue,line width=1}
  			  	\addlegendentry{$n_\mathrm{qp}=2$, Trapezoidal, linear $\mathrm{LMM}$};
  			 \end{axis}
	\end{tikzpicture}
	\\ \vspace{-3mm}
	\subfloat[\label{f:n_evl_frcsphr1}]{%
		\begin{tikzpicture}
			\def\cdot{\times}
			\begin{axis}[xmode=log,ymode=log,grid=both,xlabel={Penalty parameter $\eps_\phi\:[E_0\,L_0^3]$},ylabel={Average condition number $[-]$},width=0.45\textwidth,xmin=0.001,xmax=1000,ymin=1e7,ymax=1e12,legend cell align={left},legend pos=north east,legend style={nodes={scale=0.75, transform shape}},tick label style={font=\footnotesize},]
				\addplot[red,densely dashed,line width=1,] table [x index = {0}, y index = {1},col sep=comma,restrict expr to domain={\coordindex}{0:6}]{fig/evl/nqp1_condNmbr.csv};
				\addplot[blue,densely dashed,line width=1,] table [x index = {0}, y index = {1},col sep=comma,restrict expr to domain={\coordindex}{7:8}]{fig/evl/nqp1_condNmbr.csv};
				\addplot[red,line width=1,] table [x index = {0}, y index = {1},col sep=comma,restrict expr to domain={\coordindex}{0:4}]{fig/evl/nqp2_condNmbr.csv};
				\addplot[blue,line width=1,] table [x index = {0}, y index = {1},col sep=comma,restrict expr to domain={\coordindex}{5:6}]{fig/evl/nqp2_condNmbr.csv};
				\draw [densely dotted,line width=1] (2.347417840375587,1e-7) -- (2.347417840375587,1e12);
			\end{axis}
		\end{tikzpicture}
	}
	\quad
	\subfloat[\label{f:n_evl_frcsphr2}]{%
		\begin{tikzpicture}
			\def\cdot{\times}
			\begin{axis}[xmode=log,grid=both,xlabel={Penalty parameter $\eps_\phi\:[E_0\,L_0^3]$},ylabel={Average number of NR iterations $[-]$},width=0.45\textwidth,xmin=0.001,xmax=1000,ymin=2.7,ymax=3.4,legend cell align={left},legend pos=south west,legend style={nodes={scale=0.75, transform shape}},tick label style={font=\footnotesize},ytick={2.7,2.8,2.9,3,3.1,3.2,3.3,3.4}]
				\addplot[red,densely dashed,line width=1,] table [x index = {0}, y index = {1},col sep=comma,restrict expr to domain={\coordindex}{0:6}]{fig/evl/nqp1_NRnmbr.csv};
				\addplot[blue,densely dashed,line width=1,] table [x index = {0}, y index = {1},col sep=comma,restrict expr to domain={\coordindex}{7:8}]{fig/evl/nqp1_NRnmbr.csv};
				\addplot[red,line width=1,] table [x index = {0}, y index = {1},col sep=comma,restrict expr to domain={\coordindex}{0:4}]{fig/evl/nqp2_NRnmbr.csv};
				\addplot[blue,line width=1,] table [x index = {0}, y index = {1},col sep=comma,restrict expr to domain={\coordindex}{5:6}]{fig/evl/nqp2_NRnmbr.csv};
				\draw [densely dotted,line width=1] (2.347417840375587,2.7) -- (2.347417840375587,3.4);
			\end{axis}
		\end{tikzpicture}
	}
	\caption{Comparison of the constraint enforcement techniques: (a) Average condition number and (b) average number of Newton-Raphson iterations for the fracturing sphere example, see Figs.~\ref{f:frc_sphr_pln}--\ref{f:n_frc_sphr_gam}. The black dotted line indicates the proposed penalty parameter from \eqsref{e:p_c_epsf}.} \label{f:n_evl_frcsphr}
\end{figure}
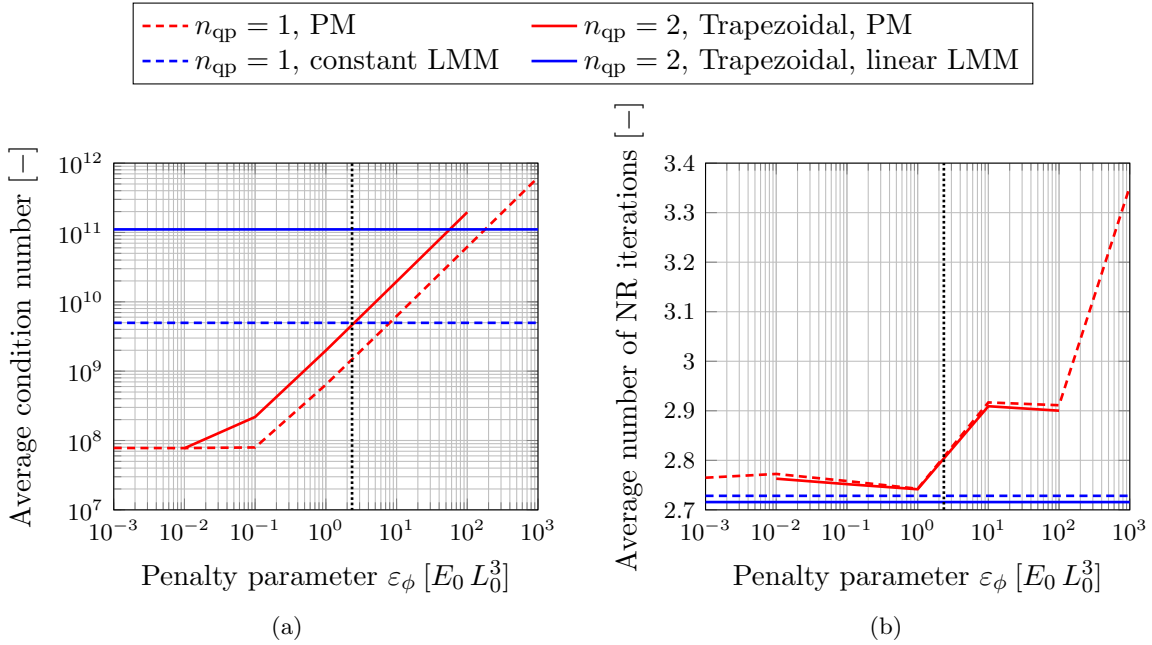

\figref{f:n_evl_frcsphr} shows the average condition number of the stiffness matrix and the average number of required Newton-Raphson iterations for the fracturing sphere example, see Figs.~\ref{f:frc_sphr_pln}--\ref{f:n_frc_sphr_gam}, over the penalty parameter $\eps_\phi$. The black dotted line marks the size of the proposed penalty parameter from \eqsref{e:p_c_epsf} ($\eps_\phi^0/\Delta x_{\min}^{\Gamma}$). \figref{f:n_evl_frcsphr1} shows that the average condition number is smaller when only one quadrature point is used for the integration along the element-wise patch interfaces. The proposed penalty parameter ensures that the condition number does not increase too much. \figref{f:n_evl_frcsphr2} shows that the Lagrange multiplier method performs better with respect to the required number of Newton-Raphson iterations. However, the difference is minor for \textit{moderate} values of the penalty parameter, especially for the proposed penalty parameter from \eqsref{e:p_c_epsf}.
As there is no noticeable difference between the results for constant and linear interpolation of the Lagrange multiplier, the authors conclude to use a constant interpolation for the coupled framework of fracture of thin shells. Further, for this case, there are no significant differences between the penalty and the Lagrange multiplier method with respect to performance (based on the data in \figref{f:n_evl_frcsphr}).

\section{Conclusion} \label{s:cncl}	
This work proposes continuity constraints for coupled fourth-order deformation and phase field models defined on isogeometric multi-patch shells. The higher continuity that stems from IGA is not automatically preserved at the patch interfaces and thus, the $G^1$- and $C^1$-continuity is recovered by enforcing patch constraints with a penalty or Lagrange multiplier approach. The interfaces between the patches are assumed to be conforming with respect to the discretization, but allow for non-smooth connections, for example to model kinks in the geometry. The constraints are fully formulated in the convective coordinate system that arises from the surface description and are thus, applicable to the majority of shell formulations.

In \secref{s:n_tot}, the purely mechanical shell framework and the coupled one of phase fields on deforming surfaces are investigated. The first is used within a quasi-static framework to study the accuracy of the $G^1$-continuity constraint on multi-patch discretizations. The proposed enforcement techniques perform well in terms of convergence rates and absolute errors. The coupled framework includes phase transitions, which are based on the Cahn-Hilliard equation, on deforming surfaces, and brittle fracture of thin shells, which is based on a higher-order phase field approach. The results for both phase field models highlight the importance of the $C^1$-continuity constraint. Comparison with fully $C^1$-continuous discretizations demonstrate the correct functioning of the proposed constraints and enforcement techniques. The examples include crack propagation across kinks showing the capability of the proposed formulation to handle non-smooth patch connections. An insufficient enforcement of the inter-patch continuity can lead to spurious stress peaks at patch interfaces. On the other hand, overconstraining can occur for the penalty method if the wrong quadrature rule is used. In this case, the phase field will deflect at patch interfaces instead of crossing it. This problem can be circumvented by using midpoint or trapezoidal quadrature. The investigations on the penalty parameter demonstrate that the proposed problem-independent penalty parameter of \eqsref{e:p_c_epsf} is a suitable choice.

Possible extensions to this work include the coupling of the $G^1$- and $C^1$-continuity constraints with each other. For example in fracture, the $G^1$-continuity may be lost in fully fractured regions. The enforcement techniques may also be extended to handle non-conforming meshes, e.g. by a modified penalty or mortar method (see \secref{s:intro}). An extension to the coupling of trimmed or unstructured splines may be of interest, especially for geometries that are created by CAD software. The present work focuses on solid materials. Liquid shells, such as lipid bilayers, that can also be described by the present shell formulation \citep{sauer2017b}, are another extension. Further, the linearization of the constraint terms can in principle be used in the framework of linear Kirchhoff-Love shells to maintain a fully linear FE formulation.

\section*{Acknowledgments}
The authors acknowledge funding by the Deutsche Forschungsgemeinschaft (DFG, German Research Foundation) – projects GSC 111 (Graduate School AICES) and 33849990/GRK2379 (IRTG Modern Inverse Problems). Simulations in \secref{s:n_f_tot} were performed with computing resources granted by RWTH Aachen University under project rwth0433. The authors also thank Deepesh Toshniwal for providing the unstructured spline discretization used in \secref{s:n_tot}.

\bigskip
\appendix
\section*{Appendix}

\section{Bending moments at patch interfaces} \label{s:b_tot}
This section derives the bending moments that are transmitted across patch interfaces for the penalty and Lagrange multiplier method. The external virtual work associated with moments is given by \citep{sauer2017a}
\eqb{l}
	\delta\Pi_\mathrm{ext}=\ds\int_{\partial\sS}\delta\bn\cdot\bM\,\dif s\,,
\eqe
with distributed moment vector $\bM$ on a cut normal to $\bnu$. The moment vector that physically acts on the element is then given by
\eqb{l}
	\bm:=\bn\times\bM\,,
\eqe
which can be decomposed into $\bm=m_\nu\,\bnu+m_\tau\,\btau$. In the following, the bending moments $m_\tau$ and $m_{\tilde\tau}$ at the patch interfaces are derived by comparing the negative virtual work $-\delta\Pi_\mrn$ with $\delta\Pi_\mathrm{ext}$. Note that in the weak form, $\delta\Pi_\mathrm{ext}$ enters with a negative sign, but $\delta\Pi_\mrn$ enters with a positive sign. Thus, $\delta\Pi_\mathrm{ext}$ needs to be compared with $-\delta\Pi_\mrn$ and not with $+\delta\Pi_\mrn$.\footnote{Note that the subsequent derivations of the bending moments have been changed compared to the journal version.}

Based on Eqs.~\eqref{e:p_crd}, \eqref{e:p_g_cos} and \eqref{e:p_g_sin} the following relations can be obtained
\eqb{l}
	\bigl(\bn\times\tilde\bn\bigr)\cdot\btau=\sin\alpha\,,\quad\mathrm{and}\quad\bigl(\bn\times\tilde\bnu\bigr)\cdot\btau=-\cos\alpha\,,\quad\mathrm{and}\quad \bigl(\tilde\bn\times\bnu\bigr)\cdot\btau=\cos\alpha\,,
\eqe
which are used in the subsequent sections.

\subsection{Penalty method} \label{s:b_pen}
Based on \eqsref{e:p_g_dpipensmpl}, the distributed moment vectors are given by
\eqb{l}
	\bM=\eps_\mrn\,\tilde\bd=\eps_\mrn\,\bigl(c_0\,\tilde\bn+s_0\,\tilde\bnu\bigr)\,,\quad\mathrm{and}\quad\tilde\bM=\eps_\mrn\,\bd=\eps_\mrn\,\bigl(c_0\,\bn+s_0\,\bnu\bigr)\,.
\eqe
The bending moment $m_\tau$ at element $\Omega^e$, which is transmitted across the patch interface, is then given by
\eqb{lll}
	m_\tau \is \bm\cdot\btau=\bigl(\bn\times\bM\bigr)\cdot\btau = \eps_\mrn\Bigl[c_0\,\bigl(\bn\times\tilde\bn\bigr)\cdot\btau+s_0\,\bigl(\bn\times\tilde\bnu\bigr)\cdot\btau\Bigr]\\[4mm]
	\is \eps_\mrn\bigl(c_0\,\sin\alpha-s_0\,\cos\alpha\bigr)=\eps_\mrn\,\sin(\alpha-\alpha_0)\,.
\eqe
In analogy, the bending moment $m_{\tilde\tau}$ at element $\tilde\Omega^e$ is given by
\eqb{lll}
	m_{\tilde\tau} \is \tilde\bm\cdot\tilde\btau=-\bigl(\tilde\bn\times\tilde\bM\bigr)\cdot\btau=-\eps_\mrn\,\Bigl[c_0\,\bigl(\tilde\bn\times\bn\bigr)\cdot\btau+s_0\,\bigl(\tilde\bn\times\bnu\bigr)\cdot\btau\Bigr]=\eps_\mrn\,\sin(\alpha-\alpha_0)\,.
\eqe

\subsection{Lagrange multiplier method} \label{s:b_lag}
Based on \eqsref{e:p_g_dpilagsmpl}, the distributed moment vectors are given by
\eqb{lll}
	\bM\is q\,\tilde\bd=q\,\bigl((s_0+c_0)\,\tilde\bn+(s_0-c_0)\,\tilde\bnu\bigr)\,,\\[2mm]
	\tilde\bM\is q\,\bd=q\,\bigl((s_0+c_0)\,\bn+(s_0-c_0)\,\bnu\bigr)\,.
\eqe
The bending moment $m_\tau$ at element $\Omega^e$, which is transmitted across the patch interface, is then given by
\eqb{lll}
	m_\tau \is \bm\cdot\btau=\bigl(\bn\times\bM\bigr)\cdot\btau = q\Bigl[(s_0+c_0)\,\bigl(\bn\times\tilde\bn\bigr)\cdot\btau+(s_0-c_0)\,\bigl(\bn\times\tilde\bnu\bigr)\cdot\btau\Bigr]\\[4mm]
	\is q\,\bigl((s_0+c_0)\,\sin\alpha-(s_0-c_0)\,\cos\alpha\bigr)=q\,\bigl(\sin(\alpha-\alpha_0\bigr)+\cos(\alpha-\alpha_0)\bigr)\,,
\eqe
which simplifies to $m_\tau=q$ since the constraint $\alpha=\alpha_0$ is fulfilled exactly for the Lagrange multiplier method. Likewise, the bending moment $m_{\tilde\tau}$ at element $\tilde\Omega^e$ is given by
\eqb{lll}
	m_{\tilde\tau} \is \tilde\bm\cdot\tilde\btau=-\bigl(\tilde\bn\times\tilde\bM\bigr)\cdot\btau = q\Bigl[-(s_0+c_0)\,\bigl(\tilde\bn\times\bn\bigr)\cdot\btau-(s_0-c_0)\,\bigl(\tilde\bn\times\bnu\bigr)\cdot\btau\Bigr]\\[4mm]
	\is q\,\bigl((s_0+c_0)\,\sin\alpha-(s_0-c_0)\,\cos\alpha\bigr)=q\,\bigl(\sin(\alpha-\alpha_0\bigr)+\cos(\alpha-\alpha_0)\bigr)\,,
\eqe
which also simplifies to $m_{\tilde\tau}=q$.

\section{Variation and linearization}\label{s:l_tot}
This section reports and derives several variations, which are required in \secref{s:p_tot}, and the linearization of the force vectors stemming from the continuity constraints in \secref{s:p_tot}. Some auxiliary variations and linearized quantities are derived in \appref{s:l_hlp}. The linearization of the force vectors for the $G^1$-continuity constraints is reported in \secref{s:l_g_tot}, and for the $C^1$-continuity constraints in \secref{s:l_c_tot}. In both cases, the penalty and Lagrange multiplier methods are considered. The linearization is reported for the most general expression of the force vectors. The linearization for the fracture model will significantly simplify, since the corresponding force vectors do not depend on $\bx$ or $\tilde\bx$, but only on $\bX$ and $\tilde\bX$. Thus, the linearization with respect to $\bx$ and $\tilde\bx$ vanishes for the brittle fracture model. Additionally, the simplification $\tilde\bnu=-\bnu$ and $\alpha=\alpha_0$ can be inserted for smooth patch connections. Here, the $G^1$-continuity constraint is not depending on any of the phase fields, so that the corresponding force vectors do not require any linearization with respect to $\phi$ or $\phitilde$.

\subsection{Variation and linearization of auxiliary quantities} \label{s:l_hlp}
The variation of \eqsref{e:p_g_pipen} requires the variations of the terms $\cos\alpha$ and $\sin\alpha$. They follow from Eqs.~\eqref{e:p_g_cos} and \eqref{e:p_g_sin} as
\eqb{l}
	\delta(\cos\alpha)=\delta\bigl(\bn\cdot\tilde\bn\bigr)=\delta\bn\cdot\tilde\bn+\bn\cdot\delta\tilde\bn\,, \label{e:p_g_dc}
\eqe
and
\eqb{lllll}
	\delta(\sin\alpha)\is\delta\Bigl(\bigl(\bn\times\tilde\bn\bigr)\cdot\btau\Bigr)=\bigl(\delta\bn\times\tilde\bn\bigr)\cdot\btau+\bigl(\bn\times\delta\tilde\bn\bigr)\cdot\btau+\bigl(\bn\times\tilde\bn\bigr)\cdot\delta\btau\\[2mm]
	\is-\bigl(\delta\bn\times\tilde\bn\bigr)\cdot\tilde\btau-\bigl(\delta\tilde\bn\times\bn\bigr)\cdot\btau+\bigl(\bn\times\tilde\bn\bigr)\cdot\delta\btau\\[2mm]
	\is-\bigl(\tilde\bn\times\tilde\btau\bigr)\cdot\delta\bn-\bigl(\bn\times\btau\bigr)\cdot\delta\tilde\bn+\bigl(\bn\times\tilde\bn\bigr)\cdot\delta\btau\\[2mm]
	\is\tilde\bnu\cdot\delta\bn+\bnu\cdot\delta\tilde\bn+\bigl(\bn\times\tilde\bn\bigr)\cdot\delta\btau\,. \label{e:p_g_ds2}
\eqe
Using Eqs.~\eqref{e:p_g_dc} and \eqref{e:p_g_ds2}, the following relations can be derived
\eqb{lll}
	\delta\bigl(\cos(\alpha-\alpha_0)\bigr) \is \delta\bigl(c_0\cos\alpha+s_0\sin\alpha\bigr) \\[2mm]
	\is c_0\Bigl(\delta\bn\cdot\tilde\bn+\bn\cdot\delta\tilde\bn\Bigr) + s_0\Bigl(\delta\btau\cdot\bigl(\bn\times\tilde\bn\bigr)+\delta\bn\cdot\tilde\bnu+\delta\tilde\bn\cdot\bnu\Bigr)\,, \label{e:p_g_dc2}
\eqe	
and
\eqb{lll}
	\delta\bigl(\sin(\alpha-\alpha_0)\bigr) \is \delta\bigl(c_0\sin\alpha-s_0\cos\alpha\bigr) \\[2mm]
	\is c_0\Bigl(\delta\btau\cdot\bigl(\bn\times\tilde\bn\bigr)+\delta\bn\cdot\tilde\bnu+\delta\tilde\bn\cdot\bnu\Bigr) - s_0\Bigl(\delta\bn\cdot\tilde\bn+\bn\cdot\delta\tilde\bn\Bigr)\,, \label{e:p_g_ds3}
\eqe
which are required in \secref{s:p_g_lag}. Note that Eqs.~\eqref{e:p_g_dc2}--\eqref{e:p_g_ds3} can be simplified using $\bn\times\tilde\bn=\sin(\alpha)\,\btau$ and $\btau\cdot\delta\btau=0$, see \eqsref{e:v_varntau}.

Further variations that are required in \secref{s:p_g_tot} are \citep{sauer2014a,duong2017}
\eqb{l}
	\delta\bn=-(\ba^\alpha\otimes\bn)\,\delta\ba_\alpha\,,\quad\delta\tilde\bn=-(\tilde\ba^\alpha\otimes\tilde\bn)\,\delta\tilde\ba_\alpha\,,\quad\delta\btau=\bM^\xi\,\delta\hat\ba_\xi\,, \label{e:v_varntau}
\eqe
with the variation of $\hat\ba_\xi$ from \eqsref{e:p_axi} and
\eqb{l}
	\bM^\xi:=\dfrac{1}{||\hat\ba_\xi||}\bigl(\bone-\btau\otimes\btau\bigr)\,. \label{e:v_ms}
\eqe

For the linearization of the force vectors in \secref{s:p_tot}, further linearizations of surface measures are required, for example \citep{sauer2017a}
\eqb{lll}
	\Delta_\mrx\,\ba^\alpha \is \bigl(a^{\alpha\beta}\,\bn\otimes\bn - \ba^\beta \otimes\ba^\alpha\bigr)\,\mN_\cb\,\Delta\mx_e\,,\\[2mm]
	\Delta_\mrx\,\bnu \is -(\btau\otimes\bnu)\,\Delta_\mrx\,\btau-(\bn\otimes\bnu)\,\Delta_\mrx\,\bn\,, \label{e:l_hlp_x1}
\eqe
with
\eqb{lll}
	\Delta_\mrx\,\bn \is -\ba^\alpha\,(\bn\cdot\Delta_\mrx\,\ba_\alpha)=-(\ba^\alpha\otimes\bn)\,\Delta_\mrx\,\ba_\alpha\,,\\[2mm]
	\Delta_\mrx\,\btau \is \bM^\xi\,\Delta_\mrx\,\hat\ba_\xi\,,  \label{e:l_hlp_x2}
\eqe
see Eqs.~\eqref{e:v_varntau}--\eqref{e:v_ms}. For simplicity, the following notation is introduced for the quantities above
\eqb{l}
	\Delta_\mrx\,\bn:=\mR\,\Delta\mx_e\,,\qquad \Delta_\mrx\,\ba^\alpha := \boldsymbol{\mathcal{A}}^{\alpha}\Delta\mx_e\,,\quad\mathrm{and}\quad\Delta_\mrx\,\bnu := \mV_1\,\Delta\mx_e+\mV_2\,\Delta\hat\mx_e\,, \label{e:l_hlp_x3}
\eqe
with the definitions
\eqb{lll}
	\mR\dis -\bigl(\ba^\alpha\otimes\bn\bigr)\,\mN_\ca\,,\\[2mm]
	 \boldsymbol{\mathcal{A}}^{\alpha} \dis \bigl(a^{\alpha\beta}\,\bn\otimes\bn - \ba^\beta \otimes\ba^\alpha\bigr)\,\mN_\cb\,,\\[2mm]
	\mV_1 \dis \nu^\alpha\,(\bn\otimes\bn)\,\mN_\ca\,,\\[2mm]
	\mV_2 \dis -(\btau\otimes\bnu)\,\bM^\xi\,\hat\mN_{,\xi}.\label{e:l_hlp_x4}
\eqe
Note that the first three quantities in \eqsref{e:l_hlp_x4} have dimension $3\times3\,n_\mathrm{e}$, while $\mV_2$ has dimension $3\times\hat{n}_\mathrm{CP}$. Here, $n_\mathrm{e}$ denotes the number of control points associated with the surface element $\Omega^e$, and $\hat{n}_\mathrm{CP}$ denotes the number of control points along the line element $\Gamma^e$. The quantity $\Delta\mx_e$ has dimension $3\,n_\mathrm{e}\times1$, and $\Delta\hat\mx_e$ has dimension $3\,\hat{n}_\mathrm{CP}\times1$ (see Eqs.~\eqref{e:p_axi}--\eqref{e:s_shpNxi}). The corresponding auxiliary variables $\tilde\mR$, $\tilde{\boldsymbol{\mathcal{A}}}^{\alpha}$, $\tilde\mV_1$ and $\tilde\mV_2$ follow in analogy to \eqsref{e:l_hlp_x4}, i.e.
\eqb{lll}
	\tilde\mR\dis -\bigl(\tilde\ba^\alpha\otimes\tilde\bn\bigr)\,\tilde\mN_\ca\,,\\[2mm]
 	\tilde{\boldsymbol{\mathcal{A}}}^{\alpha} \dis \bigl(\tilde a^{\alpha\beta}\,\tilde\bn\otimes\tilde\bn - \tilde\ba^\beta \otimes\tilde\ba^\alpha\bigr)\,\tilde\mN_\cb\,,\\[2mm]
 	\tilde\mV_1 \dis \tilde\nu^\alpha\,(\tilde\bn\otimes\tilde\bn)\,\tilde\mN_\ca\,,\\[2mm] 	
 	\tilde\mV_2 \dis -(\tilde\btau\otimes\tilde\bnu)\,\bM^{\xi}\,\hat\mN_{,\xi}\,, \label{e:l_hlp_x5}
\eqe
so that $\Delta_{\tilde\mrx}\,\tilde\bn:=\tilde\mR\,\Delta\tilde\mx_e$, $\Delta_{\tilde\mrx}\,\tilde\ba^\alpha:=\tilde{\boldsymbol{\mathcal{A}}}^{\alpha}\Delta\tilde\mx_e$ and $\Delta_{\tilde\mrx}\,\tilde\bnu:=\tilde\mV_1\,\Delta\tilde\mx_e+\tilde\mV_2\,\Delta\hat\mx_e$. Note that the simplifications $\tilde\btau=-\btau$ and $\tilde{s}=-s$ can be inserted, see \figref{f:p_vcs}.

The individual linearizations of the Lagrange multiplier $\lambda$ (see \secref{s:p_c_lag}) are
\eqb{l}
	\Delta_\phi\,\lambda=0\,,\quad\Delta_\phitilde\,\lambda=0\,,\quad\Delta_\mrx\,\lambda=0\,,\quad\Delta_\mrxtilde\,\lambda=0\,,\quad\Delta_\lambda\,\lambda=\hat\mN_\lambda\,\Delta\hat\blam_e\,, \label{e:l_hlp_l}
\eqe
with shape function array $\hat\mN_\lambda$ of dimension $1\times n_\lambda$ and corresponding linearized nodal values $\Delta\hat\blam_e$ of size $n_\lambda\times1$, see \eqsref{e:p_shpNlag}. The variable $n_\lambda$ depends on the order of approximation of the Lagrange multiplier, see \figref{f:p_css}. The linearizations of the Lagrange multiplier $q$ (see \secref{s:p_g_lag}) follow in analogy.

Based on Eqs.~\eqref{e:l_hlp_x3}--\eqref{e:l_hlp_l}, the individual linearizations of the $C^1$-continuity constraint in \eqsref{e:p_c_cnstr4} are given by
\eqb{lll}
	\Delta_\phi\, g_\nabla \is (\ba^\alpha\cdot\bnu)\,\bar\mN_\ca\,\Delta\bphi_e\,,\\[2mm]
	\Delta_\phitilde\,g_\nabla \is (\tilde\ba^\alpha\cdot\tilde\bnu)\,\tilde{\bar\mN}_\ca\,\Delta\tilde\bphi_e\,,\\[2mm]
	\Delta_\mrx\,g_\nabla \is \phi_\ca\,\bnu\cdot\boldsymbol{\mathcal{A}}^{\alpha}\Delta\mx_e\,,\\[2mm]
	\Delta_\mrxtilde\,g_\nabla \is \phitilde_\ca\,\tilde\bnu\cdot\tilde{\boldsymbol{\mathcal{A}}}^{\alpha}\Delta\tilde\mx_e\,,\\[2mm]
	\Delta_\lambda\,g_\nabla \is 0\,, \\[2mm]
	\Delta_{\hat\mrx}\,g_\nabla \is \Bigl[\phi_\ca\,\ba^\alpha\cdot\mV_2+\phitilde_\ca\,\tilde\ba^\alpha\cdot\tilde\mV_2\Bigr]\,\Delta\hat\mx_e\,. \label{e:l_hlp_cnstr}
\eqe

Additionally, the linearization of the terms $(\ba^\alpha\cdot\bnu)$ and $(\tilde\ba^\alpha\cdot\tilde\bnu)$ will be required, since both appear in the force vectors in Eqs.~\eqref{e:p_c_fpen} and \eqref{e:p_c_flag}. The linearizations are given by
\eqb{lllllll}
	\Delta_\phi\,(\ba^\alpha\cdot\bnu) &\!\!\!= 0\,, \qquad
	\Delta_\phitilde\,(\ba^\alpha\cdot\bnu) &\!\!\!= 0\,, \qquad
	\Delta_\mrx\,(\ba^\alpha\cdot\bnu) &\!\!\!= \bnu\cdot\boldsymbol{\mathcal{A}}^{\alpha}\Delta\mx_e\,,\\[2mm]
	\Delta_\mrxtilde\,(\ba^\alpha\cdot\bnu) &\!\!\!= 0\,, \qquad
	\Delta_\lambda\,(\ba^\alpha\cdot\bnu) &\!\!\!= 0\,, \qquad
	\Delta_{\hat\mrx}\,(\ba^\alpha\cdot\bnu) &\!\!\!= \ba^\alpha\cdot\mV_2\,\Delta\hat\mx_e\,,
	 \label{e:l_hlp_1}
\eqe
and
\eqb{rlrlrl}
	\Delta_\phi\,(\tilde\ba^\alpha\cdot\tilde\bnu) &\!\!\!= 0\,, \qquad
	\Delta_\phitilde\,(\tilde\ba^\alpha\cdot\tilde\bnu) = 0\,,& \quad
	\Delta_\mrx\,(\tilde\ba^\alpha\cdot\tilde\bnu) &\!\!\!= 0\,, \qquad
	\Delta_\lambda\,(\tilde\ba^\alpha\cdot\tilde\bnu) = 0\,, \\[2mm]
	\Delta_\mrxtilde\,(\tilde\ba^\alpha\cdot\tilde\bnu) &\!\!\!=\tilde\bnu\cdot\tilde{\boldsymbol{\mathcal{A}}}^{\alpha}\,\Delta\tilde\mx_e\,, &
	\Delta_{\hat\mrx}\,(\tilde\ba^\alpha\cdot\tilde\bnu) &\!\!\!=\tilde\ba^\alpha\cdot\tilde\mV_2\,\Delta\hat\mx_e\,, \label{e:l_hlp_2}
\eqe
Further, the definitions
\eqb{lll}
	\mW^\alpha:=\bnu\cdot\boldsymbol{\mathcal{A}}^{\alpha}=-\nu^\alpha\,\ba^\beta\cdot\mN_\cb\,,\qquad\tilde\mW^\alpha:=\tilde\bnu\cdot\tilde{\boldsymbol{\mathcal{A}}}^{\alpha}=-\tilde\nu^\alpha\,\tilde\ba^\beta\cdot\tilde\mN_\cb\,,
\eqe
are used to simplify the notation, such that $\Delta_\mrx\,(\ba^\alpha\cdot\bnu)=\mW^\alpha\Delta\mx_e$ and $\Delta_\mrxtilde\,(\tilde\ba^\alpha\cdot\tilde\bnu)=\tilde\mW^\alpha\Delta\tilde\mx_e$.

\subsection{Linearization for the \texorpdfstring{$G^1$}{G1}-continuity constraint}\label{s:l_g_tot}
This section reports the linearization of the force vectors that are used to enforce the $G^1$-continuity constraint, see \secref{s:p_g_tot}.

\subsubsection{Penalty method} \label{s:l_g_pen}
The force vectors in \eqsref{e:p_g_fpen} do not depend on the phase field. Thus, only their linearization with respect to $\bx$, $\tilde\bx$ and $\hat\bx$ is required. They are given by
\eqb{lll}
	\Delta_\mrx\,\mf_\mrn^e\is\ds\int_{\Gamma_0^e}\eps_\mrn\,\mN_\ca^\mrT\,\Bigl[\bn\,\bigl(\tilde\bd^{\,\mrT}\boldsymbol{\mathcal{A}}^{\alpha}\bigr)+\bigl(\tilde\bd\cdot\ba^\alpha\bigr)\,\mR\,\Bigr]\,\dif S\,\Delta\mx_e\,,\\[4mm]
	\Delta_{\tilde\mrx}\,\mf_\mrn^e\is\ds\int_{\Gamma_0^e}\eps_\mrn\,\mN_\ca^\mrT\,\bn\,\Bigl[\bigl(\ba^\alpha\bigr)^{\!\mrT}\bigl(c_0\,\tilde\mR+s_0\,\tilde\mV_1\bigr)\Bigr]\,\dif S\,\Delta\tilde\mx_e\,, \\[4mm]
	\Delta_{\hat\mrx}\,\mf_\mrn^e\is\ds\int_{\Gamma_0^e}\eps_\mrn\,\mN_\ca^\mrT\,s_0\,\bn\,\bigl(\ba^\alpha\bigr)^{\!\mrT}\,\tilde\mV_2\,\dif S\,\Delta\hat\mx_e\,, \label{e:l_g_pen_f1}
\eqe
and
\eqb{lll}
	\Delta_\mrx\,\mf_{\tilde\mrn}^e\is\ds\int_{\Gamma_0^e}\eps_\mrn\,\tilde\mN_\ca^\mrT\,\tilde\bn\,\Bigl[\bigl(\tilde\ba^\alpha\bigr)^{\!\mrT}\bigl(c_0\,\mR+s_0\,\mV_1\bigr)\Bigr]\,\dif S\,\Delta\mx_e\,,\\[4mm]
	\Delta_{\tilde\mrx}\,\mf_{\tilde\mrn}^e\is\ds\int_{\Gamma_0^e}\eps_\mrn\,\tilde\mN_\ca^\mrT\,\Bigl[\tilde\bn\,\bigl(\bd^{\,\mrT}\tilde{\boldsymbol{\mathcal{A}}}^{\alpha}\bigr)+\bigl(\bd\cdot\tilde\ba^\alpha\bigr)\,\tilde\mR\,\Bigr]\,\dif S\,\Delta\tilde\mx_e\,, \\[4mm]
		\Delta_{\hat\mrx}\,\mf_{\tilde\mrn}^e\is\ds\int_{\Gamma_0^e}\eps_\mrn\,\tilde\mN_\ca^\mrT\,s_0\,\tilde\bn\,\bigl(\tilde\ba^\alpha\bigr)^{\!\mrT}\,\mV_2\,\dif S\,\Delta\hat\mx_e\,,\\[4mm]
 \label{e:l_g_pen_f2}
\eqe
with $\bd$ and $\tilde\bd$ given in Eqs.~(\ref{e:p_g_hlp}.2)--(\ref{e:p_g_hlp}.3).

\subsubsection{Lagrange multiplier method} \label{s:l_g_lag}
The force vectors in \eqsref{e:p_g_flag} only depend on $\bx$, $\tilde\bx$ and $q$. The corresponding linearizations are given by
\eqb{lll}
	\Delta_\mrx\,\bar\mf_\mrn^e\is\ds\int_{\Gamma_0^e}q\,\mN_\ca^\mrT\,\Bigl[\bn\,\bigl(\tilde\bd^{\,\mrT}\boldsymbol{\mathcal{A}}^{\alpha}\bigr)+\bigl(\tilde\bd\cdot\ba^\alpha\bigr)\,\mR\,\Bigr]\,\dif S\,\Delta\mx_e\,,\\[4mm]
	\Delta_{\tilde\mrx}\,\bar\mf_\mrn^e\is\ds\int_{\Gamma_0^e}q\,\mN_\ca^\mrT\,\bn\,\Bigl[\bigl(\ba^\alpha\bigr)^{\!\mrT}\bigl((s_0+c_0)\,\tilde\mR+(s_0-c_0)\,\tilde\mV_1\bigr)\Bigr]\,\dif S\,\Delta\tilde\mx_e\,,\\[4mm]
	\Delta_{\hat\mrx}\,\bar\mf_\mrn^e\is\ds\int_{\Gamma_0^e}q\,\mN_\ca^\mrT\,(s_0-c_0)\,\bn\,\bigl(\ba^\alpha\bigr)^{\!\mrT}\,\tilde\mV_2\,\dif S\,\Delta\hat\mx_e\,,\\[4mm]
	\Delta_\mrq\,\bar\mf_\mrn^e\is\ds\int_{\Gamma_0^e}\mN_\ca^\mrT\,\bigl(\tilde\bd\cdot\ba^\alpha\bigr)\,\bn\,\hat\mN_\mrq\,\dif S\,\Delta\hat\mq_e\,, \label{e:l_g_lag_f1}
\eqe
and
\eqb{lll}
	\Delta_\mrx\,\bar\mf_{\tilde\mrn}^e\is\ds\int_{\Gamma_0^e}q\,\tilde\mN_\ca^\mrT\,\tilde\bn\,\Bigl[\bigl(\tilde\ba^\alpha\bigr)^{\!\mrT}\bigl((s_0+c_0)\,\mR+(s_0-c_0)\,\mV_1\bigr)\Bigr]\,\dif S\,\Delta\mx_e\,,\\[4mm]
	\Delta_{\tilde\mrx}\,\bar\mf_{\tilde\mrn}^e\is\ds\int_{\Gamma_0^e}q\,\tilde\mN_\ca^\mrT\,\Bigl[\tilde\bn\,\bigl(\bd^{\,\mrT}\tilde{\boldsymbol{\mathcal{A}}}^{\alpha}\bigr)+\bigl(\bd\cdot\tilde\ba^\alpha\bigr)\,\tilde\mR\,\Bigr]\,\dif S\,\Delta\tilde\mx_e\,,\\[4mm]
		\Delta_{\hat\mrx}\,\bar\mf_{\tilde\mrn}^e\is\ds\int_{\Gamma_0^e}q\,\tilde\mN_\ca^\mrT\,(s_0-c_0)\,\tilde\bn\,\bigl(\tilde\ba^\alpha\bigr)^\mrT\,\mV_2\,\dif S\,\Delta\hat\mx_e\,,\\[4mm]
	\Delta_\mrq\,\bar\mf_{\tilde\mrn}^e\is\ds\int_{\Gamma_0^e}\tilde\mN_\ca^\mrT\,\bigl(\bd\cdot\tilde\ba^\alpha\bigr)\,\tilde\bn\,\hat\mN_\mrq\,\dif S\,\Delta\hat\mq_e\,, \label{e:l_g_lag_f2}
\eqe
and based on Eqs.~\eqref{e:p_g_dc2}--\eqref{e:p_g_ds3},
\eqb{lll}
	\Delta_\mrx\,\bar\mf_\mrq^e\is\ds\int_{\Gamma_0^e}\hat\mN_\mrq^\mrT\,\Bigl[-c_0\,\bigl(\tilde\bn+\tilde\bnu\bigr)+s_0\,\bigl(\tilde\bn-\tilde\bnu\bigr)\Bigr]^\mrT\mR\,\dif S\,\Delta\mx_e\,,\\[4mm]
	\Delta_{\tilde\mrx}\,\bar\mf_\mrq^e\is\ds\int_{\Gamma_0^e}\hat\mN_\mrq^\mrT\,\Bigl[-c_0\,\bigl(\bn+\bnu\bigr)+s_0\,\bigl(\bn-\bnu\bigr)\Bigr]^\mrT\tilde\mR\,\dif S\,\Delta\tilde\mx_e\,,\\[4mm]
	\Delta_\mrq\,\bar\mf_\mrq^e\is\boldsymbol{0}\,, \label{e:l_g_lag_f3}
\eqe
with $\bd$ and $\tilde\bd$ given in Eqs.~(\ref{e:p_g_hlp2}.2)--(\ref{e:p_g_hlp2}.3).

\subsection{Linearization for the \texorpdfstring{$C^1$}{C1}-continuity constraint}\label{s:l_c_tot}
This section reports the linearization of the force vectors that are used to enforce the $C^1$-continuity constraint, see \secref{s:p_c_tot}.

\subsubsection{Penalty method}\label{s:l_c_pen}
This section shows the linearization of the force vectors $\mf_\phi^e$ and $\mf_\phitilde^e$ given in Eq.~\eqref{e:p_c_fpen}.  Their contributions follow from
\eqb{l}
	\Delta_\bullet\,\mf_\phi^e = \ds\int_{\Gamma^e_0}\eps_\phi\,\bar\mN_\ca^\mrT\,\Delta_\bullet\Bigl(g_\nabla\,\bigl(\ba^\alpha\cdot\bnu\bigr)\Bigr)\,\dif S\,,\quad\mathrm{and}\quad
	\Delta_\bullet\,\mf_\phitilde^e = \ds\int_{\Gamma^e_0}\eps_\phi\,\tilde{\bar\mN}_\ca^\mrT\,\Delta_\bullet\Bigl(g_\nabla\,\bigl(\tilde\ba^\alpha\cdot\tilde\bnu\bigr)\Bigr)\,\dif S\,.
\eqe
These are given by
\eqb{lll}
	\Delta_\phi\,\mf_\phi^e         &= \ds\int_{\Gamma^e_0}\eps_\phi\,\bar\mN_\ca^\mrT(\ba^\alpha\cdot\bnu)\,(\ba^\beta\cdot\bnu)\,\bar\mN_\cb\,\dif S\,\Delta\bphi_e\,,\\[4mm]
	\Delta_\phitilde\,\mf_\phi^e  &= \ds\int_{\Gamma^e_0}\eps_\phi\,\bar\mN_\ca^\mrT(\ba^\alpha\cdot\bnu)\,(\tilde\ba^\beta\cdot\tilde\bnu)\,\tilde{\bar\mN}_\cb\,\dif S\,\Delta\tilde\bphi_e\,,\\[4mm]
	\Delta_\mrx\,\mf_\phi^e        &= \ds\int_{\Gamma^e_0}\eps_\phi\,\bar\mN_\ca^\mrT\Bigl[\bigl(\ba^\alpha\cdot\bnu\bigr)\,\phi_\cb\,\mW^\beta+g_\nabla\,\mW^\alpha\Bigr]\,\dif S\,\Delta\mx_e\,,\\[4mm]
	\Delta_{\tilde\mrx}\,\mf_\phi^e        &= \ds\int_{\Gamma^e_0}\eps_\phi\,\bar\mN_\ca^\mrT\bigl(\ba^\alpha\cdot\bnu\bigr)\,\phitilde_\cb\,\tilde\mW^\beta\,\dif S\,\Delta\tilde\mx_e\,, \\[4mm]
	\Delta_{\hat\mrx}\,\mf_\phi^e        &= \ds\int_{\Gamma^e_0}\eps_\phi\,\bar\mN_\ca^\mrT\biggl[\bigl(\ba^\alpha\cdot\bnu\bigr)\Bigl(\phi_\cb\,\bigl(\ba^\beta\bigr)^{\!\mrT}\,\mV_2+\phitilde_\cb\,\bigl(\tilde\ba^\beta\bigr)^{\!\mrT}\,\tilde\mV_2\Bigr)+g_\nabla\,\bigl(\ba^\alpha\bigr)^{\!\mrT}\,\mV_2\biggr]\dif S\,\Delta\hat\mx_e\,, \label{e:l_pen_f1}
\eqe
and
\eqb{lll}
	\Delta_\phi\,\mf_\phitilde^e         &= \ds\int_{\Gamma^e_0}\eps_\phi\,\tilde{\bar\mN}_\ca^\mrT(\tilde\ba^\alpha\cdot\tilde\bnu)\,(\ba^\beta\cdot\bnu)\,\bar\mN_\cb\,\dif S\,\Delta\bphi_e\,,\\[4mm]
	\Delta_\phitilde\,\mf_\phitilde^e  &= \ds\int_{\Gamma^e_0}\eps_\phi\,\tilde{\bar\mN}_\ca^\mrT(\tilde\ba^\alpha\cdot\tilde\bnu)\,(\tilde\ba^\beta\cdot\tilde\bnu)\,\tilde{\bar\mN}_\cb\,\dif S\,\Delta\tilde\bphi_e\,,\\[4mm]
	\Delta_\mrx\,\mf_\phitilde^e		   &= \ds\int_{\Gamma^e_0}\eps_\phi\,\tilde{\bar\mN}_\ca^\mrT\bigl(\tilde\ba^\alpha\cdot\tilde\bnu\bigr)\,\phi_\cb\,\mW^\beta\,\dif S\,\Delta\mx_e\,,\\[4mm]
	\Delta_\mrxtilde\,\mf_\phitilde^e &= \ds\int_{\Gamma^e_0}\eps_\phi\,\tilde{\bar\mN}_\ca^\mrT\Bigl[\bigl(\tilde\ba^\alpha\cdot\tilde\bnu\bigr)\,\phitilde_\cb\,\tilde\mW^\beta+g_\nabla\,\tilde\mW^\alpha\Bigr]\,\dif S\,\Delta\tilde\mx_e\,, \\[4mm]
	\Delta_{\hat\mrx}\,\mf_\phitilde^e &= \ds\int_{\Gamma^e_0}\eps_\phi\,\tilde{\bar\mN}_\ca^\mrT\biggl[\bigl(\tilde\ba^\alpha\cdot\tilde\bnu\bigr)\Bigl(\phi_\cb\,\bigl(\ba^\beta\bigr)^{\!\mrT}\,\mV_2+\phitilde_\cb\,\bigl(\tilde\ba^\beta\bigr)^{\!\mrT}\,\tilde\mV_2\Bigr)+g_\nabla\,\bigl(\tilde\ba^\alpha\bigr)^{\!\mrT}\,\tilde\mV_2\biggr]\,\dif S\,\Delta\hat\mx_e\,. \label{e:l_pen_f2}
\eqe
Note that the linearizations in Eqs.~(\ref{e:l_pen_f1}.3)--(\ref{e:l_pen_f1}.5) and Eqs.~(\ref{e:l_pen_f2}.3)--(\ref{e:l_pen_f2}.5) vanish for the brittle fracture model. The element tangent matrix will have the following form
\eqb{l}
	\begin{tikzpicture}
		\matrix [matrix of math nodes,left delimiter={[},right delimiter={]},column sep=1.5em] (m)
			{
				\ds\pa{\mf_\phi^e}{\mx_e} & \ds\pa{\mf_\phi^e}{\tilde{\mx}_e} & \ds\pa{\mf_\phi^e}{\hat{\mx}_e} & \ds\pa{\mf_\phi^e}{\bphi_e} & \ds\pa{\mf_\phi^e}{\tilde{\bphi}_e} \\
				\ds\pa{\mf_\phitilde^e}{\mx_e} & \ds\pa{\mf_\phitilde^e}{\tilde{\mx}_e} & \ds\pa{\mf_\phitilde^e}{\hat{\mx}_e} & \ds\pa{\mf_\phitilde^e}{\bphi_e} & \ds\pa{\mf_\phitilde^e}{\tilde{\bphi}_e} \\
			};
		\draw[dashdotted] (m-1-1.north west) -- (m-1-3.north east) -- (m-2-3.south east) -- (m-2-1.south west) -- (m-1-1.north west);
	\end{tikzpicture}\,.
\eqe
The dot-dashed framed entries vanish for the brittle fracture model. The entries $\partial\mf_\phi^e/\partial\tilde\bphi_e$ and $\partial\mf_{\tilde\phi}^e/\partial\bphi_e$ are the transpose of each other.

\subsubsection{Lagrange multiplier method}\label{s:l_c_lag}
The contributions to the linearizations of the force vectors in \eqsref{e:p_c_flag} for the Lagrange multiplier method are computed from
\eqb{lll}
	\Delta_\bullet\,\bar\mf_\phi^e \is			\ds\int_{\Gamma^e_0} \bar\mN_\ca^\mrT\,\Delta_\bullet\Bigl(\lambda\,\bigl(\ba^\alpha\cdot\bnu\bigr)\Bigr)\,\dif S\,,\\[4mm]
	\Delta_\bullet\,\bar\mf_\phitilde^e \is	\ds\int_{\Gamma^e_0} \tilde{\bar\mN}_\ca^\mrT\,\Delta_\bullet\Bigl(\lambda\,\bigl(\tilde{\ba}^\alpha\cdot\tilde\bnu\bigr)\Bigr)\,\dif S\,,\\[4mm]
	\Delta_\bullet\,\bar\mf_\lambda^e \is	\ds\int_{\Gamma^e_0}\hat\mN_\lambda^\mrT\,\Delta_\bullet\,g_\nabla\,\dif S\,.
\label{e:linfch}\eqe
They are given by
\eqb{lll}
	\Delta_\phi\,\bar\mf_\phi^e         \is \boldsymbol{0}\,, \qquad\qquad
	\Delta_\phitilde\,\bar\mf_\phi^e  = \boldsymbol{0}\,,  \qquad\qquad
	\Delta_\mrx\,\bar\mf_\phi^e        = \ds\int_{\Gamma^e_0}\bar\mN_\ca^\mrT\,\lambda\,\mW^\alpha\,\dif S\,\Delta\mx_e\,,\\[4mm]
	\Delta_\mrxtilde\,\bar\mf_\phi^e \is \boldsymbol{0}\,, \qquad\qquad
	\Delta_{\hat\mrx}\,\bar\mf_\phi^e        = \ds\int_{\Gamma^e_0}\bar\mN_\ca^\mrT\,\lambda\,\bigl(\ba^\alpha\bigr)^{\!\mrT}\,\mV_2\,\dif S\,\Delta\hat\mx_e\,,\\[4mm]
	\Delta_\lambda\,\bar\mf_\phi^e \is \ds\int_{\Gamma^e_0}\bar\mN_\ca^\mrT\,\bigl(\ba^\alpha\cdot\bnu\bigr)\,\hat\mN_\lambda\,\dif S\,\Delta\hat\blam_e \, \label{e:l_lag_f1}
\eqe
and
\eqb{lllllll}
	\Delta_\phi\,\bar\mf_{\tilde\phi}^e         \is \boldsymbol{0}\,,\qquad\qquad
	\Delta_\phitilde\,\bar\mf_{\tilde\phi}^e  = \boldsymbol{0}\,,\qquad\qquad
	\Delta_\mrx\,\bar\mf_{\tilde\phi}^e        = \boldsymbol{0}\,,\\[4mm]
	\Delta_\mrxtilde\,\bar\mf_{\tilde\phi}^e \is \ds\int_{\Gamma^e_0}\tilde{\bar\mN}_\ca^\mrT\,\lambda\,\tilde\mW^\alpha\,\dif S\,\Delta\tilde\mx_e\,,\qquad\qquad
	\Delta_{\hat\mrx}\,\bar\mf_{\tilde\phi}^e = \ds\int_{\Gamma^e_0}\tilde{\bar\mN}_\ca^\mrT\,\lambda\,\bigl(\tilde\ba^\alpha\bigr)^{\!\mrT}\,\tilde\mV_2\,\dif S\,\Delta\hat\mx_e\,,\\[4mm]
	\Delta_\lambda\,\bar\mf_{\tilde\phi}^e  \is \ds\int_{\Gamma^e_0}\tilde{\bar\mN}_\ca^\mrT\,(\tilde\ba^\alpha\cdot\tilde\bnu)\,\hat\mN_\lambda\,\dif S\,\Delta\hat\blam_e\,, \label{e:l_lag_f2}
\eqe
and
\eqb{lll}
	\Delta_\phi\,\bar\mf_\lambda^e         \is \ds\int_{\Gamma^e_0}\hat\mN_\lambda^\mrT\,(\ba^\alpha\cdot\bnu)\,\bar\mN_\ca\,\dif S\,\Delta\bphi_e\,,\\[4mm]
	\Delta_\phitilde\,\bar\mf_\lambda^e  \is \ds\int_{\Gamma^e_0}\hat\mN_\lambda^\mrT\,(\tilde\ba^\alpha\cdot\tilde\bnu)\,\tilde{\bar\mN}_\ca\,\dif S\,\Delta\tilde\bphi_e\,,\\[4mm]
	\Delta_\mrx\,\bar\mf_\lambda^e        \is \ds\int_{\Gamma^e_0}\hat\mN_\lambda^\mrT\,\phi_\ca\,\mW^\alpha\,\dif S\,\Delta\mx_e\,,\\[4mm]
	\Delta_\mrxtilde\,\bar\mf_\lambda^e \is \ds\int_{\Gamma^e_0}\hat\mN_\lambda^\mrT\,\phitilde_\ca\,\tilde\mW^\alpha\,\dif S\,\Delta\tilde\mx_e\,,\\[4mm]
	\Delta_{\hat\mrx}\,\bar\mf_\lambda^e \is \ds\int_{\Gamma^e_0}\hat\mN_\lambda^\mrT\Bigl[\phi_\ca\,\bigl(\ba^\alpha\bigr)^{\!\mrT}\,\mV_2+\phitilde_\ca\,\bigl(\tilde\ba^\alpha\bigr)^{\!\mrT}\,\tilde\mV_2\Bigr]\,\dif S\,\Delta\hat\mx_e\,,\\[4mm]
	\Delta_\lambda\,\bar\mf_\lambda^e  \is \boldsymbol{0}\,. \label{e:l_lag_f3}
\eqe
Note that the linearizations in Eqs.~(\ref{e:l_lag_f1}.3)--(\ref{e:l_lag_f1}.5), Eqs.~(\ref{e:l_lag_f2}.3)--(\ref{e:l_lag_f2}.5) and Eqs.~(\ref{e:l_lag_f3}.3)--(\ref{e:l_lag_f3}.5) vanish for the brittle fracture model. The full element tangent matrix has the following form
\eqb{l}
	\begin{tikzpicture}
		\matrix [matrix of math nodes,left delimiter={[},right delimiter={]},column sep=1.5em] (m)
			{
				\ds\pa{\bar\mf_\phi^e}{\mx_e} & \hspace{2.4mm}\mathbf{0}\vphantom{\ds\pa{\bar\mf_\phi^e}{\mx_e}}\hspace{2.4mm} & \ds\pa{\bar\mf_\phi^e}{\hat\mx_e} & \mathbf{0}& \mathbf{0} & \ds\pa{\bar\mf_\phi^e }{\blam_e} \\
				\mathbf{0} & \ds\pa{\bar\mf_\phitilde^e }{\tilde{\mx}_e} & \ds\pa{\bar\mf_\phitilde^e }{\hat{\mx}_e} & \mathbf{0} & \mathbf{0} & \ds\pa{\bar\mf_\phitilde^e }{\blam_e} \\
				\ds\pa{\bar\mf_\lambda^e}{\mx_e\vphantom{\vphantom{\partial\tilde{\bphi}_e}}} & \ds\pa{\bar\mf_\lambda^e }{\tilde{\mx}_e\vphantom{\partial\tilde{\bphi}_e}} & \ds\pa{\bar\mf_\lambda^e}{\hat{\mx}_e\vphantom{\partial\tilde{\bphi}_e}} & \ds\pa{\bar\mf_\lambda^e }{\bphi_e\vphantom{\partial\tilde{\bphi}_e}} & \ds\pa{\bar\mf_\lambda^e}{\tilde{\bphi}_e\vphantom{\partial\tilde{\bphi}_e}} & \mathbf{0} \\
			};
		\draw[solid] (m-1-6.north west) -- (m-1-6.north east) -- (m-2-6.south east) -- (m-2-6.south west) -- (m-1-6.north west);
		\draw[solid] (m-3-4.north west) -- (m-3-5.north east) -- (m-3-5.south east) -- (m-3-4.south west) -- (m-3-4.north west);
		\draw[dashdotted] (m-1-1.north west) -- (m-1-3.north east) -- (m-3-3.south east) -- (m-3-1.south west) -- (m-1-1.north west);
	\end{tikzpicture}\,.
\eqe
Note that one solid framed block is the transpose of the other solid framed block, while the dot-dashed framed entries vanish for the fracture model.

\section{Comparison of the six-patch and unstructured spline discretization of a sphere}\label{s:sphr_tot}
\begin{figure}[!ht]
	\centering
		\begin{tikzpicture}
			\def\cdot{\times}
			\begin{axis}[xmode=log,ymode=log,grid=both,xlabel={Number of control points $n_\mathrm{CP}$ $[-]$},ylabel={Error $\epsilon_{L^2}$ $[-]$},width=0.8\textwidth,height=0.4\textwidth,
			xmin=50,xmax=1e5,ymin=1e-5,ymax=1e-1,legend cell align={left},legend pos = north east,legend style={nodes={scale=1, transform shape},cells={align=left}},tick label style={font=\footnotesize},
		]
				\addplot[red,line width=1,mark=*,mark size=2]table [x index = {0}, y index = {2},col sep=comma]{fig/l/usErrs.csv};
				\addlegendentry{Unstructured splines};
				\addplot[blue,line width=1,mark=triangle*,mark size=2]table [x index = {0}, y index = {2},col sep=comma]{fig/l/sxptchErrs.csv};
				\addlegendentry{Six-patch discretization};
			\end{axis}
		\end{tikzpicture}
	\caption{Construction of the spherical six-patch discretization: Comparison of the $L^2$-error, defined in \eqsref{e:l_sphr_l2err}, between the six-patch and unstructured spline discretization over the mesh refinement.} \label{f:l_sphr_l2err}
\end{figure}
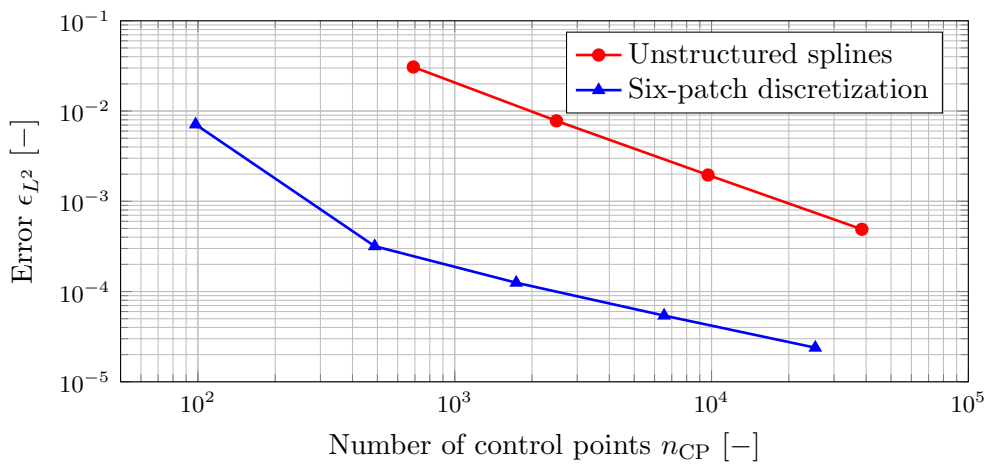

The following error is defined in order to assess the accuracy of the six-patch and unstructured spline discretizations of a sphere,
\eqb{lll}
	\epsilon_{L^2} \dis \ds\dfrac{1}{R^2}\sqrt{\int_{\sS_0}\normadjust[\Big]{\normadjust[\big]{\bX-\bX_0}_2-R}_2^2\,\dif S}\,. \label{e:l_sphr_l2err}
\eqe
Here, $R$ is the radius and $\bX_0$ is the sphere's origin. \figref{f:l_sphr_l2err} shows the $L^2$-error, defined in \eqsref{e:l_sphr_l2err}, over the mesh refinement. The error decays for both approaches as the number of control points is increased. The absolute error of the six-patch discretization is more than one order of magnitude smaller compared to the unstructured spline discretization. However, the six-patch sphere is only $C^0$-continuous at patch interfaces.

\bigskip
\bibliographystyle{apalike}
\bibliography{continuity_constraints}

\end{document}

%% file: neco.tex
\newcommand{\secref}[1]{Sec.~\ref{#1}}
\newcommand{\eqsref}[1]{Eq.~\eqref{#1}}
\newcommand{\figref}[1]{Fig.~\ref{#1}}
\newcommand{\tabref}[1]{Table~\ref{#1}}
\newcommand{\appref}[1]{Appendix~\ref{#1}}

\newcommand{\lr}[1]{\left(#1\right)}

\newcommand{\ab}{{\alpha\beta}}
\newcommand{\ag}{{\alpha\gamma}}

\newcommand{\btad}{{\beta\delta}}

\newcommand{\gd}{{\gamma\delta}}

\newcommand{\phitilde}{{\tilde{\phi}}}

\newcommand{\ca}{{,\alpha}}
\newcommand{\cb}{{,\beta}}

\newcommand{\mrxtilde}{{\tilde{\mrx}}}

\newcommand{\mrT}{\mathrm{T}}

\newcommand {\mrB}{\mathrm{B}}

\newcommand {\mrS}{\mathrm{S}}

\newcommand{\nablas}{\nabla_{\!\!\mrs}}
\newcommand{\nablao}{\nabla_{\!\mrS}}

\newcommand {\eqb}[1]{\begin{equation}\begin{array}{#1}}
\newcommand {\eqe}{\end{array}\end{equation}}

\newcommand {\esb}[1]{\begin{equation*}\begin{array}{#1}}
\newcommand {\ese}{\end{array}\end{equation*}}
\newcommand {\ds}{\displaystyle}

\newcommand {\df}[2]{\frac{\ds#1}{\ds#2}}
\newcommand {\pa}[2]{\frac{\partial{#1}}{\partial{#2}}}

\newcommand {\back}{\! \! \!}
\newcommand {\is}{\back &=& \back}
\newcommand {\dis}{\back &:=& \back}

\newcommand {\norm}[1]{\|#1\|}

\newcommand {\dif}{\mathrm{d}}

\newcommand {\II}{{I\kern-.3em I}}
\newcommand {\III}{{I\kern-.3em I\kern-.3em I}}

\newcommand {\mrb}{\mathrm{b}}
\newcommand {\mrc}{\mathrm{c}}

\newcommand {\mri}{\mathrm{i}}

\newcommand {\mrn}{\mathrm{n}}

\newcommand {\mrq}{\mathrm{q}}
\newcommand {\mrr}{\mathrm{r}}
\newcommand {\mrs}{\mathrm{s}}

\newcommand {\mrx}{\mathrm{x}}

\newcommand {\mf}{\mathbf{f}}

\newcommand {\mk}{\mathbf{k}}

\newcommand {\mm}{\mathbf{m}}

\newcommand {\mq}{\mathbf{q}}

\newcommand {\mx}{\mathbf{x}}

\newcommand {\ba}{\boldsymbol{a}}

\newcommand {\bd}{\boldsymbol{d}}

\newcommand {\bff}{\boldsymbol{f}}
\newcommand {\bg}{\boldsymbol{g}}

\newcommand {\bm}{\boldsymbol{m}}
\newcommand {\bn}{\boldsymbol{n}}

\newcommand {\bv}{\boldsymbol{v}}

\newcommand {\bx}{\boldsymbol{x}}

\newcommand {\blam}{\mbox{\boldmath$\lambda$}}
\newcommand {\bnu}{\mbox{\boldmath$\nu$}}

\newcommand {\bphi}{\mbox{\boldmath$\phi$}}

\newcommand {\btheta}{\mbox{\boldmath$\theta$}}

\newcommand {\mM}{\mathbf{M}}
\newcommand {\mN}{\mathbf{N}}

\newcommand {\mR}{\mathbf{R}}

\newcommand {\mV}{\mathbf{V}}
\newcommand {\mW}{\mathbf{W}}

\newcommand {\bA}{\boldsymbol{A}}

\newcommand {\bG}{\boldsymbol{G}}

\newcommand {\bK}{\boldsymbol{K}}

\newcommand {\bM}{\boldsymbol{M}}
\newcommand {\bN}{\boldsymbol{N}}

\newcommand {\bX}{\boldsymbol{X}}

\newcommand {\eps}{\varepsilon}
\newcommand {\sig}{\sigma}

\newcommand {\btau}{\mbox{\boldmath$\tau$}}

\newcommand {\bone}{\mathbf{1}}

\newcommand {\IR}{{\rm\kern.24em
   \vrule width.02em height1.53ex depth-.05ex
   \kern-.3em R}}
\newcommand {\ic}{{\rm\kern.20em
   \vrule width.02em height1.0ex depth-.05ex
   \kern-.22em c}}
\newcommand {\ia}{{\rm\kern.20em
   \vrule width.02em height1.05ex depth-.0ex
   \kern-.25em a}}
\newcommand {\IC}{{\rm\kern.24em
   \vrule width.02em height1.4ex depth-.05ex
   \kern-.26em C}}
\newcommand {\ID}{{\rm\kern.34em
   \vrule width.02em height1.5ex depth-.05ex
   \kern-.36em D}}
\newcommand {\IS}{{\rm\kern.24em
   \vrule width.02em height1.6ex depth.05ex
   \kern-.26em S}}
\newcommand {\IT}{{\rm\kern.50em
   \vrule width.02em height1.55ex depth-.05ex
   \kern-.52em T}}

\newcommand {\IE}{{\rm\kern.24em
   \vrule width.02em height1.55ex depth-.05ex
   \kern-.33em E}}
\newcommand {\IEa}{{\rm\kern.24em
   \vrule width.02em height1.55ex depth-.05ex
   \kern-.33em E}^{1}_{ijkl}}
\newcommand {\IEb}{{\rm\kern.24em
   \vrule width.02em height1.55ex depth-.05ex
   \kern-.33em E}^{2}_{ijkl}}

\newcommand {\sG}{\mathcal{G}}
\newcommand {\sH}{\mathcal{H}}

\newcommand {\sS}{\mathcal{S}}

\newcommand {\sU}{\mathcal{U}}
\newcommand {\sV}{\mathcal{V}}

\newcommand {\Ass}[2]{\kern 0.9ex \vrule width0.45em height0.2ex depth0ex \kern -2.1ex \bigwedge_{#1}^{#2}}
\newcommand {\ASS}[2]{\kern 1.45ex \vrule width0.5em height0.2ex depth0ex \kern -2.65ex \bigwedge_{#1}^{#2}}

